\documentclass[final,5p,times,twocolumn,nofootinbib,nopreprintline]{elsarticle}
\usepackage{amsmath,amssymb,amsfonts}
\usepackage{graphicx}
\usepackage{hyperref}
\usepackage{slashed}
\usepackage{booktabs,tabulary}
\usepackage{mciteplus}
\usepackage{color}
\usepackage{soul}

\usepackage{bibentry}
\newcommand{\ignore}[1]{}
\newcommand{\nobibentry}[1]{{\let\nocite\ignore\bibentry{#1}}}

\makeatletter
\def\bibinfo@X@title#1,{\ignorespaces}
\makeatother

\begin{document}

\begin{frontmatter}

\title{Data-driven dispersive analysis of the $\pi \pi$ and $\pi K$ scattering}
\author{Igor Danilkin}
\author{Oleksandra Deineka}
\author{Marc Vanderhaeghen}
\address{Institut f\"ur Kernphysik \& PRISMA$^+$  Cluster of Excellence, Johannes Gutenberg Universit\"at,  D-55099 Mainz, Germany}

\begin{abstract}
We present a data-driven analysis of the resonant S-wave $\pi\pi \to \pi\pi$ and $\pi K \to \pi K$ reactions using the partial-wave dispersion relation. The contributions from the left-hand cuts are accounted for using the Taylor expansion in a suitably constructed conformal variable. The fits are performed to experimental and lattice data as well as Roy analyses. For the $\pi\pi$ scattering we present both a single- and coupled-channel analysis by including additionally the $K\bar{K}$ channel. For the latter the central result is the Omn\`es matrix, which is consistent with the most recent Roy and Roy-Steiner results on $\pi\pi \to \pi\pi$ and $\pi\pi \to K\bar{K}$, respectively. By the analytic continuation to the complex plane, we found poles associated with the lightest scalar resonances $\sigma/f_0(500)$, $f_0(980)$, and $\kappa/K_0^*(700)$ for the physical pion mass value and in the case of $\sigma/f_0(500)$, $\kappa/K_0^*(700)$ also for unphysical pion mass values. 

\end{abstract}
\end{frontmatter}

\section{Introduction}\label{intro}

There is a renewed interest in the hadron spectroscopy, motivated by recent discoveries of unexpected exotic hadron resonances \cite{Aaij:2020fnh, Aaij:2019vzc,*Aaij:2015tga,Adolph:2014rpp}. Currently, LHCb, BESIII, and COMPASS collected data with unprecedented statistics, BELLE-II and GlueX just started to operate,  and more facilities are planned in the near future, such as PANDA and EIC. Besides, lattice QCD has been applied to a broad range of hadron processes and recently was able to calculate the lowest excitation spectrum with the masses of the light quarks near their physical values \cite{Briceno:2017max,*Shepherd:2016dni}.

To correctly identify resonance parameters one has to search for poles in the complex plane. This is particularly important when there is an interplay between several inelastic channels or when the pole is lying very deep in the complex plane. In these cases, the structure of the resonance is quite different from a typical Breit-Wigner behaviour. In order to determine the pole position of the resonance, one has to analytically continue the amplitude to the unphysical Riemann sheets. At this stage, the right theoretical framework has to be applied. The latter should satisfy the main principles of the S-matrix theory, namely unitarity, analyticity, and crossing symmetry. These constraints were successfully incorporated in the set of Roy or Roy-Steiner equations \cite{Roy:1971tc,*Hite:1973pm}. In a practical application, however, the rigorous implementation of these equations is almost impossible, since it requires experimental knowledge of all partial waves in the direct channel and all channels related by crossing. Therefore, the current precision studies of $\pi\pi$ \cite{Pelaez:2015qba,GarciaMartin:2011jx,*GarciaMartin:2011cn,*Pelaez:2019eqa,Ananthanarayan:2000ht,*Caprini:2005zr,*Leutwyler:2008xd,Colangelo:2001df} and $\pi K$ \cite{Buettiker:2003pp,*DescotesGenon:2006uk, Pelaez:2020uiw,*Pelaez:2020gnd} scattering are based on a finite truncation, which in turn limits the results to a given kinematic region, and require a large experimental data basis. Furthermore, applying Roy-like equations for coupled-channel cases is quite complicated and has not been achieved in the literature so far. Because of the above-mentioned difficulties, in the experimental analyses, it is a common practice to ignore the S-matrix constraints and rely on simple parameterizations. The most used ones are a superposition of Breit-Wigner resonances or the K-matrix approach. The latter implements unitarity, but ignores the existence of the left-hand cut and often leads to spurious poles in the complex plane.

A good alternative to the K-matrix approach and a complementary method to Roy analysis is the so-called $N/D$ technique \cite{Chew:1960iv}, which is based on the partial-wave dispersion relations. In this method, the dominant constraints of resonance scattering, such as unitarity and analyticity are implemented exactly. Since the time it was introduced by Chew and Mandelstam \cite{Chew:1960iv}, the $N/D$ method has been extensively studied for different processes \cite{Oller:1998zr, Szczepaniak:2010re,*Guo:2010gx, Gasparyan:2010xz,*Danilkin:2010xd,*Gasparyan:2011yw,*Gasparyan:2012km, Danilkin:2011fz,*Danilkin:2012ap}. The required input to solve the $N/D$ equation are the discontinuities along the left-hand cut, which are typically approximated one way or another using chiral perturbation theory ($\chi$PT). In the present paper, we extend the ideas of \cite{Gasparyan:2010xz,*Danilkin:2010xd,*Gasparyan:2011yw,*Gasparyan:2012km}, where the left-hand cut contributions were approximated using an expansion in powers of a suitably chosen conformal variable. In contrast to \cite{Danilkin:2011fz,*Danilkin:2012ap}, however, we follow here a data-driven approach and adjust the unknown coefficients in the expansion scheme to empirical data directly. In this way, the model dependence is avoided, and the method can also be applied to the reactions which do not include Goldstone bosons, like for instance the $J/\psi J/\psi$ scattering \cite{Aaij:2020fnh}.

In this paper, we apply the $N/D$ method to the resonant $\pi\pi$ and $\pi K$ scattering in the S-wave. There are three main reasons for this choice:

\begin{itemize} 
\item The system of two pions (or pion and kaon) shows up very often as a part of the final state of many hadronic interactions and therefore serves as input in various theoretical or experimental data analyses, like e.g. $\eta\to 3\pi$ \cite{Guo:2015zqa,*Guo:2016wsi,Colangelo:2016jmc,*Colangelo:2018jxw,Albaladejo:2017hhj}, $\eta'\to \pi\pi\eta$ \cite{Isken:2017dkw,Gonzalez-Solis:2018xnw,Gan:2020aco}, $\gamma\gamma \to \pi\pi$ \cite{GarciaMartin:2010cw,Hoferichter:2011wk,Dai:2014zta}, $e^+e^-\to J/\psi(\psi')\,\pi\pi$ \cite{Molnar:2019uos,Chen:2019mgp,Danilkin:2020kce} or $D\to K \pi\pi$ \cite{Niecknig:2015ija,*Niecknig:2017ylb}.

\item Even though the $\pi\pi \to \pi\pi$ (and to a lesser extent $\pi K \to \pi K$ and $\pi\pi \to K\bar{K}$) amplitudes are known very well from the Roy (Roy-Steiner) analyses \cite{GarciaMartin:2011jx,*GarciaMartin:2011cn,*Pelaez:2019eqa, Ananthanarayan:2000ht,*Caprini:2005zr,*Leutwyler:2008xd,Colangelo:2001df,Buettiker:2003pp,*DescotesGenon:2006uk,Pelaez:2020uiw,*Pelaez:2020gnd, Pelaez:2018qny}, in the practical dispersive applications the final state interactions (FSI) are implemented with the help of the so-called Omn\`es function, which does not have left-hand cuts. Indeed, the left-hand cuts are different for each production/decay mechanism, while the unitarity makes a connection between the production/decay and the scattering amplitudes only on the right-hand cut. In the $N/D$ ansatz, the Omn\`es functions come out naturally, as the inverse of the $D$-functions.

\item Recently, it has become possible to calculate $\pi\pi$ and $\pi K$ scattering using lattice QCD with almost physical masses \cite{Lang:2012sv,*Prelovsek:2010kg, Briceno:2016mjc, Liu:2016cba, Fu:2017apw, Guo:2018zss,*Mai:2019pqr, Wilson:2019wfr, Rendon:2020rtw}. Since, both the $\sigma/f_0(500)$ and $\kappa/K_0^*(700)$ states lie deep in the complex plane, the reliable extraction of their properties requires the use of the formalism that goes beyond the simple $K$-matrix parametrization and incorporates in addition the analyticity constraint. 
\end{itemize}

The paper is organized as follows. In the next section, we focus on the formalism that we adopt in this paper. We start with the review of the $N/D$ method in Sec \ref{subsec:N/D method}. We then  discuss the left-hand cut  contributions in Sec \ref{subsec:Left-hand cuts}. In Sec \ref{subsec:Omnes} we make the connection to the Omn\`es functions. The numerical results are presented in Sec. \ref{sec:Numerical results}. We start with $I=0$, $\pi \pi$ single-channel analysis of both experimental and lattice data, which is followed by the coupled-channel $\{\pi\pi, K\bar{K} \}$ analysis of the experimental data. These results are then used to determine the two-photon coupling of $\sigma/f_0(500)$ and $f_0(980)$. At the very end, we focus on $\pi K$, $I=1/2$ scattering of both experimental and lattice data. A summary and outlook is presented in Sec. \ref{sec:Conclusion and outlook}.

\section{Formalism}
\subsection{N/D method}
\label{subsec:N/D method}

The $s$-channel partial-wave decomposition for $2\to 2$ process is given by
\begin{equation}\label{p.w.expansion}
T_{ab}(s,t)={\cal N}_{ab}\,\sum_{J=0}^{\infty}(2J+1)\,t_{ab}^{(J)}(s)\,P_J(\cos\theta)\,,
\end{equation}
where $\theta$ is the c.m. scattering angle and $ab$ are the coupled-channel indices with $a$ and $b$ standing for the initial and final state, respectively. For the following discussion, we focus only on the S-wave $(J=0)$ and therefore will suppress the label $(J)$. The different normalization factors (${\cal N}_{\pi\pi \pi\pi}=2$, ${\cal N}_{\pi\pi K\bar{K}}=\sqrt{2}$ and ${\cal N}_{K\bar{K} K\bar{K}}={\cal N}_{\pi K \pi K}=1$) are needed to ensure that the unitarity condition for identical and non-identical two-particle states are the same and can be written in the matrix form as
\begin{align}\label{Eq:Unitarity}
\text{Disc}\,t_{ab}(s)&\equiv\frac{1}{2i}\left(t_{ab}(s+i \epsilon)-t_{ab}(s-i \epsilon)\right)\nonumber\\
&=\sum_{c} t_{ac}(s)\,\rho_{c}(s)\,t^*_{cb}(s)\,,
\end{align}
where the sum goes over all intermediate states. The phase space factor $\rho_{c}(s)$ in Eq. (\ref{Eq:Unitarity}) is given by 
\begin{align}\label{Eq:rho}
\rho_{c}(s)&=\frac{1}{8\pi}\frac{p_{c}(s)}{\sqrt{s}}\,\theta(s-s_{th})\,,
\end{align}
with $p_{c}(s)$ and $s_{th}$ being the center-of-mass three momentum and threshold of the corresponding two-meson system. Within the maximal analyticity assumption \cite{Mandelstam:1958xc,*Mandelstam:1959bc}, the partial-wave amplitudes satisfy the dispersive representation
\begin{equation}\label{DR_0}
t_{ab}(s)=\int_{-\infty}^{s_L}\frac{d s'}{\pi}\frac{\text{Disc } t_{ab}(s')}{s'-s} + \int_{s_{th}}^{\infty}\frac{d s'}{\pi}\frac{\text{Disc } t_{ab}(s')}{s'-s}\,,
\end{equation}
where $s_L$ is the position of the closest left-hand cut singularity and the discontinuity along the right-hand cut is given by (\ref{Eq:Unitarity}). For unequal masses, as in $\pi K$ scattering, the left-hand singularities of the partial-wave amplitude do not all lie on the real axis and the integration in the first term in Eq. (\ref{DR_0}) goes partly along the circle. We note, that the separation into left and right-hand cuts given in (\ref{DR_0}) is only possible for the systems where no anomalous thresholds are present \cite{Mandelstam:1960zz, Lutz:2018kaz}.

The unitarity condition (\ref{Eq:Unitarity}) guarantees that the partial-wave amplitudes at infinity approach at most constants. In accordance with that, we can make one subtraction in Eq. (\ref{DR_0}) to suppress the high-energy contribution under the dispersive integrals. Thus we rewrite Eq. (\ref{DR_0}) as
\begin{equation}\label{DR_1}
t_{ab}(s)=U_{ab}(s) + \frac{s-s_M}{\pi} \int_{s_{th}}^{\infty}\frac{d s'}{s'-s_M}\frac{\text{Disc } t_{ab}(s')}{s'-s} \,,
\end{equation}
where we combined the subtraction constant together with the left-hand cut contributions into the function $U_{ab}(s)$. The choice of the subtraction point $s_M$ will be discussed later. The solution to (\ref{DR_1}) can be written using the $N/D$ ansatz \cite{Chew:1960iv}
\begin{equation}\label{N/D}
t_{ab}(s)=\sum_c D^{-1}_{ac}(s)\,N_{cb}(s)\,,
\end{equation}
where the contributions of left- and right-hand cuts are separated into $N(s)$ and $D(s)$ functions, respectively. The discontinuity relation along the right-hand cut $\text{Disc}\, D_{ab}(s)=-N_{ab}(s)\rho_{b}(s)\,$ allows us to write a dispersive representation for the $D$-function, which up to a Castillejo-Dalitz-Dyson (CDD) ambiguity \cite{Castillejo:1955ed}\footnote{For detailed discussion of the CDD ambiguity in the $N/D$ context we refer the reader to \cite{Szczepaniak:2010re,*Guo:2010gx,Danilkin:2011fz,*Danilkin:2012ap,Oller:2019opk,*Oller:2018zts,*Guo:2013rpa}} is given by
\begin{align}\label{D-fun}
D_{ab}(s)=&\delta_{ab}- \frac{s-s_M}{\pi} \int_{s_{th}}^{\infty}\frac{d s'}{s'-s_M}\frac{N_{ab}(s')\,\rho_{b}(s')}{s'-s}\,.
\end{align}
Due to the non-uniqueness of the $N/D$ ansatz, we have normalized the $D$-function in Eq. (\ref{D-fun}) such that $D_{ab}(s_M)=\delta_{ab}$. Since $D_{ab}(s)$ is a complex matrix above the threshold, the position of $s_M$ has to be chosen such that all of its elements are real at this point, \textit{i.e.} $s_M \leq s_{th}$. To arrive at an integral equation for the $N(s)$ function, one can write a once-subtracted dispersion relation for $\sum_c D_{ac}(s)\,(t(s)-U(s))_{cb}$ and fix its subtraction constant by requiring that
\begin{align}
t_{ab}(s_M)=U_{ab}(s_M),
\end{align}
which follows from Eq. (\ref{DR_1}). As a result, it yields \cite{Luming:1964, *Johnson:1979jy}
\begin{align}\label{N-fun}
N_{ab}(s)&=U_{ab}(s)+ \\
&\frac{s-s_M}{\pi} \sum_{c} \int_{s_{th}}^{\infty}\frac{d s'}{s'-s_M}\frac{N_{ac}(s')\,\rho_{c}(s')\,(U_{cb}(s')-U_{cb}(s))}{s'-s}\,. \nonumber
\end{align}
The above integral equation can be solved numerically given the input of $U_{ab}(s)$. Knowing the $N_{ab}(s)$ function on the right-hand cut, the $D_{ab}(s)$ function is calculated by (\ref{D-fun}) and finally the partial-wave amplitude is produced with Eq. (\ref{N/D}). In other words, if the discontinuities across all the left-hand cuts were known\footnote{in that case the subtraction constant is probably unnecessary to introduce.} the exact solution can be obtained by $N/D$ method. An important property of Eq. (\ref{N-fun}) is that the input of $U(s)$ is only needed on the right-hand cut. In the case of many channels, both the diagonal and off-diagonal t-matrix elements have a right-hand cut starting at the lowest threshold $s_{th}$. However, only the input of the off-diagonal $U_{ab}(s)$ is required outside the physical region, while in order to solve (\ref{N-fun}), the input of the diagonal $U_{aa}(s)$ is needed in the physical region due to the phase space factor. It has a direct relevance for the $\{\pi \pi,K\bar{K}\}$ case, where in the $K\bar{K}\to K\bar{K}$ channel the overlap of left- and right-hand cuts happens, but only in the non-physical region, $4m_\pi^2<s<4(m_K^2-m_\pi^2)$, and therefore does not require any modifications of the dispersion integrals. We also emphasize that by means of Eq. (\ref{N/D}), the scattering amplitude can be rigorously continued into the complex plane, where one can determine pole parameters of the resonances. In our convention the scattering amplitude in the vicinity of the poles on the unphysical Riemann sheets (or physical Riemann sheet in the case of the bound state) is given by,
\begin{align}
    {\cal N}_{ab}\,t_{ab}(s)\simeq \frac{g_{pa}\,g_{pb}}{s_p-s}\,,
\end{align}
where the normalization factor ${\cal N}_{ab}$ comes from Eq. (\ref{p.w.expansion}) and $g_{pi}$ denotes the coupling of the pole at $s=s_p$ to the channel $i=a,b$.

We wish to comment on the case when there is a bound state in the system, since it happens for the relatively large unphysical pion masses. To find the binding energy $s_B$, one searches for a zero of the determinant of the $D_{ab}$ matrix for energies below threshold,
\begin{align}
\det(D_{ab}(s_B))=0,\quad s_B<s_{th}\,.
\end{align}
In this case, the solution obtained using the set of $N/D$ equations (\ref{N/D}) with input from (\ref{ConfExpansion}) satisfies the dispersion relation (\ref{DR_1}) combined with the bound state term,
\begin{align}\label{BoundState}
t_{ab}(s)=&U_{ab}(s) +\frac{s-s_M}{s_B-s_M}\frac{g_{Ba}\,g_{Bb}}{s_B-s}\nonumber\\
&+ \frac{s-s_M}{\pi} \int_{s_{th}}^{\infty}\frac{d s'}{s'-s_M}\frac{\text{Disc } t_{ab}(s')}{s'-s} \,.
\end{align}
At the same time, it is straightforward to show that including such a bound state term into the definition of $U_{ab}(s)$ does not change the solution of (\ref{N/D}) or the integral equation (\ref{N-fun}), provided that the residues $g_{Ba}\,g_{Bb}$ are dialed properly using the $\det(D_{ab}(s_B))=0$ condition.

\subsection{Left-hand cuts}
\label{subsec:Left-hand cuts}

\begin{figure}[t]
\centering
\includegraphics[width =0.45\textwidth]{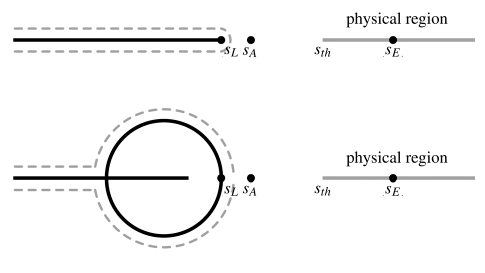}
\caption{Left-hand cut singularities (solid black curves) in the complex $s$-plane for the $\pi\pi \to \pi\pi$ (a) and $\pi K \to \pi K$ (b) scattering. In the plot we schematically show the position of the closest left-hand cut singularity ($s_L$), Adler zero ($s_A$), threshold ($s_{th}$) and the expansion point ($s_E$). Dashed lines determine the specific form of the conformal map and subsequently the domain of convergence of the conformal expansion in Eq. (\ref{ConfExpansion}).}
\label{Fig:LeftHandCutStructure}
\end{figure}

In a general scattering problem, little is known about the left-hand cuts, except their analytic structure in the complex plane. The progress has been made in \cite{Gasparyan:2010xz,*Danilkin:2010xd,*Gasparyan:2011yw,*Gasparyan:2012km}, by considering an analytic continuation of $U_{ab}(s)$ to the physical region, which is needed as input to Eq.~(\ref{N-fun}), by means of an expansion in a suitably contracted conformal mapping variable $\xi(s)$, 
\begin{equation}\label{ConfExpansion}
U(s)= \sum_{n=0}^\infty C_{n}\,\xi^n(s)\,,
\end{equation}
which is chosen such that it maps the left-hand cut plane onto the unit circle \cite{Frazer:1961zz}. The form of $\xi(s)$ depends on the cut structure of the reaction (i.e. $\{ab\}$) and specified by the position of the closest left-hand cut branching point ($s_L$) and an expansion point ($s_E$) around which the series is expanded, $\xi(s_E)=0$. Since for the $\{\pi\pi, K\bar{K}\}$ system all the left-hand cuts lie on the real axis, $-\infty<s<s_L$, one can use a simple function
\begin{equation}\label{xi-1}
\xi(s)=\frac{\sqrt{s-s_L}-\sqrt{s_E-s_L}}{\sqrt{s-s_L}+\sqrt{s_E-s_L}}\,,
\end{equation}
where $s_L(\pi\pi\to\pi\pi)=s_L(\pi\pi\to K\bar{K})=0$ and $s_L(K\bar{K}\to K\bar{K})=4\,(m_K^2-m_\pi^2)$. 
For the case of $\pi K \to \pi K$, the left-hand cut structure is a bit more complicated (see Fig.~\ref{Fig:LeftHandCutStructure}). In addition to the left-hand cut lying on the real axis $-\infty < s < (m_K - m_\pi)^2 $, there is a circular cut at $|s|=m_K^2 - m_\pi^2$. The conformal map that meets these requirements is defined as
\begin{equation}\label{xi-2}
\xi(s)=-\frac{(\sqrt{s}-\sqrt{s_E})(\sqrt{s} \sqrt{s_E}+s_L)}{(\sqrt{s}+\sqrt{s_E}) (\sqrt{s} \sqrt{s_E}-s_L)}\,,
\end{equation}
where $s_L(\pi K\to\pi K)=m_K^2 - m_\pi^2$. We note that, given the forms of $\xi(s)$ in Eqs.~(\ref{xi-1}) and (\ref{xi-2}), the series (\ref{ConfExpansion}) truncated at any finite order is bounded asymptotically. This is consistent with the assigned asymptotic behavior of $U(s)$ in the once-subtracted dispersion relation (\ref{DR_1}).

For reactions involving Goldstone bosons, in principle, $\chi$PT allows to calculate the amplitude over a finite portion of the closest left-hand cut and can be used to estimate $C_n$ in (\ref{ConfExpansion}) as it has been done for other processes in \cite{Gasparyan:2010xz,*Danilkin:2010xd,*Gasparyan:2011yw,*Gasparyan:2012km, Danilkin:2011fz,*Danilkin:2012ap}. However, it is not clear at which point $\chi$PT calculated to a given order still represents a good approximation. In addition to that, in order to merge the conformal expansion with the chiral expansion, the expansion point $s_E$ should lie within the region where $\chi$PT can be computed safely. For instance, for the elastic $\pi\pi \to \pi\pi$ scattering the natural choice would be to identify $s_E$ with the two-pion threshold. However in that case, the last data point, which can be described with the elastic unitarity, corresponds to $\xi(s^{1/2}_{max}=0.7\,\text{GeV}) \simeq 0.45$. On the other side, the faster convergence of the sum in Eq. (\ref{ConfExpansion}) can be achieved for the choice of $s_E$ in between the threshold and $s_{max}$, i.e. in the regions where $\chi$PT is at the limit of its applicability. Besides, for the coupled-channel case, one needs to rely on SU(3) $\chi$PT, which  converges slower than SU(2) version of it.

In our paper, we determine the unknown $C_n$ in Eq. (\ref{ConfExpansion}) and the optimal positions of $s_E$ directly from the data and use $\chi$PT results only as constraints for the scattering lengths, slope parameters, and Adler zero values. We note that the latter brings a stringent constraint on the scattering amplitude, since for both $\pi\pi$ and $\pi K$ scattering the Adler zero is located very close to the left-hand cut (see Fig.\ref{Fig:LeftHandCutStructure}), and cannot be determined precisely from the fit to the data. However, once the Adler zero is imposed as a constraint, it improves drastically the convergence of (\ref{ConfExpansion}) in the threshold region.

\subsection{Relation to the Omn\`es function}
\label{subsec:Omnes}

The unitarity connects the partial-wave amplitudes in production (or decay) and scattering processes. Therefore, the reactions like $\gamma p \to \pi\pi p$, $\gamma\gamma \to \pi\pi$, $J/\psi \to \pi\pi \gamma$, $\eta \to 3\pi$, etc. are very sensitive to the FSI. In a dispersive formalism, FSI are typically implemented with the help of
the so-called Omn\`es function \cite{Omnes:1958hv,*Muskhelishvili-book}, $\Omega_{ab}(s)$, that fulfills the following unitarity relation on the right-hand cut 
\begin{equation}\label{Omnes_Disc}
\text{Disc}\,\Omega_{ab}(s)=\sum_{c} t_{ac}^*(s)\,\rho_{c}(s)\,\Omega_{cb}(s)\,,
\end{equation}
and analytic everywhere else in the complex plane, i.e. it satisfies a once-subtracted dispersion relation
\begin{align}\label{Omnes_DR}
\Omega_{ab}(s)=&\delta_{ab}+ \frac{s-s_M}{\pi} \int_{s_{th}}^{\infty}\frac{d s'}{s'-s_M}\frac{\text{Disc}\,\Omega_{ab}(s')}{s'-s}\,.
\end{align}
Therefore, for the case of no bound states or CDD poles, the  $D_{ab}(s)$ function obtained in (\ref{D-fun}) can be easily related to the Omn\`es function as
\begin{equation}\label{Omnes_Dfun}
\Omega_{ab}(s)=D^{-1}_{ab}(s)\,.
\end{equation}

For the single-channel case, the Omn\`es function can be expressed in the analytic form in terms of the phase shift $\delta(s)$,
\begin{equation}\label{Omnes_phase_shift}
\Omega(s)=D^{-1}(s)=\exp\left(\frac{s-s_M}{\pi}\int_{s_{th}}^{\infty}\frac{d s'}{s'-s_M}\frac{\delta(s')}{s'-s}\right)\,.
\end{equation}
with the convention that $\delta(s_{th})=0$. Therefore, in single-channel approximations, the Omn\`es function is frequently computed directly from the existing parametrizations of the phase-shift data and various assumptions about its asymptotic behavior at infinity. The latter constrains the asymptotic behavior of the Omn\`es function: for $\delta(\infty)\to \alpha \pi$ one obtains $\Omega(\infty) \to 1/s^{\alpha}$.
In our approach, the phase shift curves are obtained from fits to the data using the $N/D$ method. The high-energy asymptotic of the phase shift is coming from the approximation of the left-hand cut by conformal expansion and subsequent solution of the once-subtracted dispersion relation. As a result, in this scheme, the obtained Omn\`es function (or its inverse) is always asymptotically bounded, if there is no bound state or CDD pole in the system. When there is a bound state in the system, the relation between the Omn\`es function and the $D(s)$ function given in Eq. (\ref{Omnes_phase_shift}) changes,
\begin{align}
\Omega(s)&=\left(\frac{s-s_B}{s_M-s_B}\right)\,D^{-1}(s)\nonumber \\
&=\exp\left(\frac{s-s_M}{\pi}\int_{s_{th}}^{\infty}\frac{d s'}{s'-s_M}\frac{\delta(s')}{s'-s}\right)\,,
\end{align}
where the extra factor $(s-s_B)/(s_M-s_B)$ removes the zero of $D(s)$. Due to this extra factor, the obtained Omn\`es function grows linearly at infinity and satisfies the twice-subtracted version of the dispersion relation given in Eq.~(\ref{Omnes_DR}).
This can also be seen from the Levinson's theorem, which relates the contribution from the number of bound states $n_B$ to the phase shift at infinity as $\delta(\infty)\to-n_B\,\pi$ (using the convention $\delta(s_{th})=0$).

For the multi-channel case, the Muskhelishvili-Omn\`es equations (\ref{Omnes_DR}) do not have analytic solutions \cite{Donoghue:1990xh, Moussallam:1999aq}, and one needs to find a numerical solution, by employing for instance a Gauss-Legendre procedure \cite{Moussallam:1999aq}. In order to achieve that, however, one needs to know the off-diagonal scattering amplitude in the unphysical region and again make the assumption about the high-energy asymptotics. On the other side, with the $N/D$ method, both the scattering amplitude and the Omn\`es function are obtained simultaneously from the fit to the available data. Additional information about the off-diagonal scattering amplitude in the unphysical region can be used as a constraint and not as a necessary requirement to obtain the Omn\`es matrix. Also, as discussed above, in most of the cases the obtained Omn\`es function (or its inverse) is asymptotically bounded. Therefore, this approach is useful in many practical applications. 

As a check of our numerical calculations, we verified that the Omn\`es functions obtained using Eqs. (\ref{D-fun}) and (\ref{Omnes_Dfun}) satisfy Eq. (\ref{Omnes_DR}).

\begin{table*}
    \begin{tabular*}{\textwidth}[t]{@{\extracolsep{\fill}}l|lllll|c@{}}
        \hline \hline
            & $\sqrt{s_E}$, MeV & $C_0$ & $C_1$ & $C_2$ & $C_3$ & $\chi^2/d.o.f$ \\
        \hline\hline
\multicolumn{7}{l}{$\pi \pi \to \pi \pi$}\\
\hline
        \textbf{Exp., SC}   & 740 & $15.9(7)$ & $51.8(1.7)$ & $58.2(1.4)$ & $24.4(3.0)$ & 0.5 \\
        \cline{2-7}
        \textbf{Exp., CC} $U_{11}(s)$   & 740 & 17.1(9)  & 52.1(2.0) &  51.1(2.2) & 17.2(3.6) & $t_{11}$: 3.4 \\
       \cline{2-6}
        \hspace{1.37cm} $U_{12}(s)$  &740 & 11.2(1.2) &  12.6(2.5) & - & - & $|t_{12}|$: 2.4 \\  
        \cline{2-6}
        \hspace{1.37cm} $U_{22}(s)$  &1095 & 70.0(6.5) & -216.2(58.0) & 321.0(53.9) & - & $\delta_{12}$: 1.8 \\

       \cline{2-7}
        \textbf{Lattice}, $m_\pi = 236$ MeV   & 646 & $13.3(2.9)$ & $64.4(1.6)$ & $64.5(5.6)$ & - & 1.2\\  
        \cline{2-7}
        \textbf{Lattice}, $m_\pi = 391$ MeV   &896 & $65.5(14.5)$ & $-293.7(47.8)$ & $409.2(35.7)$ & - & 1.2\\
        \hline
\multicolumn{7}{l}{$\pi K \to \pi K$}\\
\hline
        \textbf{Exp. SC}     & {833} & $16.1(8)$ & $-37.8(3.5)$ & $32.9(2.7)$ & $-18.6(6.0)$ & 1.2\\
        \cline{2-7}
        \textbf{Lattice}, $m_\pi=239$ MeV  & 884 & $16.8(3.6)$  & $-49.1(2.5)$ & $28.2(7.5)$ & - & 0.2\\
        \hline\hline 
    \end{tabular*}
\caption{Fit parameters entering Eq. (\ref{ConfExpansion}) which were adjusted to reproduce available experimental (whenever possible replaced by the most recent Roy-like results) or lattice data. SC and CC stand for single-channel and coupled-channel analyses, respectively. See text for more details.}
\label{tab:FitResults}
\end{table*}

\begin{table*}
    \begin{tabular*}{\textwidth}[t]{@{\extracolsep{\fill}}l|lll|lll@{}}
        \hline \hline
            & $\sqrt{s_A}$, MeV & $m_\pi\, a$ & $m_\pi^3\,b$& $\sqrt{s_A}$ ($\chi \text{PT}_{\text{NLO}}$), MeV& $m_\pi\,a$ ($\chi \text{PT}_{\text{NNLO}}$) & $m_\pi^3\,b$ ($\chi \text{PT}_{\text{NNLO}}$) \\
        \hline\hline
\multicolumn{7}{l}{$\pi \pi \to \pi \pi$}\\
        \hline
        \textbf{Exp., SC}   & ${ 90(9)}$ & $0.220(5)$ & ${0.276(6)}$ & ${ 90(9)}$ & $0.220(5)$ \cite{Colangelo:2001df}& $0.276(6)$ \cite{Colangelo:2001df}\\
        \textbf{Exp., CC}   & $ 90(15)$ &  $0.218(9)$ &  $0.278(11)$ & - & - & - \\
        \hline
        \textbf{Lattice}, $m_\pi = 236$ MeV   & $187(35)$ & $0.98(19)$ &  $0.89(43)$ & $ 150(18)$ & $0.75 - 0.87$ \cite{Colangelo:2001df} &-\\
        \textbf{Lattice}, $m_\pi = 391$ MeV   & -     & -$4.07(36)$ & $67.0(19.0)$ & -     & -&-\\
        \hline
\multicolumn{7}{l}{$\pi K \to \pi K$}\\
        \hline
        \textbf{Exp., SC}     &  480(6) & $0.219(10)$ & $0.113(10)$ & $ 480(6)$ & 0.220 \cite{Bijnens:2004bu} & 0.130 \cite{Bijnens:2004bu}\\
        \hline
        \textbf{Lattice}, $m_\pi=239$ MeV  & 472(8) & 0.426(71) & 0.277(68) &$472(9)$ & - & - \\
        \hline\hline 
    \end{tabular*}
    \caption{Fit results for the threshold parameters $a$ and $b$ defined in Eq. (\ref{eq:ThresholdPar}) and the Adler zeros $s_A$ compared to $\chi$PT values. SC and CC stand for single-channel and coupled-channel analyses, respectively. The uncertainties on NLO Adler zero positions we estimated as $|\text{NLO}-\text{LO}|$, as explained in the text.}
    \label{tab:ThresholdPar}
\end{table*}

\section{Numerical results}
\label{sec:Numerical results}

In this paper, we study the resonant $\pi\pi$ and $\pi K$ scattering in the S-wave. These are the channels where $\sigma/f_0(500)$, $f_0(980)$, and $\kappa/K_0^*(700)$ resonances reside. Both $\pi\pi$ and $\pi K$ channels have been measured experimentally \cite{Protopopescu:1973sh, Grayer:1974cr, *Kaminski:1996da,Batley:2007zz, *Batley:2010zza, Estabrooks:1977xe,Aston:1987ir}. However, throughout the whole energy range there are large differences between different data-sets and a careful choice of the data is required to achieve a controllable data-driven description of the phase shifts and inelasticity. For the $\pi\pi$ scattering, the situation is a bit better than for $\pi K$ scattering, since there is very precise low-energy data coming from $K_{l4}$ decays \cite{Batley:2007zz, *Batley:2010zza} and, in general, SU(2) $\chi$PT is a much more accurate theory than the SU(3) version of it. In order to be consistent with $\chi$PT in the threshold region, we employ the effective range expansion
\begin{align}\label{eq:ThresholdPar}
    \frac{2}{\sqrt{s}}\,\text{Re}\,\left(\frac{t(s)}{16\,\pi}
    \right)\simeq a+b\,p^2(s)+...\,,
\end{align}
where $a$ is the scattering length and $b$ is the slope parameter. For the $\pi\pi$ and $\pi K$ scattering both $a$ and $b$ have been calculated at NNLO in $\chi$PT \cite{Colangelo:2001df,Bijnens:2004bu}. As expected, for the $\pi K$ scattering, the chiral convergence is a bit worse than for the $\pi\pi$ scattering \cite{Bijnens:2004bu}, however the results for the scattering length and slope parameter do not show large discrepancies with the Roy-Steiner results \cite{Buettiker:2003pp,*DescotesGenon:2006uk, Pelaez:2020uiw,*Pelaez:2020gnd}. As for the Adler zero, we have checked that its position does not acquire large higher order corrections, and for simplicity one can take the LO result. In all numerical fits, however, we take the NLO result \cite{Gasser:1983yg,*Bernard:1990kx,*GomezNicola:2001as} as a central value, with the uncertainties from the omitted higher orders as $|\text{NLO}-\text{LO}|$, which should provide a conservative estimate. The NLO values for the low-energy constants are taken from \cite{Bijnens:2014lea}. For the case of non-physical pion masses with $m_\pi=236$ MeV and $m_\pi=239$ MeV, we only use Adler zero positions as a constraint, while for $m_\pi=391$ MeV, where $\sigma/f_0(500)$ shows up as a bound state, no constraints are imposed.

The free parameters in our approach are the conformal coefficients in (\ref{ConfExpansion}), which determine the form of the left-hand cut contribution $U_{ab}(s)$ in Eq. (\ref{DR_1}). Apart from the standard $\chi^2$ criteria, the number of parameters is chosen in a way to ensure that the series (\ref{ConfExpansion}) converges. The uncertainties are propagated using a bootstrap approach. In several cases, however, we will be fitting Roy (Roy-Steiner) solutions, which are smooth functions and their errors are fully correlated from one point to another. In these cases, $\chi^2/d.o.f$ loses its statistical meaning and can be  $<1$. In our fits, this scenario will simply indicate that we obtained the $N/D$ solution which is consistent with the Roy (Roy Steiner) solutions, and we just make sure that the obtained uncertainty is consistent with that from Roy analyses. 

\begin{table*}[t]
\renewcommand*{\arraystretch}{1.3}
\begin{tabular*}{\textwidth}[t]{@{\extracolsep{\fill}}l|ll|ll@{}}
\hline \hline
& \multicolumn{2}{c|}{Our results} & \multicolumn{2}{c}{Roy-like analyses}\\ 
& $\sqrt{s_p}$, MeV &$|g_{pa}|/\sqrt{{\cal N}_{aa}}$, GeV & $\sqrt{s_p}$, MeV & $|g_{pa}|/\sqrt{{\cal N}_{aa}}$, GeV\\ 
\hline \hline 
\multicolumn{5}{l}{$\sigma/f_0(500)$}\\
\hline
\bf{Exp., SC} & $458(7)^{+4}_{-10} - i\,245(6)^{+7}_{-10}$ & \begin{tabular}{l} 
$\gamma\gamma:  5.6(1)(0)\cdot 10^{-3}$\\ 
$\pi\pi: 3.15(5)^{+0.11}_{-0.20}$ \end{tabular}& $449^{+22}_{-16}-i\,275(15)$ \cite{Pelaez:2015qba} & 
\hspace{-0.55em}\begin{tabular}{l} $\gamma\gamma: 6.1(7) \cdot 10^{-3}$ \cite{Hoferichter:2011wk}\\ $\pi\pi: 3.45^{+0.25}_{-0.29}$ \cite{Pelaez:2015qba}\\$K\bar{K}: -$\end{tabular}
 \\
 \cline{2-3}
\bf{Exp., CC} & $458(10)^{+7}_{-15}- i\,256(9)^{+5}_{-8}$ & \begin{tabular}{l} $\gamma\gamma: 5.6(2)^{+0.1}_{-0.1}\cdot 10^{-3}$\\ $\pi\pi:  3.33(8)^{+0.12}_{-0.20}$\\ $K\bar{K}: 2.11(17)^{+0.27}_{-0.11}$ \end{tabular}&  & \\
\hline
\hspace{-0.55em}\begin{tabular}{l} \textbf{Lattice}\\ $m_\pi=236$ MeV\end{tabular} &$498(21)^{+12}_{-19}-i\,138(13)^{+5}_{-10}$ & \begin{tabular}{l} $\gamma\gamma: 10.7(9)^{+0.7}_{-0.3}\cdot 10^{-3}$\\ $\pi\pi: 2.96(5)^{+0.05}_{-0.06}$\end{tabular}&&\\
\hspace{-0.55em}\begin{tabular}{l} \textbf{Lattice}\\ $m_\pi=391$ MeV\end{tabular} & $758(5)(0)$ & \begin{tabular}{l} $\pi\pi: {3.91(26)(0)}\qquad $\end{tabular}&&\\
\hline 
\multicolumn{5}{l}{$f_0(980)$}\\
\hline
\bf{Exp., CC} & $993(2)^{+2}_{-1} - i\,21(3)^{+2}_{-4}$ &                            \begin{tabular}{l}$\gamma\gamma: 4.0(8)^{+0.3}_{-1.1}\cdot 10^{-3}$\\ $\pi\pi: 1.93(15)^{+0.07}_{-0.12}$\\ $K\bar{K}: 5.31(24)^{+0.04}_{-0.24}$\end{tabular} & $996^{+7}_{-14}-i\,25^{+11}_{-6}$ \cite{GarciaMartin:2011jx,*GarciaMartin:2011cn,*Pelaez:2019eqa,Moussallam:2011zg}& 
\hspace{-0.55em}\begin{tabular}{l} $\gamma\gamma: 3.8(1.4)\cdot 10^{-3}$ \cite{Moussallam:2011zg}\\ $\pi\pi: 2.3(2)$ \cite{GarciaMartin:2011jx,*GarciaMartin:2011cn,*Pelaez:2019eqa}\\$K\bar{K}: -$\end{tabular}\\ 
\hline
\multicolumn{5}{l}{$\kappa/K^*_0(700)$}\\
\hline
\bf{Exp. SC}    & $702(12)^{+4}_{-5}-i\,285(16)^{+8}_{-13}$ & \begin{tabular}{l}             $\pi K:  4.12(14)^{+0.13}_{-0.18}$\end{tabular} & $653^{+18}_{-12}-i\,280(16)$ \cite{ Buettiker:2003pp,*DescotesGenon:2006uk,Pelaez:2020uiw,*Pelaez:2020gnd} & \hspace{-0.55em}\begin{tabular}{l}             $\pi K: 3.81(9)$ \cite{Pelaez:2020uiw,*Pelaez:2020gnd} \end{tabular}\\
\hline
\hspace{-0.55em}\begin{tabular}{l} \textbf{Lattice}\\ $m_\pi=239$ MeV\end{tabular} & $747(39)^{+2}_{-0}-i\,265(16)_{-6}^{+7}$ & \begin{tabular}{l} $\pi K: 4.19(18)_{-0.06}^{+0.07}$\end{tabular}&&\\   
\hline \hline 
\end{tabular*}
\caption{Poles and couplings of the $\sigma/f_0(500)$, $f_0(980)$, and $\kappa/K_0^*(700)$ resonances calculated in data-driven $N/D$ approach compared with the results of Roy-like analyses. SC and CC stand for single-channel and coupled-channel analyses, respectively. For the $f_0(980)$ or $\kappa/K_0^*(700)$ poles we take a conservative dispersive average between \cite{GarciaMartin:2011jx,*GarciaMartin:2011cn,*Pelaez:2019eqa} and \cite{Moussallam:2011zg} or \cite{Buettiker:2003pp,*DescotesGenon:2006uk} and \cite{Pelaez:2020uiw,*Pelaez:2020gnd}, similar as it was done for $\sigma/f_0(500)$ in \cite{Pelaez:2015qba}. In our results, the first error is the statistical one, while the second one comes from a variation of $s_E$ and has a systematic nature.
}
\label{tab:Poles}
\end{table*}

Before entering the discussion of the results of the fits, we would like to briefly comment on the freedom of the choice of the subtraction point $s_M$ in the dispersion relation (\ref{DR_0}). The common choice in the application of the Omn\`es functions is $s_M=0$, due to its relation to scalar form factors and matching to $\chi$PT. On the other side, one can fix $s_M$ at the threshold, $s_M=s_{th}$, and then relate $\sum_{n=0}^{n_{max}} C_{n}\,\xi^{n}(s_{th})$ to the scattering length. Similarly, one can fix $s_M$ at the Adler zero\footnote{On the technical level, it may look that Adler zero could be accounted for as a CDD pole in the $D$-function \cite{Yao:2020bxx,Salas-Bernardez:2020hua}. However, every CDD pole physically corresponds to the genuine QCD state, while the existence of the Adler zero is the property of the chiral symmetry. Therefore we encode it as a zero in the $N$-function and not as a pole in the $D$-function.}, $s_M=s_A$, which would imply that $\sum_{n=0}^{n_{max}} C_{n}\,\xi^{n}(s_{A})=0$. The last two choices can therefore reduce the number of fitted parameters by one. Eventually different choices of $s_M$ redefine the fitted coefficients $C_n$ in the $U_{ab}(s)$ function and the results of the $N/D$ method are immune to that (after computing the $D$-function, it can be re-normalized to any other point below threshold). Since not in all the fits we impose threshold or Adler zero constraints, we decided to make the choice
\begin{align}
s_M=0
\end{align}
in all the cases for simplicity. As for the expansion point $s_E$, we choose it in the middle between the threshold and the energy of the last data point that is fitted,
\begin{equation}\label{Eq:s_E}
\sqrt{s_E}=\frac{1}{2}\,\left(\sqrt{s_{th}}+\sqrt{s_{max}}\right)\,.
\end{equation}
Note, that in the coupled channel case, $s_{th}$ in Eq. (\ref{Eq:s_E}) denotes the physical threshold for the diagonal terms  $U_{ab}(s)$, while for the off-diagonal terms it is the lowest threshold. We emphasize, that this particular choice guarantees a fast convergence of the conformal expansion (\ref{ConfExpansion}) in the region where the scattering amplitude is fitted to the data and also where it is needed as input to Eq. (\ref{N-fun}). 

Unlike the physical region, where the reaction models are typically fitted to data, the pole extraction may carry significant systematic uncertainties, especially if the pole lies deep in the complex plane \cite{Caprini:2008fc,Caprini:2016uxy}. To assess these, we vary the parameter $s_E$ around its central value fixed to (\ref{Eq:s_E}). We allow for a conservative variation by 25\% of the difference $\sqrt{s_{max}}-\sqrt{s_{th}}$, in order to have a compromise between $\sqrt{s_{th}}$ and $\sqrt{s_{max}}$. Note, that the extreme choice of 50\% would correspond to $s_E=s_{th}$ or $s_E=s_{max}$, which we clearly want to avoid, since it would bias the fit towards one or the other region. In the following results, the first error will indicate the statistical uncertainty ({\it i.e.} reflect the errors of the data and $\chi$PT input), while the second one will be associated with a variation of $s_E$.

All results presented below have been checked to fulfill the partial-wave dispersion relation given in Eq. (\ref{DR_1}) or Eq. (\ref{BoundState}) in the case when there is a physical bound state in the system. In addition we checked that there are no spurious poles or bound states in the considered cases\footnote{In principle, it is possible to expect the situation when det$(D_{ab}(s))$ has an unphysical zero far away from the threshold on the first Riemann sheet. To avoid this spurious bound state, one has to impose in the fit the fulfilment of p.w. dispersion relation which does not contain the bound state.}.

\subsection{Single channel $\pi\pi \to \pi\pi$ analysis of the experimental and lattice data} 
\label{subsec:SCpipi}
As a first step, we consider only the elastic $\pi\pi$ scattering, which should be enough to get a realistic estimate of the resonance position of $\sigma/f_0(500)$, which is known to be connected almost exclusively to the pion sector. The reason for that is twofold. In many practical applications it is convenient to remove the $K\bar{K}$ (or $f_0(980)$) effects, which do not influence much the $\sigma/f_0(500)$ pole parameters, but at the same time require a proper coupled-channel treatment. Additionally, the current lattice QCD result for $m_\pi=236$ MeV covers only the elastic region \cite{Briceno:2016mjc}. Therefore, as a necessary prerequisite of a meaningful $\sigma/f_0(500)$ pole extraction for unphysical pion masses, one has to test the $N/D$ formalism first for physical quark mass values, where the position of $\sigma/f_0(500)$ has already been obtained from the sophisticated Roy analyses  \cite{Pelaez:2015qba,GarciaMartin:2011jx,*GarciaMartin:2011cn,*Pelaez:2019eqa,Ananthanarayan:2000ht,*Caprini:2005zr,*Leutwyler:2008xd,Colangelo:2001df}. The inclusion of the $K\bar{K}$ channel (or $f_0(980)$ resonance) will allow for a slightly more precise evaluation of $\sigma/f_0(500)$ parameters and will be given in the next subsection. 

\begin{figure*}[t]
\centering
\includegraphics[width =0.90\textwidth ]{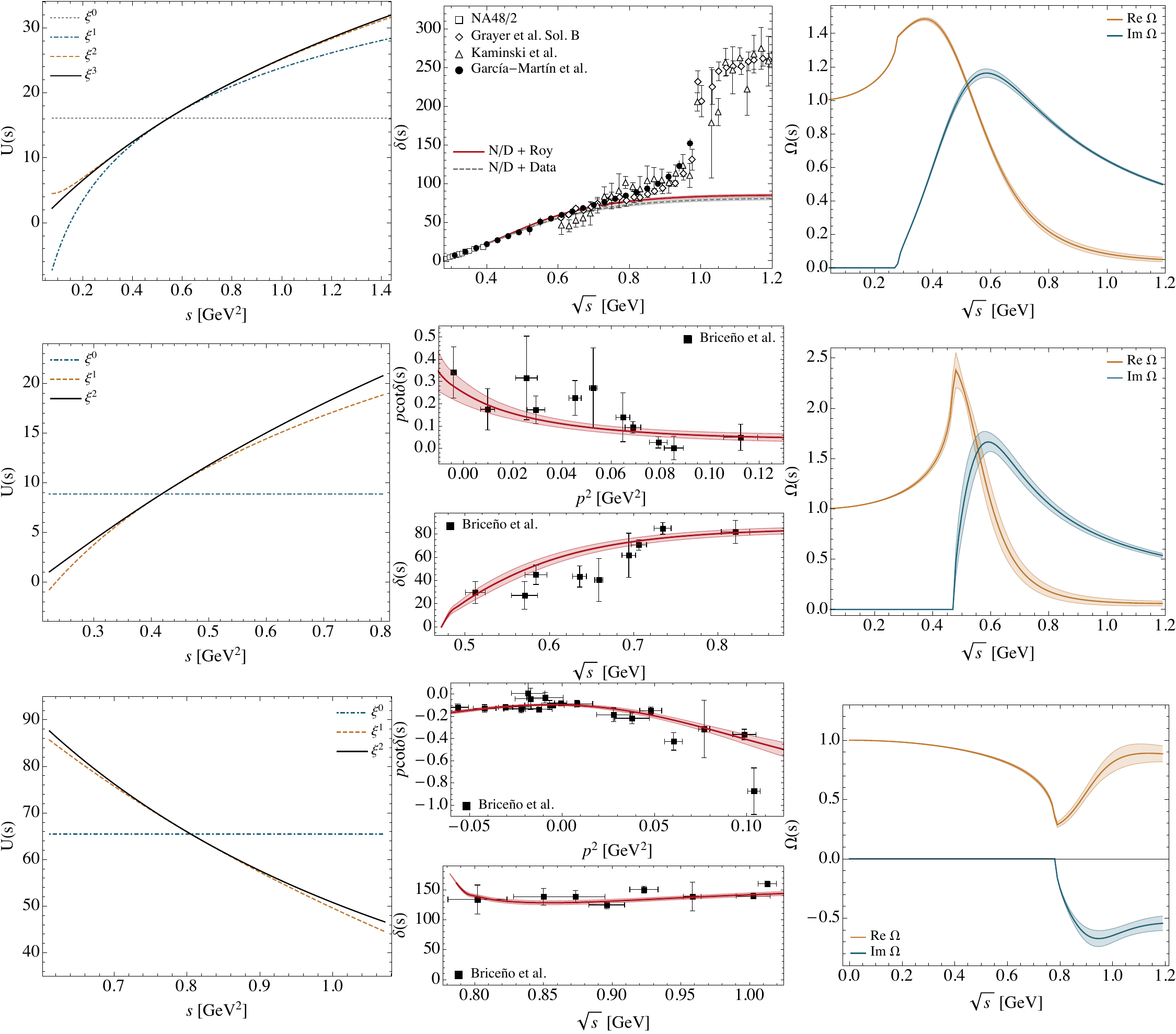}
\caption{Results for the $\pi\pi \to \pi\pi$ scattering with $J=0,\,I=0$ in the single-channel case. Top, central and bottom panels correspond to $m_\pi=\text{physical},\,236,\,391$ MeV, respectively. 
Left panels show the convergence of the conformal expansion in Eq. (\ref{ConfExpansion}), central panels show the comparison with the data and right panels show the corresponding Omn\`es functions. In the phase shift plot for the physical pion mass two curves are shown: fit to the experimental data \cite{Grayer:1974cr, *Kaminski:1996da,Batley:2007zz, *Batley:2010zza} (dashed curve) and fit to the pseudo data from Roy analysis \cite{Pelaez:2015qba,GarciaMartin:2011jx,*GarciaMartin:2011cn,*Pelaez:2019eqa} (thick curve). Note, that for the sake of comparison with the coupled-channel case (see Fig.\ref{Fig:pipiCC}), we adopted for this case $s_E$ based on $\sqrt{s_{max}} = 1.2$ GeV, as discussed in the text.
}
\label{Fig:pipiSC}
\end{figure*}

Relying only on the available data up to $\sqrt{s_{max}} = 0.7$ GeV, where a strong influence of the $K\bar{K}$ threshold is not yet expected, we obtain a decent fit even without imposing chiral constraints. The pole occurs at $\sqrt{s_{\sigma}}=463(8)^{+6}_{-7}-i\,217(6)^{+8}_{-9}$ MeV. The scattering length and slope parameters turn out to be compatible with those of $\chi$PT due to the presence of $K_{l4}$ data. As we discussed above, this is not the case for the Adler zero, which is located too close to the left hand cut, 
\begin{align}
s_A(\chi \text{PT}_{\text{LO}})=m_\pi^2/2\, ,
\end{align}
i.e. where the series (\ref{ConfExpansion}) simply converges too slow. With the additional constraints for the scattering length, slope parameter and Adler zero, the best fit result contains four parameters and leads to $\sqrt{s_{\sigma}}=435(7)^{+6}_{-8}-i\,250(5)^{+6}_{-8}$ MeV. This result is compatible with the value $\sqrt{s_{\sigma}}=446(5)^{+6}_{-9}-i\,230(5)^{+7}_{-9}$ MeV, obtained by replacing the experimental data with the pseudo data from the Roy-like analysis \cite{GarciaMartin:2011jx,*GarciaMartin:2011cn,*Pelaez:2019eqa}. As it is shown in Fig.~\ref{Fig:pipiSC} both $N/D$ fits are consistent within the error. This provides a proof for our expectation, that even in the case where there is no available Roy analyses (like lattice QCD data), we can rely on the $N/D$ approximation. For our final result of the single-channel Omn\`es function with physical pion mass, we opt for fitting the result of the Roy analysis \cite{GarciaMartin:2011jx,*GarciaMartin:2011cn,*Pelaez:2019eqa}, as the best representation of the data. The values of the fitted parameters are collected in Table \ref{tab:FitResults}, which result in the fast convergence of the conformal expansion (\ref{ConfExpansion}) as shown in the left panel of Fig.\,\ref{Fig:pipiSC}. Note, that in order to use these fit parameters as the starting values of the more complicated coupled-channel fit, we have chosen $s_E$ here to be the same as for the coupled-channel case, where we aim to describe the data up to $\sqrt{s_{max}}=1.2$ GeV. Also, this choice slightly improves the obtained $\sigma/f_0(500)$ pole positions, since it pushes $s_E$ further away from the threshold region, which is constrained accurately from $\chi$PT.
In Table \ref{tab:ThresholdPar} we compare threshold parameters and Adler zeros to $\chi$PT values, while in Table \ref{tab:Poles} poles and couplings are collected. Overall we achieve a good description of the Roy analyses results. In Fig.\,\ref{Fig:pipiSC} we also show phase shift and Omn\`es function. Note, that a similar result for the Omn\`es function can be obtained by using the phase shift from the single-channel modified Inverse Amplitude Method (mIAM) \cite{GomezNicola:2007qj,Hanhart:2008mx,Nebreda:2010wv,Salas-Bernardez:2020hua} and Eq. (\ref{Omnes_phase_shift}). In this method, the dispersion relation is written for the inverse amplitude, while the left-hand cut and subtraction constants are approximated by the chiral expansion. The result closest to the Roy analysis for the $\sigma/f_0(500)$ pole is achieved by performing two-loop mIAM fit \cite{Pelaez:2010fj}. In elastic $N/D$ and mIAM approaches the $K\bar{K}$ channel is separated naturally from the $\pi\pi$ channel, which is beneficial for the practical applications.

Apart from the experimental data, the recent lattice analysis \cite{Briceno:2016mjc} provided the results for the energy levels for pion mass values of $m_{\pi} = 236$ MeV and $m_{\pi}=391$ MeV. While the former case is much closer to the physical pion mass, the lattice result for the larger mass deserves special attention, since in that case $\sigma/f_0(500)$ shows up as a bound state. In the lattice QCD analysis, the discrete energy spectrum in a finite volume is related to the infinite-volume scattering amplitude through the L\"uscher formalism \cite{Luscher:1991cf,*Luscher:1990ck}, which was extended in \cite{Rummukainen:1995vs,*Kim:2005gf,*Christ:2005gi,*Leskovec:2012gb} to the case of moving frames. In the case of elastic scattering at low energies it gives a one-to-one relation to $p\,\cot\delta$. The lattice results for $p\,\cot\delta$ with $m_{\pi} = 236$ MeV and $m_{\pi}=391$ MeV were shown in \cite{Briceno:2016mjc}. To fit these data, we analytically continue $p\,\cot\delta$ below threshold, such that it does not produce any cusp behaviour at the threshold,
\begin{align}
   p(s)\,\cot\delta(s)=\frac{\sqrt{s}}{2}\left(\frac{1}{t(s)}+i\,\rho_0(s)\right)16\pi\, , 
\end{align}
where $\rho_0$ is the same as $\rho$ in Eq. (\ref{Eq:rho}), but without the Heaviside step function.

For both $m_{\pi} = 236$ MeV and $m_{\pi}=391$ MeV, we find that the three-parameter fit covers the data quite well (see central and bottom panels of Fig.\ref{Fig:pipiSC}). Similar to the K-matrix fits performed in \cite{Briceno:2016mjc}, we found $\sigma/f_0(500)$ as a deep pole on the second Riemann sheet for $m_{\pi} = 236$ MeV and as a bound state for $m_{\pi}=391$ MeV. In our approach, however, the obtained scattering amplitudes satisfy p.w. dispersion relations, which is a stringent constraint on the real part of the inverse of the amplitude. As a result, the pole position is determined much more precisely, see Table \ref{tab:Poles}. We also checked that the obtained scattering length $m_{\pi}\,a= 0.98(19)$ for $m_{\pi} = 236$ MeV  is consistent with the chiral extrapolation result $m_{\pi}\,a_{\text{NNLO}}= 0.75 - 0.87$ of \cite{Colangelo:2001df} and therefore including such additional constraint in the fit barely affects the results of the $\sigma/f_0(500)$ pole and coupling.

It is instructive to compare the obtained pole positions of $\sigma/f_0(500)$ for non-physical pion masses with the predictions of unitarized chiral perturbation theory (U$\chi$PT). The most popular are two approaches: mIAM \cite{Pelaez:2010fj} and Bethe-Salpeter equation (BSE) \cite{Albaladejo:2012te}. Both observe the same qualitative behaviour of the $\sigma/f_0(500)$ pole. With increasing pion mass values the imaginary part of the pole decreases, then $\sigma/f_0(500)$ becomes a virtual bound state and as $m_\pi$ increases further, one of the virtual states moves towards threshold and jumps onto the first Riemann sheet and become a real bound state. For $m_{\pi} = 236$ MeV, the extracted value from lattice data is consistent with U$\chi$PT predictions for the real part, but somewhat lower for the width,
\begin{align}
    \sqrt{s_\sigma}&=498(21)^{+12}_{-19}-i\,138(13)^{+5}_{-10} &(\text{lattice} +N/D), \nonumber\\
    \sqrt{s_\sigma}&=510-i\,175 &(\text{mIAM}_{\text{NNLO}}, \text{fit\,D}),\\
    \sqrt{s_\sigma}&=490(15)-i\,180(10)  &(\text{BSE}_{\text{NLO}}), \nonumber
\end{align}
all in units of MeV. 
For $m_{\pi} = 391$ MeV the situation is a bit different. Since it is on the edge of the applicability of $\chi$PT, the results of U$\chi$PT are very sensitive to the chiral order. Both mIAM \cite{Hanhart:2008mx} and BSE \cite{Albaladejo:2012te} at one loop found $\sigma/f_0(500)$ as a virtual bound state for $m_{\pi} = 391$ MeV. However, including the higher-order corrections (two loop) in mIAM \cite{Pelaez:2010fj} predicted the conventional bound state very close to the lattice results
\begin{align}
    \sqrt{s_\sigma}&=758(5)(0)~ \text{MeV} &(\text{lattice} +N/D),\nonumber\\
    \sqrt{s_\sigma}&=765 ~ \text{MeV} &(\text{mIAM}_{\text{NNLO}}, \text{fit\,D}),
\end{align}
confirming the proposed trajectory. However, as pointed out in \cite{Briceno:2016mjc}, it would be useful to perform lattice calculation between 236 and 391 MeV, to see what really happens in the transition region between a resonance lying deep in the second Riemann sheet and the bound state.

\begin{figure}[t]
\centering
\includegraphics[width =0.43\textwidth ]{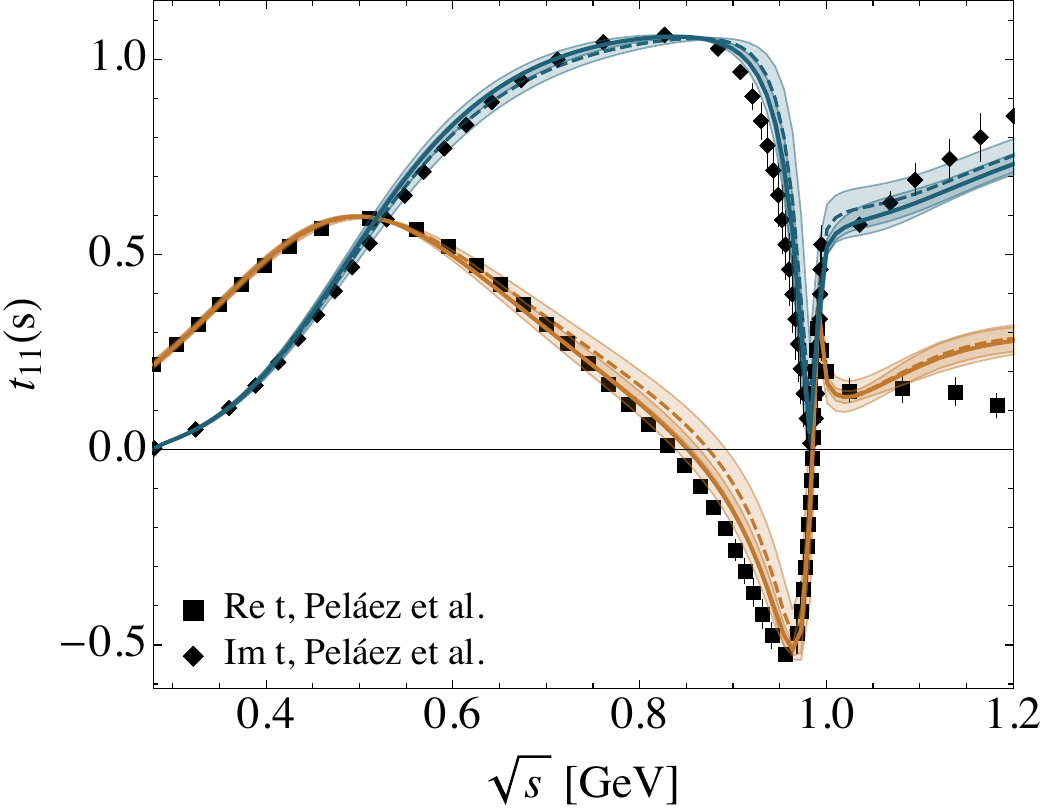}
\caption{Comparison between the coupled-channel $N/D$ fits and the Roy-like solution from \cite{GarciaMartin:2011jx,*GarciaMartin:2011cn,*Pelaez:2019eqa}. The dashed curves are the fit solely to the experimental data, while the solid curves take advantage of both the experimental data and the results of Roy (Roy-Steiner) analyses on $\pi\pi \to \pi\pi$ ( $\pi\pi \to K\bar{K}$).}
\label{Fig:ReImt11}
\end{figure}

\begin{figure*}[t]
\centering
\includegraphics[width =0.95\textwidth ]{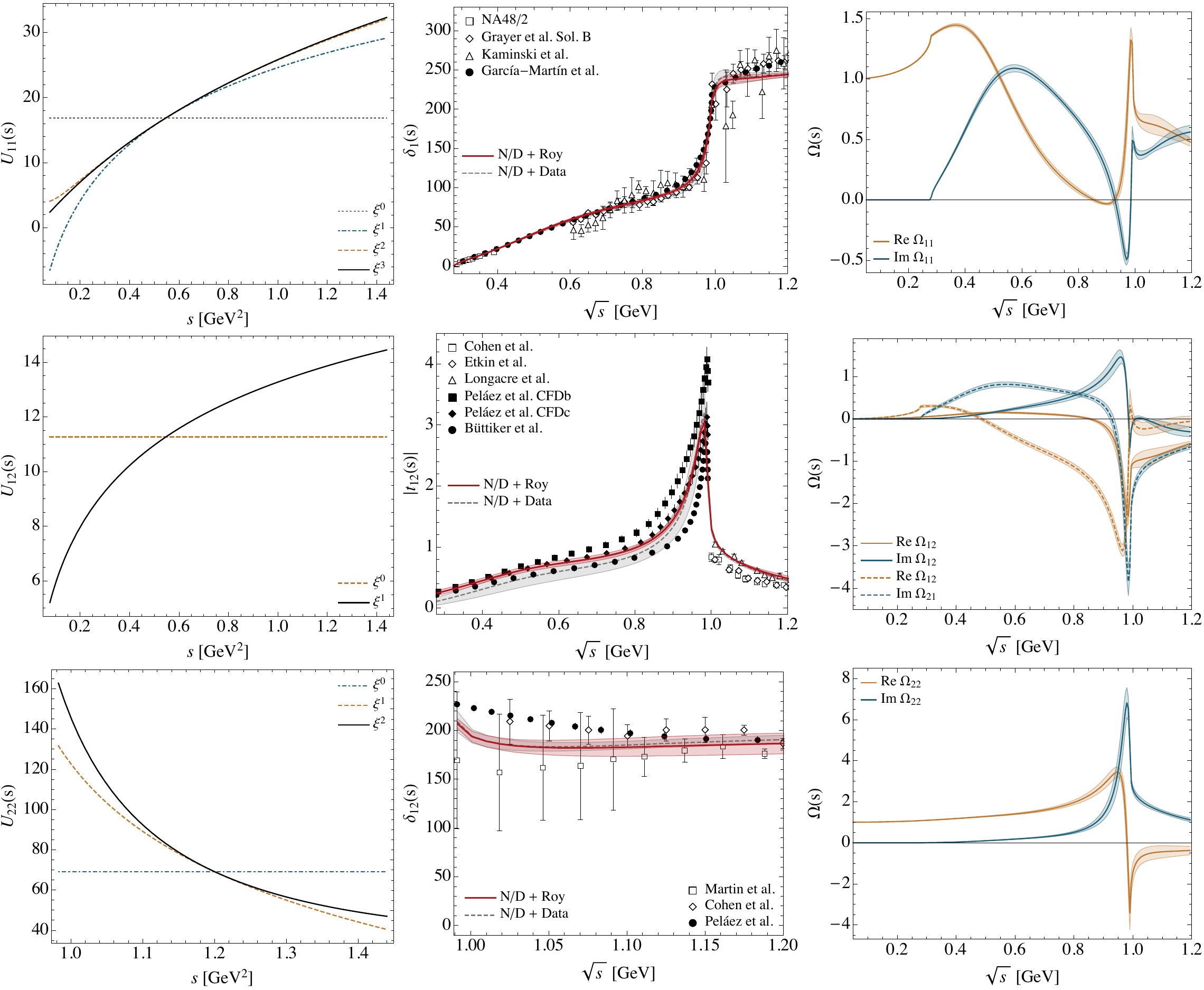}
\caption{Results for the $\pi\pi \to \pi\pi,\,K\bar{K}$ scattering with $J=0,\,I=0$ in the coupled-channel case. Top, central and bottom panels correspond to $11,\,12$ and $22$ matrix elements, respectively, with $1=\pi\pi$ and $2=K\bar{K}$. 
Left panels show the convergence of the conformal expansion in Eq. (\ref{ConfExpansion}), central panels show the comparison with the data, right panels show the elements of the Omn\`es matrix. In the central plots two curves are shown: fit to the experimental data \cite{Grayer:1974cr, *Kaminski:1996da,Batley:2007zz, *Batley:2010zza} (dashed curve) and fit to the pseudo data from Roy analyses \cite{Pelaez:2015qba,GarciaMartin:2011jx,*GarciaMartin:2011cn,*Pelaez:2019eqa} (thick curve).}
\label{Fig:pipiCC}
\end{figure*}

\subsection{Coupled-channel $\{\pi\pi,K\bar{K}\}$ analysis of the experimental data} 

While the single-channel analysis allows us to reproduce the low-energy behavior of the phase shifts and gives very reasonable values of the $\sigma/f_0(500)$ pole parameters, a comprehensive study of the region up to $\sqrt{s}=1.2$ GeV should account for the interplay between $\pi\pi$ and $K\bar{K}$ channels. In our normalization (see Eqs. (\ref{p.w.expansion}-\ref{Eq:rho})), the two-dimensional $t$-matrix, with channels denoted by $1=\pi\pi$ and $2=K\bar{K}$, is given by
\begin{equation}\label{t_CC}
t(s)=\begin{pmatrix}
\frac{\eta(s)\,e^{2\,i\,\delta_1(s)}-1}{2\,i\,\rho_{1}(s)} & |t_{12}(s)|\,e^{\delta_{12}(s)}\\
|t_{12}(s)|\,e^{\delta_{12}(s)} & \frac{\eta(s)\,e^{2\,i\,\delta_2(s)}-1}{2\,i\,\rho_{2}(s)}
\end{pmatrix}\,.
\end{equation}
Under assumption of two-channel unitarity, the inelasticity is related to $|t_{12}(s)|$ as
\begin{align}\label{inelasticity}
\eta(s)=\sqrt{1-4\,\rho_{1}(s)\,\rho_{2}(s)\,|t_{12}(s)|^2}\,,
\end{align}
and due to Watson's theorem, 
\begin{align}
\delta_{12}(s)=\delta_{1}(s)+\delta_{2}(s)\,\theta(s>4m_K^2)\,.
\end{align}
In the physical region the $t$-matrix is fully described by experimental information on the $\pi\pi$ phase shift $\delta_1(s)$ \cite{Protopopescu:1973sh, Grayer:1974cr, *Kaminski:1996da,Batley:2007zz, *Batley:2010zza}, the inelasticity $\eta(s)$ (or $|t_{12}(s)|$ for $s>4m_K^2$ \cite{Cohen:1980cq,Etkin:1981sg,Longacre:1986fh}) and the $\pi\pi \to K\bar{K}$ phase $\delta_{12}(s)$ \cite{Etkin:1981sg,Cohen:1980cq,Martin:1979gm}. 

Similar to the single-channel analysis, we first fit the available experimental data supplemented with constraints for scattering length, slope parameter and Adler zero from $\chi$PT in the $\pi\pi \to \pi\pi$ channel. As for the $\pi\pi\to K \bar{K}$ channel, the  complication stems from two facts. Firstly, the experimental data exist only in the physical region above $K\bar{K}$ threshold. Therefore, in order to stabilize the fits, we make sure that the obtained $|t_{12}(s)|$ stays small around\footnote{Specifically, at $s=m_\pi^2/2$ we impose NLO $\chi$PT with a conservative error that covers LO $\chi$PT result.} $s=0$ as a manifestation of $\chi$PT. Secondly, the existing experimental data for both $|t_{12}(s)|$ and $\delta_{12}(s)$ contains incompatible data sets and require to make some choice. Since the phase $\delta_{12}(s)$ is fully defined below $K\bar{K}$ threshold by means of Watson's theorem, we discard the data from \cite{Etkin:1981sg} as it suggests that $\pi\pi\to K \bar{K}$ phase goes much lower than it is forced by the presence of $f_0(980)$ resonance. Therefore, we fit the data from \cite{Cohen:1980cq} and \cite{Martin:1979gm} which are consistent due to the large error bars of the latter set. As for $|t_{12}(s)|$, the two data sets from \cite{Longacre:1986fh} and \cite{Cohen:1980cq,Etkin:1981sg} should in principle be treated separately. However, only the data from \cite{Longacre:1986fh} is compatible with the $\pi\pi$ inelasticity around the $K\bar{K}$ threshold. In order to describe the data from \cite{Cohen:1980cq,Etkin:1981sg}, most likely one has to include the four-pion channel, which is beyond the scope of the present paper. The best fit with $(4,4,3)$ parameters in $(11,12,22)$ channels 
\cite{Longacre:1986fh}, provides $\sigma/f_0(500)$ and $f_0(980)$ poles at $\sqrt{s_\sigma}= 454(12)^{+6}_{-7} - 262(12)^{+8}_{-12}\,i$ MeV and $\sqrt{s_{f_0}}= 990(7)^{+2}_{-4} - 17(7)^{+4}_{-1}\, i$ MeV. These results are remarkably close to the Roy (for $\pi\pi\to \pi\pi$) and Roy-Steiner solutions for ($\pi\pi \to K\bar{K}$) as shown in Figs. \ref{Fig:ReImt11} and \ref{Fig:pipiCC}. The large error bars arise from scarce experimental data around $K\bar{K}$ threshold and almost unconstrained $|t_{12}|$ in the unphysical region.

On the other side, we have at our disposal very precise $\pi\pi \to \pi\pi$ Roy-like analyses from \cite{GarciaMartin:2011jx,*GarciaMartin:2011cn,*Pelaez:2019eqa} and $\pi\pi \to K\bar{K}$ Roy-Steiner analyses from  \cite{Buettiker:2003pp,*DescotesGenon:2006uk,Pelaez:2020uiw,*Pelaez:2020gnd, Pelaez:2018qny}. Unfortunately, they do not come from the coupled-channel Roy-Steiner analyses and may display some inconsistencies between each other. In particularly, the Roy results on the real and imaginary parts of the $t_{11}(s)$ amplitude can constrain $\delta_1(s)$ and $\eta(s)$. The latter, in the two-channel approximation, is related to $|t_{12}(s)|$ by Eq. (\ref{inelasticity}) and turns out to be inconsistent with any available Roy-Steiner solution on $\pi\pi \to K\bar{K}$ \cite{Buettiker:2003pp,*DescotesGenon:2006uk,Pelaez:2020uiw,*Pelaez:2020gnd, Pelaez:2018qny}. Therefore in order to avoid possible conflict in fitting two independent analyses, we impose $\pi\pi \to K\bar{K}$ Roy-Steiner solution only as constraint on $|t_{12}(s)|$ in the unphysical region $4m_\pi^2<s<4m_K$. Currently, there are three competing solutions: one from Büttiker et al. \cite{Buettiker:2003pp} and two (CFDc and CFDb) from Peláez et al. \cite{Pelaez:2020uiw,*Pelaez:2020gnd}. We let the fit decide which solution to choose. As for the $\delta_{12}$, we take advantage of experimental data of Cohen et al. \cite{Cohen:1980cq} in the fit, which are quite precise. The good description of the data can be achieved with as low as $(4,2,3)$ parameters in $(11,12,22)$ channels, respectively. The results of the fit are collected in Tables \ref{tab:FitResults},\ref{tab:ThresholdPar} and \ref{tab:Poles} and shown in Fig.\,\ref{Fig:pipiCC}. As expected, the values for the fit parameters in the $11$-channel do not deviate much from the single-channel analysis in Sec.\ref{subsec:SCpipi}. In the coupled-channel analysis the $\sigma/f_0(500)$ pole position comes a bit closer to the Roy analysis value, than in the single-channel study. Moreover, we are now in a position to calculate its coupling to the $K\bar{K}$ channel, which we include in Table \ref{tab:Poles}. By inspecting Table \ref{tab:FitResults}, one can also see the striking similarity between the fit parameters in the $22$ channel and the fit to lattice $\pi\pi\to \pi\pi$ data with $m_\pi=391$ MeV, for which there is a bound state. Similarly, $f_0(980)$ will be a bound state in the $22$ channel, if we eliminate its connection to the $11$ channel, i.e. by putting $U_{12}=0$. This feature is not new and has already been observed in U$\chi$PT calculations, see for instance \cite{Oller:1997ti}. As for the $12$ channel, the fit clearly favours CFDc solution of \cite{Pelaez:2020uiw,*Pelaez:2020gnd}. This is also consistent with our previous "free" fit to to the experimental data, as shown by the dashed curves in Fig.~\ref{Fig:pipiCC}. On the right panels of Fig.~\ref{Fig:pipiCC} we show the elements of the Omn\`es matrix calculated using Eq.~(\ref{Omnes_Dfun}). The previous version of them, with the fit to \cite{Buettiker:2003pp,Pelaez:2018qny} has already been successfully applied for the dispersive coupled-channel study of $\gamma^{(*)}\gamma^{*} \to \pi \pi(K\bar{K})$ \cite{Danilkin:2018qfn,*Danilkin:2019opj,*Deineka:2019bey} and $e^+e^- \to J/\psi \pi \pi (K\bar{K})$ \cite{Danilkin:2020kce}.

\begin{figure*}[t]
\centering
\includegraphics[width =0.43\textwidth ]{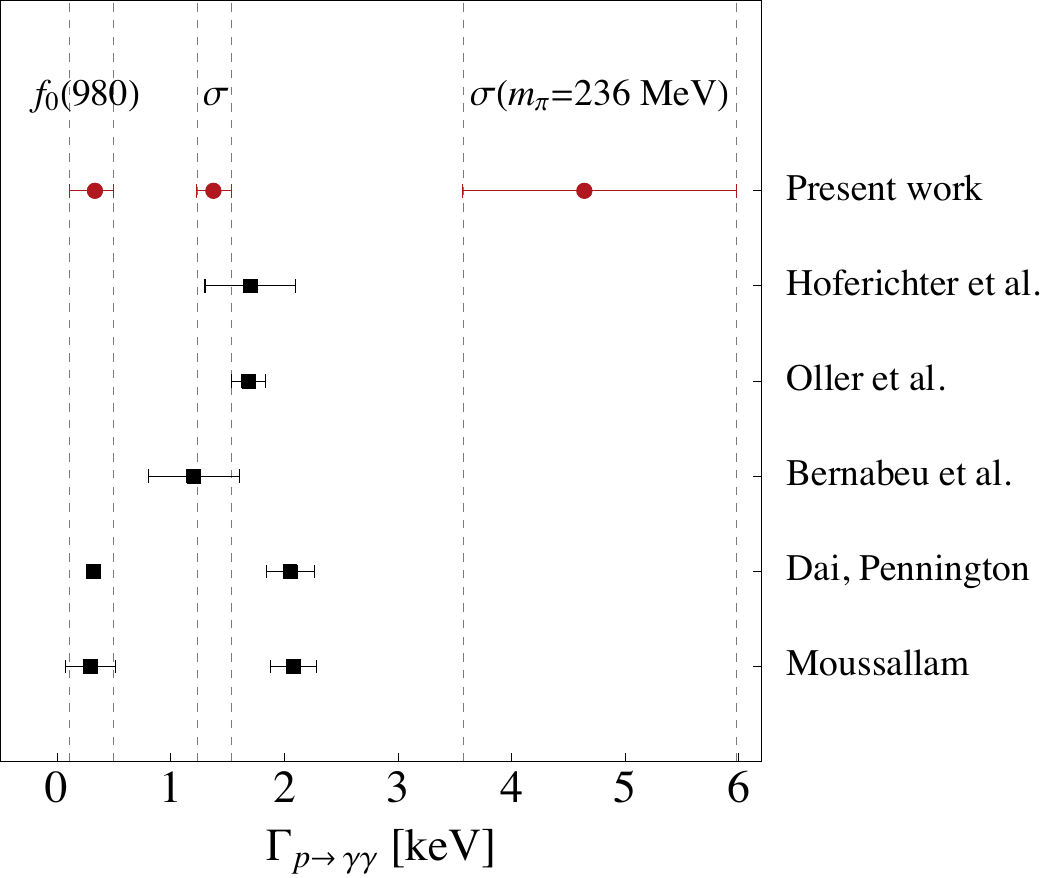}\quad\quad
\includegraphics[width =0.43\textwidth ]{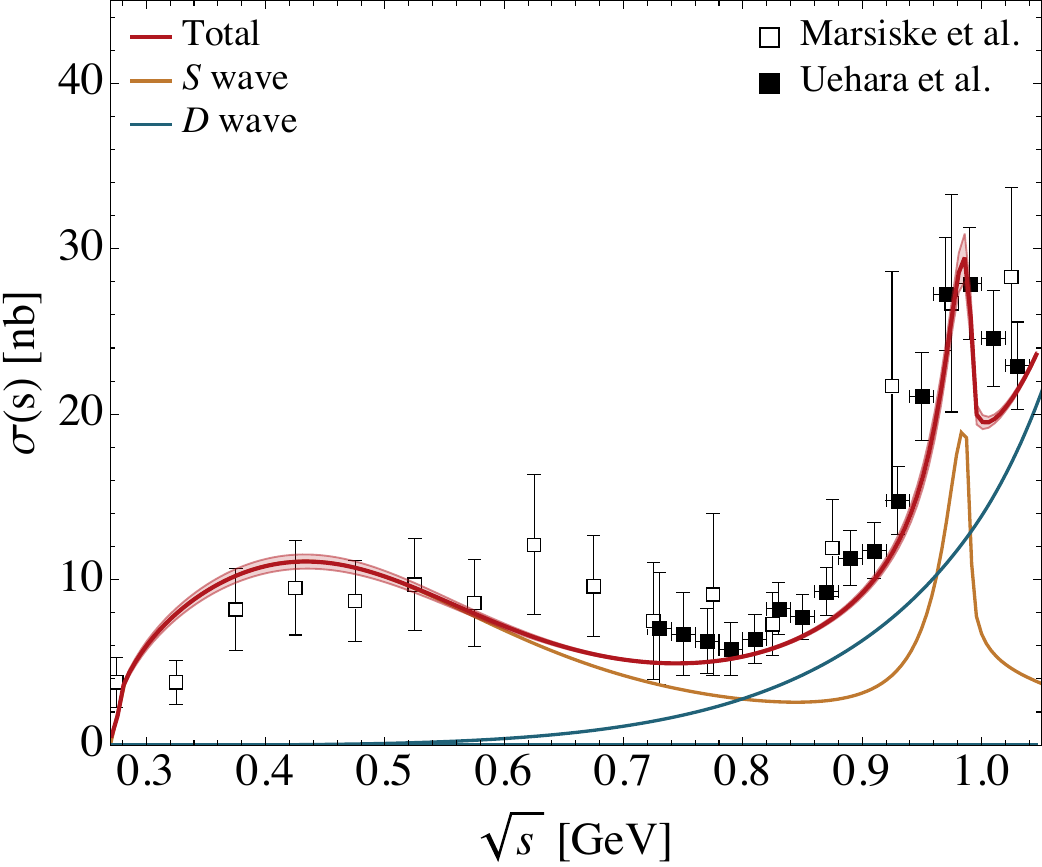}
\caption{Left panel: two-photon decay width of $\sigma/f_0(500)$ and $f_0(980)$ compared to the recent dispersive estimations from \cite{Bernabeu:2008wt,Oller:2008kf,Hoferichter:2011wk,Moussallam:2011zg,Dai:2014lza}. Right panel: total cross section of $\gamma\gamma \to \pi^0\pi^0$ $(|\cos\theta|<0.8)$ from \cite{Danilkin:2018qfn} with updated $I=0, J=0$ contribution. The data for the cross section are taken from \cite{Uehara:2009cka,*Marsiske:1990hx}.}
\label{Fig:ggTopipi}
\end{figure*}

We leave the coupled-channel study of the existing lattice data on $\{\pi\pi,K\bar{K}\}$ \cite{Briceno:2017qmb} with $m_\pi=391$ MeV for a future work. In our opinion, this channel has to be analysed together with $\{\pi\eta,K\bar{K}\}$ lattice data \cite{Dudek:2016cru}, to shed more light onto the differences between the light scalar resonances $f_0(980)$ and $a_0(980)$.

\subsection{Two-photon couplings of $\sigma/f_0(500)$ and $f_0(980)$} 
As an application of the obtained Omn\`es functions in the $N/D$ approach, we would like to extract the two-photon couplings of $\sigma/f_0(500)$ and $f_0(980)$. In principle, the coupling to the external currents has the potential to infer the scalar meson composition. Furthermore, it characterizes the interaction strength of  $\sigma/f_0(500)$ and $f_0(980)$ in the two-photon channel. The latter is important for the light-by-light sum rule applications \cite{Pascalutsa:2010sj,Pascalutsa:2012pr,Danilkin:2016hnh,Dai:2017cvz} and serves as a key input to estimate the isoscalar two-pion (kaon) contribution to the hadronic light-by-light scattering for $(g-2)$ of the muon \cite{Aoyama:2020ynm,*Danilkin:2019mhd}. The central result in this section will be obtained using a coupled-channel dispersive representation, however, for $\sigma/f_0(500)$ we will employ as well the single-channel representation both for physical and non-physical pion masses.

\begin{figure*}[t]
\centering
\includegraphics[width =0.95\textwidth ]{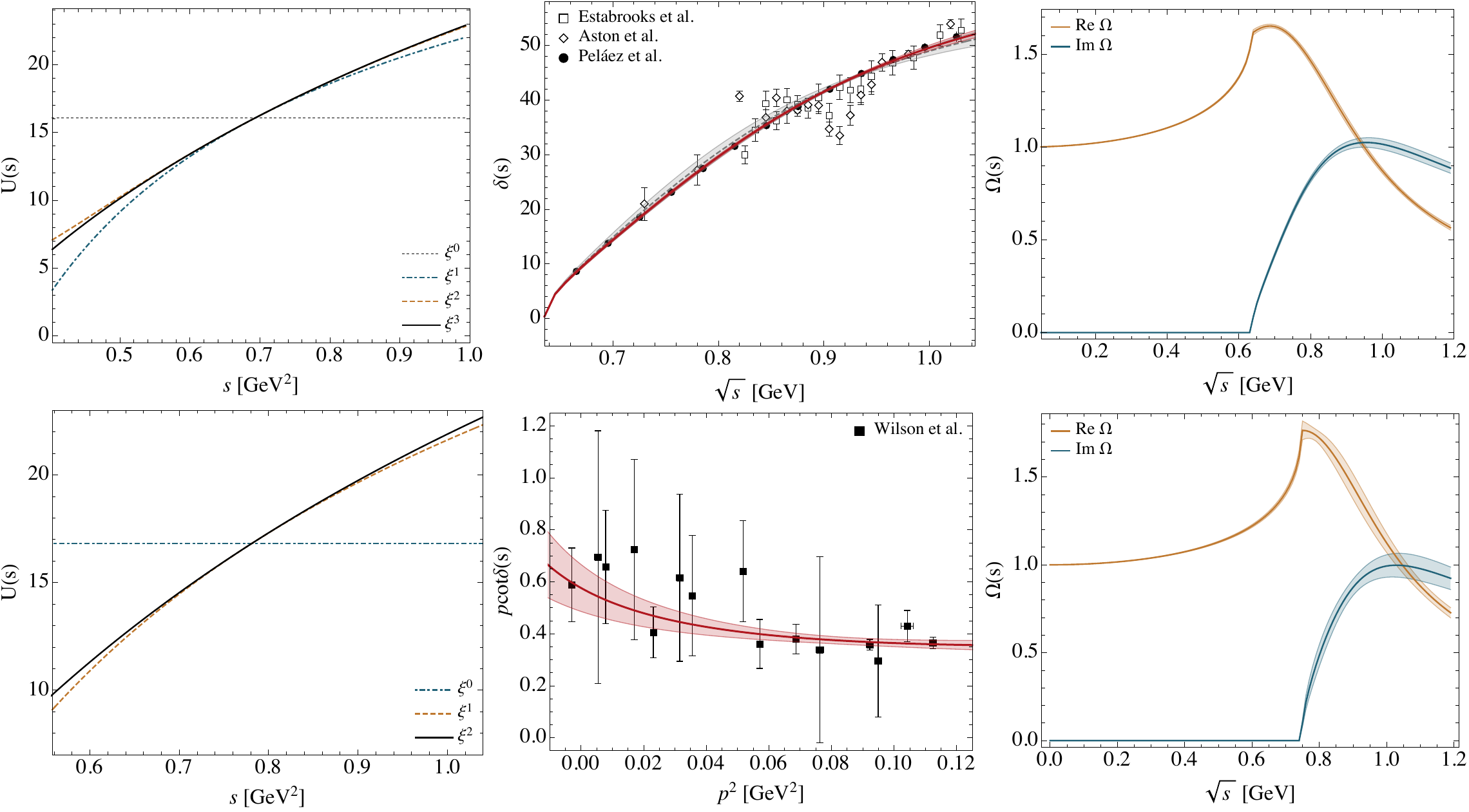}
\caption{Results for the $\pi K \to \pi K$ scattering with $J=0,\,I=1/2$ in the single-channel approximation. Top and bottom panels correspond to $m_\pi=\text{physical},\,239$ MeV, respectively. Left panels show the convergence of the conformal expansion in Eq. (\ref{ConfExpansion}), central panels show the comparison with the data and right panels show the corresponding Omn\`es functions. In the phase shift plot for the physical pion mass two curves are shown: fit to the experimental data \cite{Estabrooks:1977xe,Aston:1987ir} (dashed curve) and fit to the pseudo-data from Roy-Steiner analysis \cite{Pelaez:2020uiw,*Pelaez:2020gnd} (thick curve).}
\label{Fig:piK}
\end{figure*}

The photon-fusion partial-wave amplitude $\gamma\gamma \to \pi\pi$, which we denote by $h_{I,\lambda_1,\lambda_2}^{(J)}$, is the off-diagonal element of the $\gamma\gamma, \pi \pi, K\bar{K}$ channels. Since the intermediate states with two photons are proportional to $e^4$, they are suppressed, and one can reduce the $(3\times 3)$ matrix dispersion relation down to the $(2\times1)$ form, which require the hadronic rescattering part, $\Omega(s)$, and the left-hand cuts as input \cite{GarciaMartin:2010cw, Hoferichter:2011wk, Danilkin:2012ua, Dai:2014zta, Oller:2007sh, Oller:2008kf}. For the low-energies around $\sigma/f_0(500)$ (and to lesser extent around $f_0(980)$) the contribution from the left-hand cuts is dominated by the pion-pole contribution (Born term), which is exactly calculable. 
Therefore, in this approximation there is no need of modeling left-hand cuts in one way or another or introducing any subtractions. The photon-fusion p.w. amplitudes are readily obtained using the Muskhelishvili-Omn\`es representation. For more details, we refer to Ref. \cite{Danilkin:2018qfn,*Danilkin:2019opj,*Deineka:2019bey}. The two-photon couplings are extracted by calculating the residue of $h^{(0)}_{0,++}(s)$ at the pole positions, $s_p$. Following \cite{Pennington:2006dg,Oller:2007sh}, in our convention it is given by
\begin{align}
\frac{g_{p\gamma\gamma}^2}{g^2_{p\pi\pi}}=-\left(\rho_{0}(s_p)\, h^{(0)}_{0,++}(s_p)\right)^2 \, ,
\end{align}
where $h^{(0)}_{0,++}(s)$ is evaluated on the first Riemann sheet for $p=\sigma/f_0(500),\,f_0(980)$. An intuitive way of re-expressing the two-photon couplings, shown in Table \ref{tab:Poles}, is by using the formal definition of the corresponding two-photon decay widths
\begin{align}\label{Eq:Gamma}
\Gamma_{p\to\gamma\gamma}=\frac{|g_{p\gamma\gamma}|^2}{16\,\pi\,\text{Re}\sqrt{s_p}}\,.
\end{align}
Converted to (\ref{Eq:Gamma}), our results read
\begin{align}\label{Eq:GammaResults}
\Gamma_{\sigma \to\gamma\gamma} &=1.37(13)_{-0.06}^{+0.09}\,\left[1.38(9)^{+0.01}_{-0.01}\right]\, \rm{keV}, \nonumber\\
\Gamma_{f_0(980)\to\gamma\gamma}&=0.33(16)^{+0.04}_{-0.16} \,\rm{keV},\\
\Gamma^{m_\pi=236\,\rm{MeV} }_{\sigma\to\gamma\gamma} &=4.64(1.01)_{-0.35}^{+0.88}\, \rm{keV}, \nonumber
\end{align}
where in square brackets the single-channel approximation is shown. As expected, its $\Gamma_{\sigma \to\gamma\gamma}$ is almost indistinguishable from the coupled-channel case. In Fig.~\ref{Fig:ggTopipi} we compare our results with the recent dispersive estimates \cite{Bernabeu:2008wt,Oller:2008kf,Hoferichter:2011wk,Moussallam:2011zg,Dai:2014lza}. While the two-photon decay width of $f_0(980)$ is consistent with the coupled-channel amplitude analysis of \cite{Dai:2014lza} and the over-subtracted coupled-channel Muskhelishvili-Omn\`es analysis \cite{Moussallam:2011zg}, the two-photon width of $\sigma/f_0(500)$ is about 25\% smaller than their values. On the other hand, the obtained two-photon width of $\sigma/f_0(500)$ is consistent with the sophisticated Roy-Steiner analysis \cite{Hoferichter:2011wk} and other dispersive analyses from \cite{Bernabeu:2008wt,Oller:2008kf}. Finally, we also predicted $\sigma/f_0(500)$ two-photon coupling/width for the unphysical $m_\pi=236$ MeV, which would be interesting to confront with the direct lattice calculations.

We note, that the errors quoted in Eq. (\ref{Eq:GammaResults}) correspond solely to the uncertainties in the Omn\`es matrix. In principle, one can perform a more comprehensive study of the theoretical uncertainties, by the inclusion of more distant left-hand cuts in $\gamma\gamma \to \pi\pi (K\bar{K})$. This would require introducing subtraction constants which can be either fixed from the pion dipole polarizabilities or fitted directly to the cross-section data. Doing so would likely enlarge the error, but we do not expect a significant change of the central values, since the current parameter-free description of the cross-section data (see Fig.\ref{Fig:ggTopipi}) is quite impressive. The advantage of the approach that accounts only for pion pole left-hand contribution, is that in the absence of any single-virtual data one can predict the behavior of the p.w. helicity amplitudes for finite virtualities \cite{Colangelo:2017qdm,*Colangelo:2017fiz,Danilkin:2018qfn,*Danilkin:2019opj,*Deineka:2019bey}, which are needed as input for $(g-2)_\mu$ \cite{Aoyama:2020ynm,*Danilkin:2019mhd}.

\subsection{$I=1/2$ single-channel: data and lattice} 
For the $\pi K\to \pi K$ single channel analysis we begin by fitting the experimental data and imposing constraints from $\chi$PT for the scattering length, slope parameter, and Adler zero. The latter at LO is given by a simple relation,
\begin{align}
s_A(\chi \text{PT}_{\text{LO}})=\frac{1}{5} \left(m_\pi^2+m_K^2+2 \sqrt{4\,m_\pi^4-7 \,m_\pi^2\, m_K^2 +4\, m_K^4}\right)\,.
\end{align}
The most precise calculation of the scattering length and slope parameter in $\chi$PT has been performed at NNLO in \cite{Bijnens:2004bu}. While the result for the scattering length $m_\pi\, a=0.22$ is consistent with the recent Roy-Steiner predictions $m_\pi\, a=0.223(9)$ \cite{Pelaez:2020uiw,*Pelaez:2020gnd}, it seems that there is a small tension in the slope parameter value $m_\pi^3\,b=0.13$ compared to $m_\pi^3\,b=0.108(8)$ from \cite{Pelaez:2020uiw,*Pelaez:2020gnd}. The calculation of uncertainties is a bit cumbersome at NNLO and has not been presented in \cite{Bijnens:2004bu}. Therefore in our fits we take NNLO $\chi$PT values as central results, but include the conservative error-bar, such that it covers the recent Roy-Steiner results \cite{Pelaez:2020uiw,*Pelaez:2020gnd}. As for the Adler zero, we take the NLO value, as explained at the beginning of Sec.\ref{sec:Numerical results}. The available experimental data for this process is scarce in the region close to the $\pi K$ threshold, and often contains the discrepancies even within one dataset \cite{Estabrooks:1977xe}. Since we consider only the single-channel approximation, we perform the fit till $\eta K$ threshold of the data from  \cite{Estabrooks:1977xe,Aston:1987ir}. In this way we also exclude the influence of the $K_0^*(1430)$ resonance.  We observe a similar situation as for the $\pi \pi\to \pi\pi$ single-channel analysis, that fitting the experimental data \cite{Estabrooks:1977xe,Aston:1987ir} or Roy-Steiner analysis of \cite{Pelaez:2020uiw,*Pelaez:2020gnd} provides equivalent four parameter fits with $\kappa/K_0^*(700)$ pole positions at $689(24)^{+3}_{-2}- i\,263(33)^{+5}_{-8}$ MeV and  $702(12)^{+4}_{-5}-i\,285(16)^{+8}_{-13}$ MeV, respectively. In general, these results compare well with the Roy-Steiner pole position $653^{+18}_{-12}-i\,280(16)$ MeV which we take as a conservative average between \cite{ Buettiker:2003pp,*DescotesGenon:2006uk} and \cite{Pelaez:2020uiw,*Pelaez:2020gnd}. The one-sigma difference in the resonance mass can be attributed to the fact, that we are fitting Roy-Steiner solution only in the elastic region. We also look forward to the results of the KLF Collaboration, which plans to study $\pi K$ scattering using a secondary $K_L$ beam at Jefferson Lab \cite{Amaryan:2020xhw}. It will further improve the position of the $\kappa/K_0^*(700)$ resonance.

For the unphysical pion mass, we again use recent lattice data from the Hadron Spectrum Collaboration \cite{Wilson:2019wfr}. We analyse the data for $m_\pi=239$ MeV, where an evidence of $\kappa/K_0^*(700)$ was observed in the $p\,\cot\delta$ distribution. Due to large uncertainties, the pole position was not determined by the lattice collaboration, calling for more sophisticated approaches that include in addition to unitarity also the analyticity constraint. By employing the data-driven $N/D$ approach, the present data can be easily described with the two-parameter fit, leading to $\chi^2/d.o.f=0.4$. In this case, however, the Adler zero of the amplitude is located relatively far from the $\chi$PT value, since the lattice data in the low $p^2$ region suffers from the large uncertainties. Also, as we discussed before, in the Adler zero region the conformal expansion (\ref{ConfExpansion}) does not converge well by construction and one has to impose Adler zero as a constraint, which effectively calls for one additional parameter. In this way, the impact of two points with prominently small errors at $p^2\sim0.09$ and $\sim0.11$ GeV$^2$ is balanced out. The results of the fit are collected in Tables \ref{tab:FitResults},\ref{tab:ThresholdPar} and \ref{tab:Poles}.

Again we would like to compare our results for the pole position and coupling with predictions of mIAM. According to [53], at $m_{\pi} = 239$ MeV, the imaginary part of the pole decreases by $\sim 17\%$, while the real part and coupling slowly increase by $\sim4\%$ and $\sim8\%$ respectively.  Our values  extracted  from  the lattice data show a similar  behavior, with the decrease in the imaginary part of $7.0(7.7)\%$, increase in the real part and coupling of $6.4(5.8)\%$ and $1.6(6.4)\%$, respectively. 

\subsection{Systematic uncertainties}
In the end, we wish to comment on the size of systematic uncertainties of our results. As it can be seen in Table \ref{tab:Poles}, under the change of $\sqrt{s_E}$ by 25\% of the difference $\sqrt{s_{max}}-\sqrt{s_{th}}$ around the central value (\ref{Eq:s_E}), the $\sigma/f_0(500)$ and $\kappa/K^*_0(700)$ poles acquire noticeable systematic errors which are of the size of statistical ones. We admit that $s_E$ variation only accounts for the dominant part of the systematic uncertainty and therefore only provides a lower bound on the systematic error.
However, even if we go to the extreme case of 50\%, which corresponds either to $s_E=s_{th}$ or $s_E=s_{max}$, the statistical error will grow only by a factor of two, compared to the case of 25\%. This is different from the K-matrix fits (see for instance \cite{Caprini:2008fc}), which cannot extract accurately the pole parameters. 
We remind, that in our approach, as opposed to K-matrix models, the obtained amplitudes satisfy p.w. dispersion relations, which is an additional constraint on the amplitude both on the real axis and in the complex plane.

\section{Conclusion and outlook}
\label{sec:Conclusion and outlook}

In this work, we presented a data-driven analysis of the resonant S-wave $\pi\pi \to \pi\pi$ and $\pi K \to \pi K$ reactions using the p.w. dispersion relation. In this approach unitarity and analyticity constraints are implemented exactly. We accounted for the contributions from the left-hand cuts using the Taylor expansion in a conformal variable, which maps the left-hand cut plane onto the unit circle. Then, the once subtracted p.w. dispersion relation was solved numerically by means of the $N/D$ method.

Using existing experimental information and threshold constraints from $\chi$PT we tested the single-channel $N/D$ formalism for the physical pion mass, where the positions of $\sigma/f_0(500)$ and $\kappa/K_0^*(700)$ have already been obtained from the sophisticated Roy and Roy-Steiner analyses. We demonstrated that the results for the pole parameters are stable and almost do not change if we replace the existing experimental data with the very precise pseudo data generated by Roy and Roy-Steiner solutions in the physical region. As a next step, we performed the fits to the lattice data of the Hadron Spectrum Collaboration for $m_\pi=236, 391$ MeV in the case of $\pi\pi \to \pi \pi $ and for $m_\pi=239$ MeV in the case of $\pi K \to \pi K$. We provided an improved determination of the $\sigma/f_0(500)$ and $\kappa/K_0^*(700)$ pole parameters compared to the simplistic $K$-matrix approach and also compared them with U$\chi$PT predictions.

An important feature of the $N/D$ method is that the Omn\`es function comes out naturally, as the inverse of the $D$-function. The knowledge of the Omn\`es function, in turn, allows employing the Muskhelishvili-Omn\`es representation for the vast majority of production/decay reactions involving two pions (or pion and kaon) in the final state. While for the single-channel case, the Omn\`es function can be obtained analytically from the parametrisation of the phase shift, this is not the case for the coupled-channel case. In order to cover the $f_0(980)$ region we extended our analysis for the coupled-channel $\{\pi\pi, K\bar{K}\}$ case and extracted the corresponding Omn\`es matrix. In our construction it is asymptotically bounded (i.e. it satisfies once-subtracted dispersion relation) and therefore useful in many dispersive applications. The unknown coefficients from the conformal expansion were adjusted to reproduce existing Roy and Roy-Steiner analyses. As a straightforward application of the Muskhelishvili-Omn\`es representation, we estimated the two-photon decay widths of the $\sigma/f_0(500)$ and $f_0(980)$ resonances, which turned out to be consistent with the previous dispersive results. The obtained Omn\`es matrix serves as an important building block, which allows for the dispersive calculation of the isoscalar two pion/kaon contribution to the hadronic light-by-light part \cite{Colangelo:2014dfa,Colangelo:2017qdm,*Colangelo:2017fiz,Pauk:2014rfa} of the anomalous magnetic moment of the muon $(g-2)_\mu$ \cite{Aoyama:2020ynm,*Danilkin:2019mhd}. In particularly, with the input from $\gamma^*\gamma^* \to \pi\pi,KK$ \cite{Danilkin:2018qfn,*Danilkin:2019opj,*Deineka:2019bey} one can estimate dispersively the contribution from the $f_0(980)$ resonance, and compare it with narrow resonance results \cite{Danilkin:2016hnh}.

The proposed method is not only limited to the $\pi\pi$ and $\pi K$ scattering. We considered these reactions in the present paper because they show up as building blocks in many hadronic reactions/decays and have been calculated recently using lattice QCD. In principle, the $N/D$ method combined with the conformal expansion for the left-hand cuts can be applied to any hadronic reaction where there is data (experimental or lattice) which possesses a broad (or coupled-channel) resonance that does not have a genuine QCD nature. For the latter (like for instance $\rho$ or $K^*$ resonances) one needs to extend the formalism to allow for CDD poles. Also, it has to be modified in the presence of anomalous thresholds.

\section*{Acknowledgements}
We thank Arkaitz Rodas for providing the results of \cite{Pelaez:2020uiw,*Pelaez:2020gnd}. I.D. acknowledges useful discussions with Cesar Fern\'andez-Ram\'irez and Daniel Mohler. This work was supported by the Deutsche Forschungsgemeinschaft (DFG, German Research Foundation), in part through the Collaborative Research Center [The Low-Energy Frontier of the Standard Model, Projektnummer 204404729 - SFB 1044], and in part through the Cluster of Excellence [Precision Physics, Fundamental Interactions, and Structure of Matter] (PRISMA$^+$ EXC 2118/1) within the German Excellence Strategy (Project ID 39083149).  O.D. acknowledges funding by DAAD.

\bibliographystyle{apsrevM}
\bibliography{PRD}

\ifx\mcitethebibliography\mciteundefinedmacro
\PackageError{apsrevM.bst}{mciteplus.sty has not been loaded}
{This bibstyle requires the use of the mciteplus package.}\fi
\begin{mcitethebibliography}{128}
\expandafter\ifx\csname natexlab\endcsname\relax\def\natexlab#1{#1}\fi
\expandafter\ifx\csname bibnamefont\endcsname\relax
  \def\bibnamefont#1{#1}\fi
\expandafter\ifx\csname bibfnamefont\endcsname\relax
  \def\bibfnamefont#1{#1}\fi
\expandafter\ifx\csname citenamefont\endcsname\relax
  \def\citenamefont#1{#1}\fi
\expandafter\ifx\csname url\endcsname\relax
  \def\url#1{\texttt{#1}}\fi
\expandafter\ifx\csname urlprefix\endcsname\relax\def\urlprefix{URL }\fi
\providecommand{\bibinfo}[2]{#2}
\providecommand{\eprint}[2][]{\url{#2}}

\bibitem[{\citenamefont{Aaij et~al.}(2020)}]{Aaij:2020fnh}
\bibinfo{author}{\bibfnamefont{R.}~\bibnamefont{Aaij}} \bibnamefont{et~al.}
  (\bibinfo{collaboration}{LHCb}), \bibinfo{journal}{Sci. Bull.}
  \textbf{\bibinfo{volume}{65}}, \bibinfo{pages}{1983}
  (\bibinfo{year}{2020})\relax
\mciteBstWouldAddEndPuncttrue
\mciteSetBstMidEndSepPunct{\mcitedefaultmidpunct}
{\mcitedefaultendpunct}{\mcitedefaultseppunct}\relax
\EndOfBibitem
\bibitem[{\citenamefont{Aaij et~al.}(2019)}]{Aaij:2019vzc}
\bibinfo{author}{\bibfnamefont{R.}~\bibnamefont{Aaij}} \bibnamefont{et~al.},
  \bibinfo{journal}{Phys. Rev. Lett.} \textbf{\bibinfo{volume}{122}},
  \bibinfo{pages}{222001} (\bibinfo{year}{2019})\relax
\mciteBstWouldAddEndPuncttrue
\mciteSetBstMidEndSepPunct{\mcitedefaultmidpunct}
{\mcitedefaultendpunct}{\mcitedefaultseppunct}\relax
\EndOfBibitem
\bibitem[{\citenamefont{Aaij et~al.}(2015)}]{Aaij:2015tga}
\bibinfo{author}{\bibfnamefont{R.}~\bibnamefont{Aaij}} \bibnamefont{et~al.},
  \bibinfo{journal}{Phys. Rev. Lett.} \textbf{\bibinfo{volume}{115}},
  \bibinfo{pages}{072001} (\bibinfo{year}{2015})\relax
\mciteBstWouldAddEndPuncttrue
\mciteSetBstMidEndSepPunct{\mcitedefaultmidpunct}
{\mcitedefaultendpunct}{\mcitedefaultseppunct}\relax
\EndOfBibitem
\bibitem[{\citenamefont{Adolph et~al.}(2015)}]{Adolph:2014rpp}
\bibinfo{author}{\bibfnamefont{C.}~\bibnamefont{Adolph}} \bibnamefont{et~al.}
  (\bibinfo{collaboration}{COMPASS}), \bibinfo{journal}{Phys. Lett. B}
  \textbf{\bibinfo{volume}{740}}, \bibinfo{pages}{303}
  (\bibinfo{year}{2015})\relax
\mciteBstWouldAddEndPuncttrue
\mciteSetBstMidEndSepPunct{\mcitedefaultmidpunct}
{\mcitedefaultendpunct}{\mcitedefaultseppunct}\relax
\EndOfBibitem
\bibitem[{\citenamefont{Briceno
  et~al.}(2018{\natexlab{a}})\citenamefont{Briceno, Dudek, and
  Young}}]{Briceno:2017max}
\bibinfo{author}{\bibfnamefont{R.~A.} \bibnamefont{Briceno}},
  \bibinfo{author}{\bibfnamefont{J.~J.} \bibnamefont{Dudek}}, \bibnamefont{and}
  \bibinfo{author}{\bibfnamefont{R.~D.} \bibnamefont{Young}},
  \bibinfo{journal}{Rev. Mod. Phys.} \textbf{\bibinfo{volume}{90}},
  \bibinfo{pages}{025001} (\bibinfo{year}{2018}{\natexlab{a}})\relax
\mciteBstWouldAddEndPuncttrue
\mciteSetBstMidEndSepPunct{\mcitedefaultmidpunct}
{\mcitedefaultendpunct}{\mcitedefaultseppunct}\relax
\EndOfBibitem
\bibitem[{\citenamefont{Shepherd et~al.}(2016)\citenamefont{Shepherd, Dudek,
  and Mitchell}}]{Shepherd:2016dni}
\bibinfo{author}{\bibfnamefont{M.~R.} \bibnamefont{Shepherd}},
  \bibinfo{author}{\bibfnamefont{J.~J.} \bibnamefont{Dudek}}, \bibnamefont{and}
  \bibinfo{author}{\bibfnamefont{R.~E.} \bibnamefont{Mitchell}},
  \bibinfo{journal}{Nature} \textbf{\bibinfo{volume}{534}},
  \bibinfo{pages}{487} (\bibinfo{year}{2016})\relax
\mciteBstWouldAddEndPuncttrue
\mciteSetBstMidEndSepPunct{\mcitedefaultmidpunct}
{\mcitedefaultendpunct}{\mcitedefaultseppunct}\relax
\EndOfBibitem
\bibitem[{\citenamefont{Roy}(1971)}]{Roy:1971tc}
\bibinfo{author}{\bibfnamefont{S.}~\bibnamefont{Roy}},
  \bibinfo{journal}{Phys.Lett.} \textbf{\bibinfo{volume}{B36}},
  \bibinfo{pages}{353} (\bibinfo{year}{1971})\relax
\mciteBstWouldAddEndPuncttrue
\mciteSetBstMidEndSepPunct{\mcitedefaultmidpunct}
{\mcitedefaultendpunct}{\mcitedefaultseppunct}\relax
\EndOfBibitem
\bibitem[{\citenamefont{Hite and Steiner}(1973)}]{Hite:1973pm}
\bibinfo{author}{\bibfnamefont{G.~E.} \bibnamefont{Hite}} \bibnamefont{and}
  \bibinfo{author}{\bibfnamefont{F.}~\bibnamefont{Steiner}},
  \bibinfo{journal}{Nuovo Cim. A} \textbf{\bibinfo{volume}{18}},
  \bibinfo{pages}{237} (\bibinfo{year}{1973})\relax
\mciteBstWouldAddEndPuncttrue
\mciteSetBstMidEndSepPunct{\mcitedefaultmidpunct}
{\mcitedefaultendpunct}{\mcitedefaultseppunct}\relax
\EndOfBibitem
\bibitem[{\citenamefont{Pelaez}(2016)}]{Pelaez:2015qba}
\bibinfo{author}{\bibfnamefont{J.~R.} \bibnamefont{Pelaez}},
  \bibinfo{journal}{Phys. Rept.} \textbf{\bibinfo{volume}{658}},
  \bibinfo{pages}{1} (\bibinfo{year}{2016})\relax
\mciteBstWouldAddEndPuncttrue
\mciteSetBstMidEndSepPunct{\mcitedefaultmidpunct}
{\mcitedefaultendpunct}{\mcitedefaultseppunct}\relax
\EndOfBibitem
\bibitem[{\citenamefont{Garcia-Martin
  et~al.}(2011{\natexlab{a}})\citenamefont{Garcia-Martin, Kaminski, Pelaez, and
  Ruiz~de Elvira}}]{GarciaMartin:2011jx}
\bibinfo{author}{\bibfnamefont{R.}~\bibnamefont{Garcia-Martin}},
  \bibinfo{author}{\bibfnamefont{R.}~\bibnamefont{Kaminski}},
  \bibinfo{author}{\bibfnamefont{J.}~\bibnamefont{Pelaez}}, \bibnamefont{and}
  \bibinfo{author}{\bibfnamefont{J.}~\bibnamefont{Ruiz~de Elvira}},
  \bibinfo{journal}{Phys.Rev.Lett.} \textbf{\bibinfo{volume}{107}},
  \bibinfo{pages}{072001} (\bibinfo{year}{2011}{\natexlab{a}})\relax
\mciteBstWouldAddEndPuncttrue
\mciteSetBstMidEndSepPunct{\mcitedefaultmidpunct}
{\mcitedefaultendpunct}{\mcitedefaultseppunct}\relax
\EndOfBibitem
\bibitem[{\citenamefont{Garcia-Martin
  et~al.}(2011{\natexlab{b}})\citenamefont{Garcia-Martin, Kaminski, Pelaez,
  Ruiz~de Elvira, and Yndurain}}]{GarciaMartin:2011cn}
\bibinfo{author}{\bibfnamefont{R.}~\bibnamefont{Garcia-Martin}},
  \bibinfo{author}{\bibfnamefont{R.}~\bibnamefont{Kaminski}},
  \bibinfo{author}{\bibfnamefont{J.}~\bibnamefont{Pelaez}},
  \bibinfo{author}{\bibfnamefont{J.}~\bibnamefont{Ruiz~de Elvira}},
  \bibnamefont{and} \bibinfo{author}{\bibfnamefont{F.}~\bibnamefont{Yndurain}},
  \bibinfo{journal}{Phys. Rev. D} \textbf{\bibinfo{volume}{83}},
  \bibinfo{pages}{074004} (\bibinfo{year}{2011}{\natexlab{b}})\relax
\mciteBstWouldAddEndPuncttrue
\mciteSetBstMidEndSepPunct{\mcitedefaultmidpunct}
{\mcitedefaultendpunct}{\mcitedefaultseppunct}\relax
\EndOfBibitem
\bibitem[{\citenamefont{Pelaez et~al.}(2019)\citenamefont{Pelaez, Rodas, and
  Ruiz De~Elvira}}]{Pelaez:2019eqa}
\bibinfo{author}{\bibfnamefont{J.~R.} \bibnamefont{Pelaez}},
  \bibinfo{author}{\bibfnamefont{A.}~\bibnamefont{Rodas}}, \bibnamefont{and}
  \bibinfo{author}{\bibfnamefont{J.}~\bibnamefont{Ruiz De~Elvira}},
  \bibinfo{journal}{Eur. Phys. J.} \textbf{\bibinfo{volume}{C79}},
  \bibinfo{pages}{1008} (\bibinfo{year}{2019})\relax
\mciteBstWouldAddEndPuncttrue
\mciteSetBstMidEndSepPunct{\mcitedefaultmidpunct}
{\mcitedefaultendpunct}{\mcitedefaultseppunct}\relax
\EndOfBibitem
\bibitem[{\citenamefont{Ananthanarayan
  et~al.}(2001)\citenamefont{Ananthanarayan, Colangelo, Gasser, and
  Leutwyler}}]{Ananthanarayan:2000ht}
\bibinfo{author}{\bibfnamefont{B.}~\bibnamefont{Ananthanarayan}},
  \bibinfo{author}{\bibfnamefont{G.}~\bibnamefont{Colangelo}},
  \bibinfo{author}{\bibfnamefont{J.}~\bibnamefont{Gasser}}, \bibnamefont{and}
  \bibinfo{author}{\bibfnamefont{H.}~\bibnamefont{Leutwyler}},
  \bibinfo{journal}{Phys.Rept.} \textbf{\bibinfo{volume}{353}},
  \bibinfo{pages}{207} (\bibinfo{year}{2001})\relax
\mciteBstWouldAddEndPuncttrue
\mciteSetBstMidEndSepPunct{\mcitedefaultmidpunct}
{\mcitedefaultendpunct}{\mcitedefaultseppunct}\relax
\EndOfBibitem
\bibitem[{\citenamefont{Caprini et~al.}(2006)\citenamefont{Caprini, Colangelo,
  and Leutwyler}}]{Caprini:2005zr}
\bibinfo{author}{\bibfnamefont{I.}~\bibnamefont{Caprini}},
  \bibinfo{author}{\bibfnamefont{G.}~\bibnamefont{Colangelo}},
  \bibnamefont{and}
  \bibinfo{author}{\bibfnamefont{H.}~\bibnamefont{Leutwyler}},
  \bibinfo{journal}{Phys. Rev. Lett.} \textbf{\bibinfo{volume}{96}},
  \bibinfo{pages}{132001} (\bibinfo{year}{2006})\relax
\mciteBstWouldAddEndPuncttrue
\mciteSetBstMidEndSepPunct{\mcitedefaultmidpunct}
{\mcitedefaultendpunct}{\mcitedefaultseppunct}\relax
\EndOfBibitem
\bibitem[{\citenamefont{Leutwyler}(2008)}]{Leutwyler:2008xd}
\bibinfo{author}{\bibfnamefont{H.}~\bibnamefont{Leutwyler}},
  \bibinfo{journal}{AIP Conf. Proc.} \textbf{\bibinfo{volume}{1030}},
  \bibinfo{pages}{46} (\bibinfo{year}{2008})\relax
\mciteBstWouldAddEndPuncttrue
\mciteSetBstMidEndSepPunct{\mcitedefaultmidpunct}
{\mcitedefaultendpunct}{\mcitedefaultseppunct}\relax
\EndOfBibitem
\bibitem[{\citenamefont{Colangelo et~al.}(2001)\citenamefont{Colangelo, Gasser,
  and Leutwyler}}]{Colangelo:2001df}
\bibinfo{author}{\bibfnamefont{G.}~\bibnamefont{Colangelo}},
  \bibinfo{author}{\bibfnamefont{J.}~\bibnamefont{Gasser}}, \bibnamefont{and}
  \bibinfo{author}{\bibfnamefont{H.}~\bibnamefont{Leutwyler}},
  \bibinfo{journal}{Nucl. Phys.} \textbf{\bibinfo{volume}{B603}},
  \bibinfo{pages}{125} (\bibinfo{year}{2001})\relax
\mciteBstWouldAddEndPuncttrue
\mciteSetBstMidEndSepPunct{\mcitedefaultmidpunct}
{\mcitedefaultendpunct}{\mcitedefaultseppunct}\relax
\EndOfBibitem
\bibitem[{\citenamefont{Buettiker et~al.}(2004)\citenamefont{Buettiker,
  Descotes-Genon, and Moussallam}}]{Buettiker:2003pp}
\bibinfo{author}{\bibfnamefont{P.}~\bibnamefont{Buettiker}},
  \bibinfo{author}{\bibfnamefont{S.}~\bibnamefont{Descotes-Genon}},
  \bibnamefont{and}
  \bibinfo{author}{\bibfnamefont{B.}~\bibnamefont{Moussallam}},
  \bibinfo{journal}{Eur.Phys.J.} \textbf{\bibinfo{volume}{C33}},
  \bibinfo{pages}{409} (\bibinfo{year}{2004})\relax
\mciteBstWouldAddEndPuncttrue
\mciteSetBstMidEndSepPunct{\mcitedefaultmidpunct}
{\mcitedefaultendpunct}{\mcitedefaultseppunct}\relax
\EndOfBibitem
\bibitem[{\citenamefont{Descotes-Genon and
  Moussallam}(2006)}]{DescotesGenon:2006uk}
\bibinfo{author}{\bibfnamefont{S.}~\bibnamefont{Descotes-Genon}}
  \bibnamefont{and}
  \bibinfo{author}{\bibfnamefont{B.}~\bibnamefont{Moussallam}},
  \bibinfo{journal}{Eur. Phys. J.} \textbf{\bibinfo{volume}{C48}},
  \bibinfo{pages}{553} (\bibinfo{year}{2006})\relax
\mciteBstWouldAddEndPuncttrue
\mciteSetBstMidEndSepPunct{\mcitedefaultmidpunct}
{\mcitedefaultendpunct}{\mcitedefaultseppunct}\relax
\EndOfBibitem
\bibitem[{\citenamefont{Pelaez and Rodas}(2020)}]{Pelaez:2020uiw}
\bibinfo{author}{\bibfnamefont{J.~R.} \bibnamefont{Pelaez}} \bibnamefont{and}
  \bibinfo{author}{\bibfnamefont{A.}~\bibnamefont{Rodas}},
  \bibinfo{journal}{Phys. Rev. Lett.} \textbf{\bibinfo{volume}{124}},
  \bibinfo{pages}{172001} (\bibinfo{year}{2020})\relax
\mciteBstWouldAddEndPuncttrue
\mciteSetBstMidEndSepPunct{\mcitedefaultmidpunct}
{\mcitedefaultendpunct}{\mcitedefaultseppunct}\relax
\EndOfBibitem
\bibitem[{\citenamefont{Pel\'aez and Rodas}(2020)}]{Pelaez:2020gnd}
\bibinfo{author}{\bibfnamefont{J.}~\bibnamefont{Pel\'aez}} \bibnamefont{and}
  \bibinfo{author}{\bibfnamefont{A.}~\bibnamefont{Rodas}},
  \bibinfo{journal}{arXiv: 2010.11222}  (\bibinfo{year}{2020})\relax
\mciteBstWouldAddEndPuncttrue
\mciteSetBstMidEndSepPunct{\mcitedefaultmidpunct}
{\mcitedefaultendpunct}{\mcitedefaultseppunct}\relax
\EndOfBibitem
\bibitem[{\citenamefont{Chew and Mandelstam}(1960)}]{Chew:1960iv}
\bibinfo{author}{\bibfnamefont{G.~F.} \bibnamefont{Chew}} \bibnamefont{and}
  \bibinfo{author}{\bibfnamefont{S.}~\bibnamefont{Mandelstam}},
  \bibinfo{journal}{Phys.Rev.} \textbf{\bibinfo{volume}{119}},
  \bibinfo{pages}{467} (\bibinfo{year}{1960})\relax
\mciteBstWouldAddEndPuncttrue
\mciteSetBstMidEndSepPunct{\mcitedefaultmidpunct}
{\mcitedefaultendpunct}{\mcitedefaultseppunct}\relax
\EndOfBibitem
\bibitem[{\citenamefont{Oller and Oset}(1999)}]{Oller:1998zr}
\bibinfo{author}{\bibfnamefont{J.}~\bibnamefont{Oller}} \bibnamefont{and}
  \bibinfo{author}{\bibfnamefont{E.}~\bibnamefont{Oset}},
  \bibinfo{journal}{Phys.Rev.} \textbf{\bibinfo{volume}{D60}},
  \bibinfo{pages}{074023} (\bibinfo{year}{1999})\relax
\mciteBstWouldAddEndPuncttrue
\mciteSetBstMidEndSepPunct{\mcitedefaultmidpunct}
{\mcitedefaultendpunct}{\mcitedefaultseppunct}\relax
\EndOfBibitem
\bibitem[{\citenamefont{Szczepaniak et~al.}(2010)\citenamefont{Szczepaniak,
  Guo, Battaglieri, and De~Vita}}]{Szczepaniak:2010re}
\bibinfo{author}{\bibfnamefont{A.~P.} \bibnamefont{Szczepaniak}},
  \bibinfo{author}{\bibfnamefont{P.}~\bibnamefont{Guo}},
  \bibinfo{author}{\bibfnamefont{M.}~\bibnamefont{Battaglieri}},
  \bibnamefont{and} \bibinfo{author}{\bibfnamefont{R.}~\bibnamefont{De~Vita}},
  \bibinfo{journal}{Phys.Rev.} \textbf{\bibinfo{volume}{D82}},
  \bibinfo{pages}{036006} (\bibinfo{year}{2010})\relax
\mciteBstWouldAddEndPuncttrue
\mciteSetBstMidEndSepPunct{\mcitedefaultmidpunct}
{\mcitedefaultendpunct}{\mcitedefaultseppunct}\relax
\EndOfBibitem
\bibitem[{\citenamefont{Guo et~al.}(2010)\citenamefont{Guo, Mitchell, and
  Szczepaniak}}]{Guo:2010gx}
\bibinfo{author}{\bibfnamefont{P.}~\bibnamefont{Guo}},
  \bibinfo{author}{\bibfnamefont{R.}~\bibnamefont{Mitchell}}, \bibnamefont{and}
  \bibinfo{author}{\bibfnamefont{A.~P.} \bibnamefont{Szczepaniak}},
  \bibinfo{journal}{Phys.Rev.} \textbf{\bibinfo{volume}{D82}},
  \bibinfo{pages}{094002} (\bibinfo{year}{2010})\relax
\mciteBstWouldAddEndPuncttrue
\mciteSetBstMidEndSepPunct{\mcitedefaultmidpunct}
{\mcitedefaultendpunct}{\mcitedefaultseppunct}\relax
\EndOfBibitem
\bibitem[{\citenamefont{Gasparyan and Lutz}(2010)}]{Gasparyan:2010xz}
\bibinfo{author}{\bibfnamefont{A.}~\bibnamefont{Gasparyan}} \bibnamefont{and}
  \bibinfo{author}{\bibfnamefont{M.~F.~M.} \bibnamefont{Lutz}},
  \bibinfo{journal}{Nucl.Phys.} \textbf{\bibinfo{volume}{A848}},
  \bibinfo{pages}{126} (\bibinfo{year}{2010})\relax
\mciteBstWouldAddEndPuncttrue
\mciteSetBstMidEndSepPunct{\mcitedefaultmidpunct}
{\mcitedefaultendpunct}{\mcitedefaultseppunct}\relax
\EndOfBibitem
\bibitem[{\citenamefont{Danilkin
  et~al.}(2011{\natexlab{a}})\citenamefont{Danilkin, Gasparyan, and
  Lutz}}]{Danilkin:2010xd}
\bibinfo{author}{\bibfnamefont{I.~V.} \bibnamefont{Danilkin}},
  \bibinfo{author}{\bibfnamefont{A.~M.} \bibnamefont{Gasparyan}},
  \bibnamefont{and} \bibinfo{author}{\bibfnamefont{M.~F.~M.}
  \bibnamefont{Lutz}}, \bibinfo{journal}{Phys.Lett.}
  \textbf{\bibinfo{volume}{B697}}, \bibinfo{pages}{147}
  (\bibinfo{year}{2011}{\natexlab{a}})\relax
\mciteBstWouldAddEndPuncttrue
\mciteSetBstMidEndSepPunct{\mcitedefaultmidpunct}
{\mcitedefaultendpunct}{\mcitedefaultseppunct}\relax
\EndOfBibitem
\bibitem[{\citenamefont{Gasparyan et~al.}(2011)\citenamefont{Gasparyan, Lutz,
  and Pasquini}}]{Gasparyan:2011yw}
\bibinfo{author}{\bibfnamefont{A.}~\bibnamefont{Gasparyan}},
  \bibinfo{author}{\bibfnamefont{M.}~\bibnamefont{Lutz}}, \bibnamefont{and}
  \bibinfo{author}{\bibfnamefont{B.}~\bibnamefont{Pasquini}},
  \bibinfo{journal}{Nucl.Phys.} \textbf{\bibinfo{volume}{A866}},
  \bibinfo{pages}{79} (\bibinfo{year}{2011})\relax
\mciteBstWouldAddEndPuncttrue
\mciteSetBstMidEndSepPunct{\mcitedefaultmidpunct}
{\mcitedefaultendpunct}{\mcitedefaultseppunct}\relax
\EndOfBibitem
\bibitem[{\citenamefont{Gasparyan et~al.}(2013)\citenamefont{Gasparyan, Lutz,
  and Epelbaum}}]{Gasparyan:2012km}
\bibinfo{author}{\bibfnamefont{A.~M.} \bibnamefont{Gasparyan}},
  \bibinfo{author}{\bibfnamefont{M.~F.~M.} \bibnamefont{Lutz}},
  \bibnamefont{and} \bibinfo{author}{\bibfnamefont{E.}~\bibnamefont{Epelbaum}},
  \bibinfo{journal}{Eur.Phys.J.} \textbf{\bibinfo{volume}{A49}},
  \bibinfo{pages}{115} (\bibinfo{year}{2013})\relax
\mciteBstWouldAddEndPuncttrue
\mciteSetBstMidEndSepPunct{\mcitedefaultmidpunct}
{\mcitedefaultendpunct}{\mcitedefaultseppunct}\relax
\EndOfBibitem
\bibitem[{\citenamefont{Danilkin
  et~al.}(2011{\natexlab{b}})\citenamefont{Danilkin, Gil, and
  Lutz}}]{Danilkin:2011fz}
\bibinfo{author}{\bibfnamefont{I.~V.} \bibnamefont{Danilkin}},
  \bibinfo{author}{\bibfnamefont{L.~I.~R.} \bibnamefont{Gil}},
  \bibnamefont{and} \bibinfo{author}{\bibfnamefont{M.~F.~M.}
  \bibnamefont{Lutz}}, \bibinfo{journal}{Phys.Lett.}
  \textbf{\bibinfo{volume}{B703}}, \bibinfo{pages}{504}
  (\bibinfo{year}{2011}{\natexlab{b}})\relax
\mciteBstWouldAddEndPuncttrue
\mciteSetBstMidEndSepPunct{\mcitedefaultmidpunct}
{\mcitedefaultendpunct}{\mcitedefaultseppunct}\relax
\EndOfBibitem
\bibitem[{\citenamefont{Danilkin and Lutz}(2012)}]{Danilkin:2012ap}
\bibinfo{author}{\bibfnamefont{I.}~\bibnamefont{Danilkin}} \bibnamefont{and}
  \bibinfo{author}{\bibfnamefont{M.}~\bibnamefont{Lutz}}, \bibinfo{journal}{EPJ
  Web Conf.} \textbf{\bibinfo{volume}{37}}, \bibinfo{pages}{08007}
  (\bibinfo{year}{2012})\relax
\mciteBstWouldAddEndPuncttrue
\mciteSetBstMidEndSepPunct{\mcitedefaultmidpunct}
{\mcitedefaultendpunct}{\mcitedefaultseppunct}\relax
\EndOfBibitem
\bibitem[{\citenamefont{Guo et~al.}(2015)\citenamefont{Guo, Danilkin, Schott,
  Fern\'andez-Ram\'\i{}rez, Mathieu, and Szczepaniak}}]{Guo:2015zqa}
\bibinfo{author}{\bibfnamefont{P.}~\bibnamefont{Guo}},
  \bibinfo{author}{\bibfnamefont{I.~V.} \bibnamefont{Danilkin}},
  \bibinfo{author}{\bibfnamefont{D.}~\bibnamefont{Schott}},
  \bibinfo{author}{\bibfnamefont{C.}~\bibnamefont{Fern\'andez-Ram\'\i{}rez}},
  \bibinfo{author}{\bibfnamefont{V.}~\bibnamefont{Mathieu}}, \bibnamefont{and}
  \bibinfo{author}{\bibfnamefont{A.~P.} \bibnamefont{Szczepaniak}},
  \bibinfo{journal}{Phys. Rev. D} \textbf{\bibinfo{volume}{92}},
  \bibinfo{pages}{054016} (\bibinfo{year}{2015})\relax
\mciteBstWouldAddEndPuncttrue
\mciteSetBstMidEndSepPunct{\mcitedefaultmidpunct}
{\mcitedefaultendpunct}{\mcitedefaultseppunct}\relax
\EndOfBibitem
\bibitem[{\citenamefont{Guo et~al.}(2017)\citenamefont{Guo, Danilkin,
  Fern\'andez-Ram\'\i{}rez, Mathieu, and Szczepaniak}}]{Guo:2016wsi}
\bibinfo{author}{\bibfnamefont{P.}~\bibnamefont{Guo}},
  \bibinfo{author}{\bibfnamefont{I.}~\bibnamefont{Danilkin}},
  \bibinfo{author}{\bibfnamefont{C.}~\bibnamefont{Fern\'andez-Ram\'\i{}rez}},
  \bibinfo{author}{\bibfnamefont{V.}~\bibnamefont{Mathieu}}, \bibnamefont{and}
  \bibinfo{author}{\bibfnamefont{A.}~\bibnamefont{Szczepaniak}},
  \bibinfo{journal}{Phys. Lett. B} \textbf{\bibinfo{volume}{771}},
  \bibinfo{pages}{497} (\bibinfo{year}{2017})\relax
\mciteBstWouldAddEndPuncttrue
\mciteSetBstMidEndSepPunct{\mcitedefaultmidpunct}
{\mcitedefaultendpunct}{\mcitedefaultseppunct}\relax
\EndOfBibitem
\bibitem[{\citenamefont{Colangelo
  et~al.}(2017{\natexlab{a}})\citenamefont{Colangelo, Lanz, Leutwyler, and
  Passemar}}]{Colangelo:2016jmc}
\bibinfo{author}{\bibfnamefont{G.}~\bibnamefont{Colangelo}},
  \bibinfo{author}{\bibfnamefont{S.}~\bibnamefont{Lanz}},
  \bibinfo{author}{\bibfnamefont{H.}~\bibnamefont{Leutwyler}},
  \bibnamefont{and} \bibinfo{author}{\bibfnamefont{E.}~\bibnamefont{Passemar}},
  \bibinfo{journal}{Phys. Rev. Lett.} \textbf{\bibinfo{volume}{118}},
  \bibinfo{pages}{022001} (\bibinfo{year}{2017}{\natexlab{a}})\relax
\mciteBstWouldAddEndPuncttrue
\mciteSetBstMidEndSepPunct{\mcitedefaultmidpunct}
{\mcitedefaultendpunct}{\mcitedefaultseppunct}\relax
\EndOfBibitem
\bibitem[{\citenamefont{Colangelo et~al.}(2018)\citenamefont{Colangelo, Lanz,
  Leutwyler, and Passemar}}]{Colangelo:2018jxw}
\bibinfo{author}{\bibfnamefont{G.}~\bibnamefont{Colangelo}},
  \bibinfo{author}{\bibfnamefont{S.}~\bibnamefont{Lanz}},
  \bibinfo{author}{\bibfnamefont{H.}~\bibnamefont{Leutwyler}},
  \bibnamefont{and} \bibinfo{author}{\bibfnamefont{E.}~\bibnamefont{Passemar}},
  \bibinfo{journal}{Eur. Phys. J. C} \textbf{\bibinfo{volume}{78}},
  \bibinfo{pages}{947} (\bibinfo{year}{2018})\relax
\mciteBstWouldAddEndPuncttrue
\mciteSetBstMidEndSepPunct{\mcitedefaultmidpunct}
{\mcitedefaultendpunct}{\mcitedefaultseppunct}\relax
\EndOfBibitem
\bibitem[{\citenamefont{Albaladejo and Moussallam}(2017)}]{Albaladejo:2017hhj}
\bibinfo{author}{\bibfnamefont{M.}~\bibnamefont{Albaladejo}} \bibnamefont{and}
  \bibinfo{author}{\bibfnamefont{B.}~\bibnamefont{Moussallam}},
  \bibinfo{journal}{Eur. Phys. J. C} \textbf{\bibinfo{volume}{77}},
  \bibinfo{pages}{508} (\bibinfo{year}{2017})\relax
\mciteBstWouldAddEndPuncttrue
\mciteSetBstMidEndSepPunct{\mcitedefaultmidpunct}
{\mcitedefaultendpunct}{\mcitedefaultseppunct}\relax
\EndOfBibitem
\bibitem[{\citenamefont{Isken et~al.}(2017)\citenamefont{Isken, Kubis,
  Schneider, and Stoffer}}]{Isken:2017dkw}
\bibinfo{author}{\bibfnamefont{T.}~\bibnamefont{Isken}},
  \bibinfo{author}{\bibfnamefont{B.}~\bibnamefont{Kubis}},
  \bibinfo{author}{\bibfnamefont{S.~P.} \bibnamefont{Schneider}},
  \bibnamefont{and} \bibinfo{author}{\bibfnamefont{P.}~\bibnamefont{Stoffer}},
  \bibinfo{journal}{Eur. Phys. J. C} \textbf{\bibinfo{volume}{77}},
  \bibinfo{pages}{489} (\bibinfo{year}{2017})\relax
\mciteBstWouldAddEndPuncttrue
\mciteSetBstMidEndSepPunct{\mcitedefaultmidpunct}
{\mcitedefaultendpunct}{\mcitedefaultseppunct}\relax
\EndOfBibitem
\bibitem[{\citenamefont{Gonz\`alez-Sol\'\i{}s and
  Passemar}(2018)}]{Gonzalez-Solis:2018xnw}
\bibinfo{author}{\bibfnamefont{S.}~\bibnamefont{Gonz\`alez-Sol\'\i{}s}}
  \bibnamefont{and} \bibinfo{author}{\bibfnamefont{E.}~\bibnamefont{Passemar}},
  \bibinfo{journal}{Eur. Phys. J. C} \textbf{\bibinfo{volume}{78}},
  \bibinfo{pages}{758} (\bibinfo{year}{2018})\relax
\mciteBstWouldAddEndPuncttrue
\mciteSetBstMidEndSepPunct{\mcitedefaultmidpunct}
{\mcitedefaultendpunct}{\mcitedefaultseppunct}\relax
\EndOfBibitem
\bibitem[{\citenamefont{Gan et~al.}(2020)\citenamefont{Gan, Kubis, Passemar,
  and Tulin}}]{Gan:2020aco}
\bibinfo{author}{\bibfnamefont{L.}~\bibnamefont{Gan}},
  \bibinfo{author}{\bibfnamefont{B.}~\bibnamefont{Kubis}},
  \bibinfo{author}{\bibfnamefont{E.}~\bibnamefont{Passemar}}, \bibnamefont{and}
  \bibinfo{author}{\bibfnamefont{S.}~\bibnamefont{Tulin}},
  \bibinfo{journal}{2007.00664 [hep-ph]}  (\bibinfo{year}{2020})\relax
\mciteBstWouldAddEndPuncttrue
\mciteSetBstMidEndSepPunct{\mcitedefaultmidpunct}
{\mcitedefaultendpunct}{\mcitedefaultseppunct}\relax
\EndOfBibitem
\bibitem[{\citenamefont{Garcia-Martin and
  Moussallam}(2010)}]{GarciaMartin:2010cw}
\bibinfo{author}{\bibfnamefont{R.}~\bibnamefont{Garcia-Martin}}
  \bibnamefont{and}
  \bibinfo{author}{\bibfnamefont{B.}~\bibnamefont{Moussallam}},
  \bibinfo{journal}{Eur. Phys. J.} \textbf{\bibinfo{volume}{C70}},
  \bibinfo{pages}{155} (\bibinfo{year}{2010})\relax
\mciteBstWouldAddEndPuncttrue
\mciteSetBstMidEndSepPunct{\mcitedefaultmidpunct}
{\mcitedefaultendpunct}{\mcitedefaultseppunct}\relax
\EndOfBibitem
\bibitem[{\citenamefont{Hoferichter et~al.}(2011)\citenamefont{Hoferichter,
  Phillips, and Schat}}]{Hoferichter:2011wk}
\bibinfo{author}{\bibfnamefont{M.}~\bibnamefont{Hoferichter}},
  \bibinfo{author}{\bibfnamefont{D.~R.} \bibnamefont{Phillips}},
  \bibnamefont{and} \bibinfo{author}{\bibfnamefont{C.}~\bibnamefont{Schat}},
  \bibinfo{journal}{Eur. Phys. J.} \textbf{\bibinfo{volume}{C71}},
  \bibinfo{pages}{1743} (\bibinfo{year}{2011})\relax
\mciteBstWouldAddEndPuncttrue
\mciteSetBstMidEndSepPunct{\mcitedefaultmidpunct}
{\mcitedefaultendpunct}{\mcitedefaultseppunct}\relax
\EndOfBibitem
\bibitem[{\citenamefont{Dai and Pennington}(2014{\natexlab{a}})}]{Dai:2014zta}
\bibinfo{author}{\bibfnamefont{L.-Y.} \bibnamefont{Dai}} \bibnamefont{and}
  \bibinfo{author}{\bibfnamefont{M.~R.} \bibnamefont{Pennington}},
  \bibinfo{journal}{Phys. Rev.} \textbf{\bibinfo{volume}{D90}},
  \bibinfo{pages}{036004} (\bibinfo{year}{2014}{\natexlab{a}})\relax
\mciteBstWouldAddEndPuncttrue
\mciteSetBstMidEndSepPunct{\mcitedefaultmidpunct}
{\mcitedefaultendpunct}{\mcitedefaultseppunct}\relax
\EndOfBibitem
\bibitem[{\citenamefont{Molnar et~al.}(2019)\citenamefont{Molnar, Danilkin, and
  Vanderhaeghen}}]{Molnar:2019uos}
\bibinfo{author}{\bibfnamefont{D.~A.} \bibnamefont{Molnar}},
  \bibinfo{author}{\bibfnamefont{I.}~\bibnamefont{Danilkin}}, \bibnamefont{and}
  \bibinfo{author}{\bibfnamefont{M.}~\bibnamefont{Vanderhaeghen}},
  \bibinfo{journal}{Phys. Lett. B} \textbf{\bibinfo{volume}{797}},
  \bibinfo{pages}{134851} (\bibinfo{year}{2019})\relax
\mciteBstWouldAddEndPuncttrue
\mciteSetBstMidEndSepPunct{\mcitedefaultmidpunct}
{\mcitedefaultendpunct}{\mcitedefaultseppunct}\relax
\EndOfBibitem
\bibitem[{\citenamefont{Chen et~al.}(2019)\citenamefont{Chen, Dai, Guo, and
  Kubis}}]{Chen:2019mgp}
\bibinfo{author}{\bibfnamefont{Y.-H.} \bibnamefont{Chen}},
  \bibinfo{author}{\bibfnamefont{L.-Y.} \bibnamefont{Dai}},
  \bibinfo{author}{\bibfnamefont{F.-K.} \bibnamefont{Guo}}, \bibnamefont{and}
  \bibinfo{author}{\bibfnamefont{B.}~\bibnamefont{Kubis}},
  \bibinfo{journal}{Phys. Rev. D} \textbf{\bibinfo{volume}{99}},
  \bibinfo{pages}{074016} (\bibinfo{year}{2019})\relax
\mciteBstWouldAddEndPuncttrue
\mciteSetBstMidEndSepPunct{\mcitedefaultmidpunct}
{\mcitedefaultendpunct}{\mcitedefaultseppunct}\relax
\EndOfBibitem
\bibitem[{\citenamefont{Danilkin
  et~al.}(2020{\natexlab{a}})\citenamefont{Danilkin, Molnar, and
  Vanderhaeghen}}]{Danilkin:2020kce}
\bibinfo{author}{\bibfnamefont{I.}~\bibnamefont{Danilkin}},
  \bibinfo{author}{\bibfnamefont{D.~A.} \bibnamefont{Molnar}},
  \bibnamefont{and}
  \bibinfo{author}{\bibfnamefont{M.}~\bibnamefont{Vanderhaeghen}},
  \bibinfo{journal}{Phys. Rev. D} \textbf{\bibinfo{volume}{102}},
  \bibinfo{pages}{016019} (\bibinfo{year}{2020}{\natexlab{a}})\relax
\mciteBstWouldAddEndPuncttrue
\mciteSetBstMidEndSepPunct{\mcitedefaultmidpunct}
{\mcitedefaultendpunct}{\mcitedefaultseppunct}\relax
\EndOfBibitem
\bibitem[{\citenamefont{Niecknig and Kubis}(2015)}]{Niecknig:2015ija}
\bibinfo{author}{\bibfnamefont{F.}~\bibnamefont{Niecknig}} \bibnamefont{and}
  \bibinfo{author}{\bibfnamefont{B.}~\bibnamefont{Kubis}},
  \bibinfo{journal}{JHEP} \textbf{\bibinfo{volume}{10}}, \bibinfo{pages}{142}
  (\bibinfo{year}{2015})\relax
\mciteBstWouldAddEndPuncttrue
\mciteSetBstMidEndSepPunct{\mcitedefaultmidpunct}
{\mcitedefaultendpunct}{\mcitedefaultseppunct}\relax
\EndOfBibitem
\bibitem[{\citenamefont{Niecknig and Kubis}(2018)}]{Niecknig:2017ylb}
\bibinfo{author}{\bibfnamefont{F.}~\bibnamefont{Niecknig}} \bibnamefont{and}
  \bibinfo{author}{\bibfnamefont{B.}~\bibnamefont{Kubis}},
  \bibinfo{journal}{Phys. Lett. B} \textbf{\bibinfo{volume}{780}},
  \bibinfo{pages}{471} (\bibinfo{year}{2018})\relax
\mciteBstWouldAddEndPuncttrue
\mciteSetBstMidEndSepPunct{\mcitedefaultmidpunct}
{\mcitedefaultendpunct}{\mcitedefaultseppunct}\relax
\EndOfBibitem
\bibitem[{\citenamefont{Pelaez and Rodas}(2018)}]{Pelaez:2018qny}
\bibinfo{author}{\bibfnamefont{J.~R.} \bibnamefont{Pelaez}} \bibnamefont{and}
  \bibinfo{author}{\bibfnamefont{A.}~\bibnamefont{Rodas}},
  \bibinfo{journal}{Eur. Phys. J.} \textbf{\bibinfo{volume}{C78}},
  \bibinfo{pages}{897} (\bibinfo{year}{2018})\relax
\mciteBstWouldAddEndPuncttrue
\mciteSetBstMidEndSepPunct{\mcitedefaultmidpunct}
{\mcitedefaultendpunct}{\mcitedefaultseppunct}\relax
\EndOfBibitem
\bibitem[{\citenamefont{Lang et~al.}(2012)\citenamefont{Lang, Leskovec, Mohler,
  and Prelovsek}}]{Lang:2012sv}
\bibinfo{author}{\bibfnamefont{C.}~\bibnamefont{Lang}},
  \bibinfo{author}{\bibfnamefont{L.}~\bibnamefont{Leskovec}},
  \bibinfo{author}{\bibfnamefont{D.}~\bibnamefont{Mohler}}, \bibnamefont{and}
  \bibinfo{author}{\bibfnamefont{S.}~\bibnamefont{Prelovsek}},
  \bibinfo{journal}{Phys. Rev. D} \textbf{\bibinfo{volume}{86}},
  \bibinfo{pages}{054508} (\bibinfo{year}{2012})\relax
\mciteBstWouldAddEndPuncttrue
\mciteSetBstMidEndSepPunct{\mcitedefaultmidpunct}
{\mcitedefaultendpunct}{\mcitedefaultseppunct}\relax
\EndOfBibitem
\bibitem[{\citenamefont{Prelovsek et~al.}(2010)\citenamefont{Prelovsek, Draper,
  Lang, Limmer, Liu, Mathur, and Mohler}}]{Prelovsek:2010kg}
\bibinfo{author}{\bibfnamefont{S.}~\bibnamefont{Prelovsek}},
  \bibinfo{author}{\bibfnamefont{T.}~\bibnamefont{Draper}},
  \bibinfo{author}{\bibfnamefont{C.~B.} \bibnamefont{Lang}},
  \bibinfo{author}{\bibfnamefont{M.}~\bibnamefont{Limmer}},
  \bibinfo{author}{\bibfnamefont{K.-F.} \bibnamefont{Liu}},
  \bibinfo{author}{\bibfnamefont{N.}~\bibnamefont{Mathur}}, \bibnamefont{and}
  \bibinfo{author}{\bibfnamefont{D.}~\bibnamefont{Mohler}},
  \bibinfo{journal}{Phys. Rev. D} \textbf{\bibinfo{volume}{82}},
  \bibinfo{pages}{094507} (\bibinfo{year}{2010})\relax
\mciteBstWouldAddEndPuncttrue
\mciteSetBstMidEndSepPunct{\mcitedefaultmidpunct}
{\mcitedefaultendpunct}{\mcitedefaultseppunct}\relax
\EndOfBibitem
\bibitem[{\citenamefont{Briceno et~al.}(2017)\citenamefont{Briceno, Dudek,
  Edwards, and Wilson}}]{Briceno:2016mjc}
\bibinfo{author}{\bibfnamefont{R.~A.} \bibnamefont{Briceno}},
  \bibinfo{author}{\bibfnamefont{J.~J.} \bibnamefont{Dudek}},
  \bibinfo{author}{\bibfnamefont{R.~G.} \bibnamefont{Edwards}},
  \bibnamefont{and} \bibinfo{author}{\bibfnamefont{D.~J.}
  \bibnamefont{Wilson}}, \bibinfo{journal}{Phys. Rev. Lett.}
  \textbf{\bibinfo{volume}{118}}, \bibinfo{pages}{022002}
  (\bibinfo{year}{2017})\relax
\mciteBstWouldAddEndPuncttrue
\mciteSetBstMidEndSepPunct{\mcitedefaultmidpunct}
{\mcitedefaultendpunct}{\mcitedefaultseppunct}\relax
\EndOfBibitem
\bibitem[{\citenamefont{Liu et~al.}(2017)}]{Liu:2016cba}
\bibinfo{author}{\bibfnamefont{L.}~\bibnamefont{Liu}} \bibnamefont{et~al.},
  \bibinfo{journal}{Phys. Rev.} \textbf{\bibinfo{volume}{D96}},
  \bibinfo{pages}{054516} (\bibinfo{year}{2017})\relax
\mciteBstWouldAddEndPuncttrue
\mciteSetBstMidEndSepPunct{\mcitedefaultmidpunct}
{\mcitedefaultendpunct}{\mcitedefaultseppunct}\relax
\EndOfBibitem
\bibitem[{\citenamefont{Fu and Chen}(2018)}]{Fu:2017apw}
\bibinfo{author}{\bibfnamefont{Z.}~\bibnamefont{Fu}} \bibnamefont{and}
  \bibinfo{author}{\bibfnamefont{X.}~\bibnamefont{Chen}},
  \bibinfo{journal}{Phys. Rev.} \textbf{\bibinfo{volume}{D98}},
  \bibinfo{pages}{014514} (\bibinfo{year}{2018})\relax
\mciteBstWouldAddEndPuncttrue
\mciteSetBstMidEndSepPunct{\mcitedefaultmidpunct}
{\mcitedefaultendpunct}{\mcitedefaultseppunct}\relax
\EndOfBibitem
\bibitem[{\citenamefont{Guo et~al.}(2018)\citenamefont{Guo, Alexandru, Molina,
  Mai, and Döring}}]{Guo:2018zss}
\bibinfo{author}{\bibfnamefont{D.}~\bibnamefont{Guo}},
  \bibinfo{author}{\bibfnamefont{A.}~\bibnamefont{Alexandru}},
  \bibinfo{author}{\bibfnamefont{R.}~\bibnamefont{Molina}},
  \bibinfo{author}{\bibfnamefont{M.}~\bibnamefont{Mai}}, \bibnamefont{and}
  \bibinfo{author}{\bibfnamefont{M.}~\bibnamefont{Döring}},
  \bibinfo{journal}{Phys. Rev.} \textbf{\bibinfo{volume}{D98}},
  \bibinfo{pages}{014507} (\bibinfo{year}{2018})\relax
\mciteBstWouldAddEndPuncttrue
\mciteSetBstMidEndSepPunct{\mcitedefaultmidpunct}
{\mcitedefaultendpunct}{\mcitedefaultseppunct}\relax
\EndOfBibitem
\bibitem[{\citenamefont{Mai et~al.}(2019)\citenamefont{Mai, Culver, Alexandru,
  D\"oring, and Lee}}]{Mai:2019pqr}
\bibinfo{author}{\bibfnamefont{M.}~\bibnamefont{Mai}},
  \bibinfo{author}{\bibfnamefont{C.}~\bibnamefont{Culver}},
  \bibinfo{author}{\bibfnamefont{A.}~\bibnamefont{Alexandru}},
  \bibinfo{author}{\bibfnamefont{M.}~\bibnamefont{D\"oring}}, \bibnamefont{and}
  \bibinfo{author}{\bibfnamefont{F.~X.} \bibnamefont{Lee}},
  \bibinfo{journal}{Phys. Rev. D} \textbf{\bibinfo{volume}{100}},
  \bibinfo{pages}{114514} (\bibinfo{year}{2019})\relax
\mciteBstWouldAddEndPuncttrue
\mciteSetBstMidEndSepPunct{\mcitedefaultmidpunct}
{\mcitedefaultendpunct}{\mcitedefaultseppunct}\relax
\EndOfBibitem
\bibitem[{\citenamefont{Wilson et~al.}(2019)\citenamefont{Wilson, Briceno,
  Dudek, Edwards, and Thomas}}]{Wilson:2019wfr}
\bibinfo{author}{\bibfnamefont{D.~J.} \bibnamefont{Wilson}},
  \bibinfo{author}{\bibfnamefont{R.~A.} \bibnamefont{Briceno}},
  \bibinfo{author}{\bibfnamefont{J.~J.} \bibnamefont{Dudek}},
  \bibinfo{author}{\bibfnamefont{R.~G.} \bibnamefont{Edwards}},
  \bibnamefont{and} \bibinfo{author}{\bibfnamefont{C.~E.}
  \bibnamefont{Thomas}}, \bibinfo{journal}{Phys. Rev. Lett.}
  \textbf{\bibinfo{volume}{123}}, \bibinfo{pages}{042002}
  (\bibinfo{year}{2019})\relax
\mciteBstWouldAddEndPuncttrue
\mciteSetBstMidEndSepPunct{\mcitedefaultmidpunct}
{\mcitedefaultendpunct}{\mcitedefaultseppunct}\relax
\EndOfBibitem
\bibitem[{\citenamefont{Rendon et~al.}(2020)\citenamefont{Rendon, Leskovec,
  Meinel, Negele, Paul, Petschlies, Pochinsky, Silvi, and
  Syritsyn}}]{Rendon:2020rtw}
\bibinfo{author}{\bibfnamefont{G.}~\bibnamefont{Rendon}},
  \bibinfo{author}{\bibfnamefont{L.}~\bibnamefont{Leskovec}},
  \bibinfo{author}{\bibfnamefont{S.}~\bibnamefont{Meinel}},
  \bibinfo{author}{\bibfnamefont{J.}~\bibnamefont{Negele}},
  \bibinfo{author}{\bibfnamefont{S.}~\bibnamefont{Paul}},
  \bibinfo{author}{\bibfnamefont{M.}~\bibnamefont{Petschlies}},
  \bibinfo{author}{\bibfnamefont{A.}~\bibnamefont{Pochinsky}},
  \bibinfo{author}{\bibfnamefont{G.}~\bibnamefont{Silvi}}, \bibnamefont{and}
  \bibinfo{author}{\bibfnamefont{S.}~\bibnamefont{Syritsyn}},
  \bibinfo{journal}{Phys. Rev. D} \textbf{\bibinfo{volume}{102}},
  \bibinfo{pages}{114520} (\bibinfo{year}{2020})\relax
\mciteBstWouldAddEndPuncttrue
\mciteSetBstMidEndSepPunct{\mcitedefaultmidpunct}
{\mcitedefaultendpunct}{\mcitedefaultseppunct}\relax
\EndOfBibitem
\bibitem[{\citenamefont{Mandelstam}(1958)}]{Mandelstam:1958xc}
\bibinfo{author}{\bibfnamefont{S.}~\bibnamefont{Mandelstam}},
  \bibinfo{journal}{Phys.Rev.} \textbf{\bibinfo{volume}{112}},
  \bibinfo{pages}{1344} (\bibinfo{year}{1958})\relax
\mciteBstWouldAddEndPuncttrue
\mciteSetBstMidEndSepPunct{\mcitedefaultmidpunct}
{\mcitedefaultendpunct}{\mcitedefaultseppunct}\relax
\EndOfBibitem
\bibitem[{\citenamefont{Mandelstam}(1959)}]{Mandelstam:1959bc}
\bibinfo{author}{\bibfnamefont{S.}~\bibnamefont{Mandelstam}},
  \bibinfo{journal}{Phys.Rev.} \textbf{\bibinfo{volume}{115}},
  \bibinfo{pages}{1741} (\bibinfo{year}{1959})\relax
\mciteBstWouldAddEndPuncttrue
\mciteSetBstMidEndSepPunct{\mcitedefaultmidpunct}
{\mcitedefaultendpunct}{\mcitedefaultseppunct}\relax
\EndOfBibitem
\bibitem[{\citenamefont{Mandelstam}(1960)}]{Mandelstam:1960zz}
\bibinfo{author}{\bibfnamefont{S.}~\bibnamefont{Mandelstam}},
  \bibinfo{journal}{Phys.Rev.Lett.} \textbf{\bibinfo{volume}{4}},
  \bibinfo{pages}{84} (\bibinfo{year}{1960})\relax
\mciteBstWouldAddEndPuncttrue
\mciteSetBstMidEndSepPunct{\mcitedefaultmidpunct}
{\mcitedefaultendpunct}{\mcitedefaultseppunct}\relax
\EndOfBibitem
\bibitem[{\citenamefont{Lutz and Korpa}(2018)}]{Lutz:2018kaz}
\bibinfo{author}{\bibfnamefont{M.~F.~M.} \bibnamefont{Lutz}} \bibnamefont{and}
  \bibinfo{author}{\bibfnamefont{C.~L.} \bibnamefont{Korpa}},
  \bibinfo{journal}{Phys. Rev.} \textbf{\bibinfo{volume}{D98}},
  \bibinfo{pages}{076003} (\bibinfo{year}{2018})\relax
\mciteBstWouldAddEndPuncttrue
\mciteSetBstMidEndSepPunct{\mcitedefaultmidpunct}
{\mcitedefaultendpunct}{\mcitedefaultseppunct}\relax
\EndOfBibitem
\bibitem[{\citenamefont{Castillejo et~al.}(1956)\citenamefont{Castillejo,
  Dalitz, and Dyson}}]{Castillejo:1955ed}
\bibinfo{author}{\bibfnamefont{L.}~\bibnamefont{Castillejo}},
  \bibinfo{author}{\bibfnamefont{R.~H.} \bibnamefont{Dalitz}},
  \bibnamefont{and} \bibinfo{author}{\bibfnamefont{F.~J.} \bibnamefont{Dyson}},
  \bibinfo{journal}{Phys. Rev.} \textbf{\bibinfo{volume}{101}},
  \bibinfo{pages}{453} (\bibinfo{year}{1956})\relax
\mciteBstWouldAddEndPuncttrue
\mciteSetBstMidEndSepPunct{\mcitedefaultmidpunct}
{\mcitedefaultendpunct}{\mcitedefaultseppunct}\relax
\EndOfBibitem
\bibitem[{\citenamefont{Oller}(2020)}]{Oller:2019opk}
\bibinfo{author}{\bibfnamefont{J.~A.} \bibnamefont{Oller}},
  \bibinfo{journal}{Prog. Part. Nucl. Phys.} \textbf{\bibinfo{volume}{110}},
  \bibinfo{pages}{103728} (\bibinfo{year}{2020}), \eprint{1909.00370}\relax
\mciteBstWouldAddEndPuncttrue
\mciteSetBstMidEndSepPunct{\mcitedefaultmidpunct}
{\mcitedefaultendpunct}{\mcitedefaultseppunct}\relax
\EndOfBibitem
\bibitem[{\citenamefont{Oller and Entem}(2019)}]{Oller:2018zts}
\bibinfo{author}{\bibfnamefont{J.~A.} \bibnamefont{Oller}} \bibnamefont{and}
  \bibinfo{author}{\bibfnamefont{D.~R.} \bibnamefont{Entem}},
  \bibinfo{journal}{Annals Phys.} \textbf{\bibinfo{volume}{411}},
  \bibinfo{pages}{167965} (\bibinfo{year}{2019}), \eprint{1810.12242}\relax
\mciteBstWouldAddEndPuncttrue
\mciteSetBstMidEndSepPunct{\mcitedefaultmidpunct}
{\mcitedefaultendpunct}{\mcitedefaultseppunct}\relax
\EndOfBibitem
\bibitem[{\citenamefont{Guo et~al.}(2014)\citenamefont{Guo, Oller, and
  R\'\i{}os}}]{Guo:2013rpa}
\bibinfo{author}{\bibfnamefont{Z.-H.} \bibnamefont{Guo}},
  \bibinfo{author}{\bibfnamefont{J.~A.} \bibnamefont{Oller}}, \bibnamefont{and}
  \bibinfo{author}{\bibfnamefont{G.}~\bibnamefont{R\'\i{}os}},
  \bibinfo{journal}{Phys. Rev. C} \textbf{\bibinfo{volume}{89}},
  \bibinfo{pages}{014002} (\bibinfo{year}{2014}), \eprint{1305.5790}\relax
\mciteBstWouldAddEndPuncttrue
\mciteSetBstMidEndSepPunct{\mcitedefaultmidpunct}
{\mcitedefaultendpunct}{\mcitedefaultseppunct}\relax
\EndOfBibitem
\bibitem[{\citenamefont{Luming}(1964)}]{Luming:1964}
\bibinfo{author}{\bibfnamefont{M.}~\bibnamefont{Luming}},
  \bibinfo{journal}{Phys. Rev.} \textbf{\bibinfo{volume}{136}},
  \bibinfo{pages}{B1120} (\bibinfo{year}{1964})\relax
\mciteBstWouldAddEndPuncttrue
\mciteSetBstMidEndSepPunct{\mcitedefaultmidpunct}
{\mcitedefaultendpunct}{\mcitedefaultseppunct}\relax
\EndOfBibitem
\bibitem[{\citenamefont{Johnson and Warnock}(1981)}]{Johnson:1979jy}
\bibinfo{author}{\bibfnamefont{P.~W.} \bibnamefont{Johnson}} \bibnamefont{and}
  \bibinfo{author}{\bibfnamefont{R.~L.} \bibnamefont{Warnock}},
  \bibinfo{journal}{J.Math.Phys.} \textbf{\bibinfo{volume}{22}},
  \bibinfo{pages}{385} (\bibinfo{year}{1981})\relax
\mciteBstWouldAddEndPuncttrue
\mciteSetBstMidEndSepPunct{\mcitedefaultmidpunct}
{\mcitedefaultendpunct}{\mcitedefaultseppunct}\relax
\EndOfBibitem
\bibitem[{\citenamefont{Frazer}(1961)}]{Frazer:1961zz}
\bibinfo{author}{\bibfnamefont{W.~R.} \bibnamefont{Frazer}},
  \bibinfo{journal}{Phys. Rev.} \textbf{\bibinfo{volume}{123}},
  \bibinfo{pages}{2180} (\bibinfo{year}{1961})\relax
\mciteBstWouldAddEndPuncttrue
\mciteSetBstMidEndSepPunct{\mcitedefaultmidpunct}
{\mcitedefaultendpunct}{\mcitedefaultseppunct}\relax
\EndOfBibitem
\bibitem[{\citenamefont{Omnes}(1958)}]{Omnes:1958hv}
\bibinfo{author}{\bibfnamefont{R.}~\bibnamefont{Omnes}},
  \bibinfo{journal}{Nuovo Cim.} \textbf{\bibinfo{volume}{8}},
  \bibinfo{pages}{316} (\bibinfo{year}{1958})\relax
\mciteBstWouldAddEndPuncttrue
\mciteSetBstMidEndSepPunct{\mcitedefaultmidpunct}
{\mcitedefaultendpunct}{\mcitedefaultseppunct}\relax
\EndOfBibitem
\bibitem[{\citenamefont{Muskhelishvili}(1953)}]{Muskhelishvili-book}
\bibinfo{author}{\bibfnamefont{N.~I.} \bibnamefont{Muskhelishvili}},
  \bibinfo{journal}{Singular Integral Equations, Wolters-Noordhoff Publishing,
  Groningen}  (\bibinfo{year}{1953})\relax
\mciteBstWouldAddEndPuncttrue
\mciteSetBstMidEndSepPunct{\mcitedefaultmidpunct}
{\mcitedefaultendpunct}{\mcitedefaultseppunct}\relax
\EndOfBibitem
\bibitem[{\citenamefont{Donoghue et~al.}(1990)\citenamefont{Donoghue, Gasser,
  and Leutwyler}}]{Donoghue:1990xh}
\bibinfo{author}{\bibfnamefont{J.~F.} \bibnamefont{Donoghue}},
  \bibinfo{author}{\bibfnamefont{J.}~\bibnamefont{Gasser}}, \bibnamefont{and}
  \bibinfo{author}{\bibfnamefont{H.}~\bibnamefont{Leutwyler}},
  \bibinfo{journal}{Nucl. Phys.} \textbf{\bibinfo{volume}{B343}},
  \bibinfo{pages}{341} (\bibinfo{year}{1990})\relax
\mciteBstWouldAddEndPuncttrue
\mciteSetBstMidEndSepPunct{\mcitedefaultmidpunct}
{\mcitedefaultendpunct}{\mcitedefaultseppunct}\relax
\EndOfBibitem
\bibitem[{\citenamefont{Moussallam}(2000)}]{Moussallam:1999aq}
\bibinfo{author}{\bibfnamefont{B.}~\bibnamefont{Moussallam}},
  \bibinfo{journal}{Eur. Phys. J.} \textbf{\bibinfo{volume}{C14}},
  \bibinfo{pages}{111} (\bibinfo{year}{2000})\relax
\mciteBstWouldAddEndPuncttrue
\mciteSetBstMidEndSepPunct{\mcitedefaultmidpunct}
{\mcitedefaultendpunct}{\mcitedefaultseppunct}\relax
\EndOfBibitem
\bibitem[{\citenamefont{Bijnens et~al.}(2004)\citenamefont{Bijnens, Dhonte, and
  Talavera}}]{Bijnens:2004bu}
\bibinfo{author}{\bibfnamefont{J.}~\bibnamefont{Bijnens}},
  \bibinfo{author}{\bibfnamefont{P.}~\bibnamefont{Dhonte}}, \bibnamefont{and}
  \bibinfo{author}{\bibfnamefont{P.}~\bibnamefont{Talavera}},
  \bibinfo{journal}{JHEP} \textbf{\bibinfo{volume}{05}}, \bibinfo{pages}{036}
  (\bibinfo{year}{2004})\relax
\mciteBstWouldAddEndPuncttrue
\mciteSetBstMidEndSepPunct{\mcitedefaultmidpunct}
{\mcitedefaultendpunct}{\mcitedefaultseppunct}\relax
\EndOfBibitem
\bibitem[{\citenamefont{Protopopescu et~al.}(1973)\citenamefont{Protopopescu,
  Alston-Garnjost, Barbaro-Galtieri, Flatte, Friedman, Lasinski, Lynch, Rabin,
  and Solmitz}}]{Protopopescu:1973sh}
\bibinfo{author}{\bibfnamefont{S.}~\bibnamefont{Protopopescu}},
  \bibinfo{author}{\bibfnamefont{M.}~\bibnamefont{Alston-Garnjost}},
  \bibinfo{author}{\bibfnamefont{A.}~\bibnamefont{Barbaro-Galtieri}},
  \bibinfo{author}{\bibfnamefont{S.~M.} \bibnamefont{Flatte}},
  \bibinfo{author}{\bibfnamefont{J.}~\bibnamefont{Friedman}},
  \bibinfo{author}{\bibfnamefont{T.}~\bibnamefont{Lasinski}},
  \bibinfo{author}{\bibfnamefont{G.}~\bibnamefont{Lynch}},
  \bibinfo{author}{\bibfnamefont{M.}~\bibnamefont{Rabin}}, \bibnamefont{and}
  \bibinfo{author}{\bibfnamefont{F.}~\bibnamefont{Solmitz}},
  \bibinfo{journal}{Phys. Rev. D} \textbf{\bibinfo{volume}{7}},
  \bibinfo{pages}{1279} (\bibinfo{year}{1973})\relax
\mciteBstWouldAddEndPuncttrue
\mciteSetBstMidEndSepPunct{\mcitedefaultmidpunct}
{\mcitedefaultendpunct}{\mcitedefaultseppunct}\relax
\EndOfBibitem
\bibitem[{\citenamefont{Grayer et~al.}(1974)}]{Grayer:1974cr}
\bibinfo{author}{\bibfnamefont{G.}~\bibnamefont{Grayer}} \bibnamefont{et~al.},
  \bibinfo{journal}{Nucl. Phys. B} \textbf{\bibinfo{volume}{75}},
  \bibinfo{pages}{189} (\bibinfo{year}{1974})\relax
\mciteBstWouldAddEndPuncttrue
\mciteSetBstMidEndSepPunct{\mcitedefaultmidpunct}
{\mcitedefaultendpunct}{\mcitedefaultseppunct}\relax
\EndOfBibitem
\bibitem[{\citenamefont{Kaminski et~al.}(1997)\citenamefont{Kaminski, Lesniak,
  and Rybicki}}]{Kaminski:1996da}
\bibinfo{author}{\bibfnamefont{R.}~\bibnamefont{Kaminski}},
  \bibinfo{author}{\bibfnamefont{L.}~\bibnamefont{Lesniak}}, \bibnamefont{and}
  \bibinfo{author}{\bibfnamefont{K.}~\bibnamefont{Rybicki}},
  \bibinfo{journal}{Z. Phys. C} \textbf{\bibinfo{volume}{74}},
  \bibinfo{pages}{79} (\bibinfo{year}{1997})\relax
\mciteBstWouldAddEndPuncttrue
\mciteSetBstMidEndSepPunct{\mcitedefaultmidpunct}
{\mcitedefaultendpunct}{\mcitedefaultseppunct}\relax
\EndOfBibitem
\bibitem[{\citenamefont{Batley et~al.}(2008)}]{Batley:2007zz}
\bibinfo{author}{\bibfnamefont{J.}~\bibnamefont{Batley}} \bibnamefont{et~al.}
  (\bibinfo{collaboration}{NA48/2}), \bibinfo{journal}{Eur. Phys. J. C}
  \textbf{\bibinfo{volume}{54}}, \bibinfo{pages}{411}
  (\bibinfo{year}{2008})\relax
\mciteBstWouldAddEndPuncttrue
\mciteSetBstMidEndSepPunct{\mcitedefaultmidpunct}
{\mcitedefaultendpunct}{\mcitedefaultseppunct}\relax
\EndOfBibitem
\bibitem[{\citenamefont{Batley et~al.}(2010)}]{Batley:2010zza}
\bibinfo{author}{\bibfnamefont{J.}~\bibnamefont{Batley}} \bibnamefont{et~al.}
  (\bibinfo{collaboration}{NA48/2}), \bibinfo{journal}{Eur. Phys. J. C}
  \textbf{\bibinfo{volume}{70}}, \bibinfo{pages}{635}
  (\bibinfo{year}{2010})\relax
\mciteBstWouldAddEndPuncttrue
\mciteSetBstMidEndSepPunct{\mcitedefaultmidpunct}
{\mcitedefaultendpunct}{\mcitedefaultseppunct}\relax
\EndOfBibitem
\bibitem[{\citenamefont{Estabrooks et~al.}(1978)\citenamefont{Estabrooks,
  Carnegie, Martin, Dunwoodie, Lasinski, and Leith}}]{Estabrooks:1977xe}
\bibinfo{author}{\bibfnamefont{P.}~\bibnamefont{Estabrooks}},
  \bibinfo{author}{\bibfnamefont{R.}~\bibnamefont{Carnegie}},
  \bibinfo{author}{\bibfnamefont{A.~D.} \bibnamefont{Martin}},
  \bibinfo{author}{\bibfnamefont{W.}~\bibnamefont{Dunwoodie}},
  \bibinfo{author}{\bibfnamefont{T.}~\bibnamefont{Lasinski}}, \bibnamefont{and}
  \bibinfo{author}{\bibfnamefont{D.~W.} \bibnamefont{Leith}},
  \bibinfo{journal}{Nucl. Phys. B} \textbf{\bibinfo{volume}{133}},
  \bibinfo{pages}{490} (\bibinfo{year}{1978})\relax
\mciteBstWouldAddEndPuncttrue
\mciteSetBstMidEndSepPunct{\mcitedefaultmidpunct}
{\mcitedefaultendpunct}{\mcitedefaultseppunct}\relax
\EndOfBibitem
\bibitem[{\citenamefont{Aston et~al.}(1988)}]{Aston:1987ir}
\bibinfo{author}{\bibfnamefont{D.}~\bibnamefont{Aston}} \bibnamefont{et~al.},
  \bibinfo{journal}{Nucl. Phys. B} \textbf{\bibinfo{volume}{296}},
  \bibinfo{pages}{493} (\bibinfo{year}{1988})\relax
\mciteBstWouldAddEndPuncttrue
\mciteSetBstMidEndSepPunct{\mcitedefaultmidpunct}
{\mcitedefaultendpunct}{\mcitedefaultseppunct}\relax
\EndOfBibitem
\bibitem[{\citenamefont{Gasser and Leutwyler}(1984)}]{Gasser:1983yg}
\bibinfo{author}{\bibfnamefont{J.}~\bibnamefont{Gasser}} \bibnamefont{and}
  \bibinfo{author}{\bibfnamefont{H.}~\bibnamefont{Leutwyler}},
  \bibinfo{journal}{Annals Phys.} \textbf{\bibinfo{volume}{158}},
  \bibinfo{pages}{142} (\bibinfo{year}{1984})\relax
\mciteBstWouldAddEndPuncttrue
\mciteSetBstMidEndSepPunct{\mcitedefaultmidpunct}
{\mcitedefaultendpunct}{\mcitedefaultseppunct}\relax
\EndOfBibitem
\bibitem[{\citenamefont{Bernard et~al.}(1991)\citenamefont{Bernard, Kaiser, and
  Meissner}}]{Bernard:1990kx}
\bibinfo{author}{\bibfnamefont{V.}~\bibnamefont{Bernard}},
  \bibinfo{author}{\bibfnamefont{N.}~\bibnamefont{Kaiser}}, \bibnamefont{and}
  \bibinfo{author}{\bibfnamefont{U.~G.} \bibnamefont{Meissner}},
  \bibinfo{journal}{Phys. Rev.} \textbf{\bibinfo{volume}{D43}},
  \bibinfo{pages}{2757} (\bibinfo{year}{1991})\relax
\mciteBstWouldAddEndPuncttrue
\mciteSetBstMidEndSepPunct{\mcitedefaultmidpunct}
{\mcitedefaultendpunct}{\mcitedefaultseppunct}\relax
\EndOfBibitem
\bibitem[{\citenamefont{Gomez~Nicola and Pelaez}(2002)}]{GomezNicola:2001as}
\bibinfo{author}{\bibfnamefont{A.}~\bibnamefont{Gomez~Nicola}}
  \bibnamefont{and} \bibinfo{author}{\bibfnamefont{J.~R.}
  \bibnamefont{Pelaez}}, \bibinfo{journal}{Phys. Rev.}
  \textbf{\bibinfo{volume}{D65}}, \bibinfo{pages}{054009}
  (\bibinfo{year}{2002})\relax
\mciteBstWouldAddEndPuncttrue
\mciteSetBstMidEndSepPunct{\mcitedefaultmidpunct}
{\mcitedefaultendpunct}{\mcitedefaultseppunct}\relax
\EndOfBibitem
\bibitem[{\citenamefont{Bijnens and Ecker}(2014)}]{Bijnens:2014lea}
\bibinfo{author}{\bibfnamefont{J.}~\bibnamefont{Bijnens}} \bibnamefont{and}
  \bibinfo{author}{\bibfnamefont{G.}~\bibnamefont{Ecker}},
  \bibinfo{journal}{Ann. Rev. Nucl. Part. Sci.} \textbf{\bibinfo{volume}{64}},
  \bibinfo{pages}{149} (\bibinfo{year}{2014})\relax
\mciteBstWouldAddEndPuncttrue
\mciteSetBstMidEndSepPunct{\mcitedefaultmidpunct}
{\mcitedefaultendpunct}{\mcitedefaultseppunct}\relax
\EndOfBibitem
\bibitem[{\citenamefont{Moussallam}(2011)}]{Moussallam:2011zg}
\bibinfo{author}{\bibfnamefont{B.}~\bibnamefont{Moussallam}},
  \bibinfo{journal}{Eur. Phys. J.} \textbf{\bibinfo{volume}{C71}},
  \bibinfo{pages}{1814} (\bibinfo{year}{2011})\relax
\mciteBstWouldAddEndPuncttrue
\mciteSetBstMidEndSepPunct{\mcitedefaultmidpunct}
{\mcitedefaultendpunct}{\mcitedefaultseppunct}\relax
\EndOfBibitem
\bibitem[{\citenamefont{Yao et~al.}(2020)\citenamefont{Yao, Dai, Zheng, and
  Zhou}}]{Yao:2020bxx}
\bibinfo{author}{\bibfnamefont{D.-L.} \bibnamefont{Yao}},
  \bibinfo{author}{\bibfnamefont{L.-Y.} \bibnamefont{Dai}},
  \bibinfo{author}{\bibfnamefont{H.-Q.} \bibnamefont{Zheng}}, \bibnamefont{and}
  \bibinfo{author}{\bibfnamefont{Z.-Y.} \bibnamefont{Zhou}},
  \bibinfo{journal}{arXiv: 2009.13495}  (\bibinfo{year}{2020})\relax
\mciteBstWouldAddEndPuncttrue
\mciteSetBstMidEndSepPunct{\mcitedefaultmidpunct}
{\mcitedefaultendpunct}{\mcitedefaultseppunct}\relax
\EndOfBibitem
\bibitem[{\citenamefont{Salas-Bern\'ardez
  et~al.}(2020)\citenamefont{Salas-Bern\'ardez, Llanes-Estrada,
  Escudero-Pedrosa, and Oller}}]{Salas-Bernardez:2020hua}
\bibinfo{author}{\bibfnamefont{A.}~\bibnamefont{Salas-Bern\'ardez}},
  \bibinfo{author}{\bibfnamefont{F.~J.} \bibnamefont{Llanes-Estrada}},
  \bibinfo{author}{\bibfnamefont{J.}~\bibnamefont{Escudero-Pedrosa}},
  \bibnamefont{and} \bibinfo{author}{\bibfnamefont{J.~A.} \bibnamefont{Oller}},
  \bibinfo{journal}{arXiv: 2010.13709}  (\bibinfo{year}{2020})\relax
\mciteBstWouldAddEndPuncttrue
\mciteSetBstMidEndSepPunct{\mcitedefaultmidpunct}
{\mcitedefaultendpunct}{\mcitedefaultseppunct}\relax
\EndOfBibitem
\bibitem[{\citenamefont{Caprini}(2008)}]{Caprini:2008fc}
\bibinfo{author}{\bibfnamefont{I.}~\bibnamefont{Caprini}},
  \bibinfo{journal}{Phys. Rev. D} \textbf{\bibinfo{volume}{77}},
  \bibinfo{pages}{114019} (\bibinfo{year}{2008})\relax
\mciteBstWouldAddEndPuncttrue
\mciteSetBstMidEndSepPunct{\mcitedefaultmidpunct}
{\mcitedefaultendpunct}{\mcitedefaultseppunct}\relax
\EndOfBibitem
\bibitem[{\citenamefont{Caprini et~al.}(2016)\citenamefont{Caprini, Masjuan,
  Ruiz~de Elvira, and Sanz-Cillero}}]{Caprini:2016uxy}
\bibinfo{author}{\bibfnamefont{I.}~\bibnamefont{Caprini}},
  \bibinfo{author}{\bibfnamefont{P.}~\bibnamefont{Masjuan}},
  \bibinfo{author}{\bibfnamefont{J.}~\bibnamefont{Ruiz~de Elvira}},
  \bibnamefont{and} \bibinfo{author}{\bibfnamefont{J.~J.}
  \bibnamefont{Sanz-Cillero}}, \bibinfo{journal}{Phys. Rev. D}
  \textbf{\bibinfo{volume}{93}}, \bibinfo{pages}{076004}
  (\bibinfo{year}{2016})\relax
\mciteBstWouldAddEndPuncttrue
\mciteSetBstMidEndSepPunct{\mcitedefaultmidpunct}
{\mcitedefaultendpunct}{\mcitedefaultseppunct}\relax
\EndOfBibitem
\bibitem[{\citenamefont{Gomez~Nicola et~al.}(2008)\citenamefont{Gomez~Nicola,
  Pelaez, and Rios}}]{GomezNicola:2007qj}
\bibinfo{author}{\bibfnamefont{A.}~\bibnamefont{Gomez~Nicola}},
  \bibinfo{author}{\bibfnamefont{J.}~\bibnamefont{Pelaez}}, \bibnamefont{and}
  \bibinfo{author}{\bibfnamefont{G.}~\bibnamefont{Rios}},
  \bibinfo{journal}{Phys. Rev. D} \textbf{\bibinfo{volume}{77}},
  \bibinfo{pages}{056006} (\bibinfo{year}{2008})\relax
\mciteBstWouldAddEndPuncttrue
\mciteSetBstMidEndSepPunct{\mcitedefaultmidpunct}
{\mcitedefaultendpunct}{\mcitedefaultseppunct}\relax
\EndOfBibitem
\bibitem[{\citenamefont{Hanhart et~al.}(2008)\citenamefont{Hanhart, Pelaez, and
  Rios}}]{Hanhart:2008mx}
\bibinfo{author}{\bibfnamefont{C.}~\bibnamefont{Hanhart}},
  \bibinfo{author}{\bibfnamefont{J.~R.} \bibnamefont{Pelaez}},
  \bibnamefont{and} \bibinfo{author}{\bibfnamefont{G.}~\bibnamefont{Rios}},
  \bibinfo{journal}{Phys. Rev. Lett.} \textbf{\bibinfo{volume}{100}},
  \bibinfo{pages}{152001} (\bibinfo{year}{2008})\relax
\mciteBstWouldAddEndPuncttrue
\mciteSetBstMidEndSepPunct{\mcitedefaultmidpunct}
{\mcitedefaultendpunct}{\mcitedefaultseppunct}\relax
\EndOfBibitem
\bibitem[{\citenamefont{Nebreda and Pelaez.}(2010)}]{Nebreda:2010wv}
\bibinfo{author}{\bibfnamefont{J.}~\bibnamefont{Nebreda}} \bibnamefont{and}
  \bibinfo{author}{\bibfnamefont{J.~R.} \bibnamefont{Pelaez.}},
  \bibinfo{journal}{Phys. Rev.} \textbf{\bibinfo{volume}{D81}},
  \bibinfo{pages}{054035} (\bibinfo{year}{2010})\relax
\mciteBstWouldAddEndPuncttrue
\mciteSetBstMidEndSepPunct{\mcitedefaultmidpunct}
{\mcitedefaultendpunct}{\mcitedefaultseppunct}\relax
\EndOfBibitem
\bibitem[{\citenamefont{Pelaez and Rios}(2010)}]{Pelaez:2010fj}
\bibinfo{author}{\bibfnamefont{J.}~\bibnamefont{Pelaez}} \bibnamefont{and}
  \bibinfo{author}{\bibfnamefont{G.}~\bibnamefont{Rios}},
  \bibinfo{journal}{Phys. Rev. D} \textbf{\bibinfo{volume}{82}},
  \bibinfo{pages}{114002} (\bibinfo{year}{2010})\relax
\mciteBstWouldAddEndPuncttrue
\mciteSetBstMidEndSepPunct{\mcitedefaultmidpunct}
{\mcitedefaultendpunct}{\mcitedefaultseppunct}\relax
\EndOfBibitem
\bibitem[{\citenamefont{Luscher}(1991)}]{Luscher:1991cf}
\bibinfo{author}{\bibfnamefont{M.}~\bibnamefont{Luscher}},
  \bibinfo{journal}{Nucl. Phys.} \textbf{\bibinfo{volume}{B364}},
  \bibinfo{pages}{237} (\bibinfo{year}{1991})\relax
\mciteBstWouldAddEndPuncttrue
\mciteSetBstMidEndSepPunct{\mcitedefaultmidpunct}
{\mcitedefaultendpunct}{\mcitedefaultseppunct}\relax
\EndOfBibitem
\bibitem[{\citenamefont{Luscher and Wolff}(1990)}]{Luscher:1990ck}
\bibinfo{author}{\bibfnamefont{M.}~\bibnamefont{Luscher}} \bibnamefont{and}
  \bibinfo{author}{\bibfnamefont{U.}~\bibnamefont{Wolff}},
  \bibinfo{journal}{Nucl. Phys.} \textbf{\bibinfo{volume}{B339}},
  \bibinfo{pages}{222} (\bibinfo{year}{1990})\relax
\mciteBstWouldAddEndPuncttrue
\mciteSetBstMidEndSepPunct{\mcitedefaultmidpunct}
{\mcitedefaultendpunct}{\mcitedefaultseppunct}\relax
\EndOfBibitem
\bibitem[{\citenamefont{Rummukainen and Gottlieb}(1995)}]{Rummukainen:1995vs}
\bibinfo{author}{\bibfnamefont{K.}~\bibnamefont{Rummukainen}} \bibnamefont{and}
  \bibinfo{author}{\bibfnamefont{S.~A.} \bibnamefont{Gottlieb}},
  \bibinfo{journal}{Nucl. Phys.} \textbf{\bibinfo{volume}{B450}},
  \bibinfo{pages}{397} (\bibinfo{year}{1995})\relax
\mciteBstWouldAddEndPuncttrue
\mciteSetBstMidEndSepPunct{\mcitedefaultmidpunct}
{\mcitedefaultendpunct}{\mcitedefaultseppunct}\relax
\EndOfBibitem
\bibitem[{\citenamefont{Kim et~al.}(2005)\citenamefont{Kim, Sachrajda, and
  Sharpe}}]{Kim:2005gf}
\bibinfo{author}{\bibfnamefont{C.~h.} \bibnamefont{Kim}},
  \bibinfo{author}{\bibfnamefont{C.~T.} \bibnamefont{Sachrajda}},
  \bibnamefont{and} \bibinfo{author}{\bibfnamefont{S.~R.}
  \bibnamefont{Sharpe}}, \bibinfo{journal}{Nucl. Phys.}
  \textbf{\bibinfo{volume}{B727}}, \bibinfo{pages}{218}
  (\bibinfo{year}{2005})\relax
\mciteBstWouldAddEndPuncttrue
\mciteSetBstMidEndSepPunct{\mcitedefaultmidpunct}
{\mcitedefaultendpunct}{\mcitedefaultseppunct}\relax
\EndOfBibitem
\bibitem[{\citenamefont{Christ et~al.}(2005)\citenamefont{Christ, Kim, and
  Yamazaki}}]{Christ:2005gi}
\bibinfo{author}{\bibfnamefont{N.~H.} \bibnamefont{Christ}},
  \bibinfo{author}{\bibfnamefont{C.}~\bibnamefont{Kim}}, \bibnamefont{and}
  \bibinfo{author}{\bibfnamefont{T.}~\bibnamefont{Yamazaki}},
  \bibinfo{journal}{Phys. Rev.} \textbf{\bibinfo{volume}{D72}},
  \bibinfo{pages}{114506} (\bibinfo{year}{2005})\relax
\mciteBstWouldAddEndPuncttrue
\mciteSetBstMidEndSepPunct{\mcitedefaultmidpunct}
{\mcitedefaultendpunct}{\mcitedefaultseppunct}\relax
\EndOfBibitem
\bibitem[{\citenamefont{Leskovec and Prelovsek}(2012)}]{Leskovec:2012gb}
\bibinfo{author}{\bibfnamefont{L.}~\bibnamefont{Leskovec}} \bibnamefont{and}
  \bibinfo{author}{\bibfnamefont{S.}~\bibnamefont{Prelovsek}},
  \bibinfo{journal}{Phys. Rev.} \textbf{\bibinfo{volume}{D85}},
  \bibinfo{pages}{114507} (\bibinfo{year}{2012})\relax
\mciteBstWouldAddEndPuncttrue
\mciteSetBstMidEndSepPunct{\mcitedefaultmidpunct}
{\mcitedefaultendpunct}{\mcitedefaultseppunct}\relax
\EndOfBibitem
\bibitem[{\citenamefont{Albaladejo and Oller}(2012)}]{Albaladejo:2012te}
\bibinfo{author}{\bibfnamefont{M.}~\bibnamefont{Albaladejo}} \bibnamefont{and}
  \bibinfo{author}{\bibfnamefont{J.}~\bibnamefont{Oller}},
  \bibinfo{journal}{Phys. Rev. D} \textbf{\bibinfo{volume}{86}},
  \bibinfo{pages}{034003} (\bibinfo{year}{2012})\relax
\mciteBstWouldAddEndPuncttrue
\mciteSetBstMidEndSepPunct{\mcitedefaultmidpunct}
{\mcitedefaultendpunct}{\mcitedefaultseppunct}\relax
\EndOfBibitem
\bibitem[{\citenamefont{Cohen et~al.}(1980)\citenamefont{Cohen, Ayres, Diebold,
  Kramer, Pawlicki, and Wicklund}}]{Cohen:1980cq}
\bibinfo{author}{\bibfnamefont{D.~H.} \bibnamefont{Cohen}},
  \bibinfo{author}{\bibfnamefont{D.}~\bibnamefont{Ayres}},
  \bibinfo{author}{\bibfnamefont{R.}~\bibnamefont{Diebold}},
  \bibinfo{author}{\bibfnamefont{S.}~\bibnamefont{Kramer}},
  \bibinfo{author}{\bibfnamefont{A.}~\bibnamefont{Pawlicki}}, \bibnamefont{and}
  \bibinfo{author}{\bibfnamefont{A.}~\bibnamefont{Wicklund}},
  \bibinfo{journal}{Phys. Rev. D} \textbf{\bibinfo{volume}{22}},
  \bibinfo{pages}{2595} (\bibinfo{year}{1980})\relax
\mciteBstWouldAddEndPuncttrue
\mciteSetBstMidEndSepPunct{\mcitedefaultmidpunct}
{\mcitedefaultendpunct}{\mcitedefaultseppunct}\relax
\EndOfBibitem
\bibitem[{\citenamefont{Etkin et~al.}(1982)}]{Etkin:1981sg}
\bibinfo{author}{\bibfnamefont{A.}~\bibnamefont{Etkin}} \bibnamefont{et~al.},
  \bibinfo{journal}{Phys. Rev. D} \textbf{\bibinfo{volume}{25}},
  \bibinfo{pages}{1786} (\bibinfo{year}{1982})\relax
\mciteBstWouldAddEndPuncttrue
\mciteSetBstMidEndSepPunct{\mcitedefaultmidpunct}
{\mcitedefaultendpunct}{\mcitedefaultseppunct}\relax
\EndOfBibitem
\bibitem[{\citenamefont{Longacre et~al.}(1986)}]{Longacre:1986fh}
\bibinfo{author}{\bibfnamefont{R.}~\bibnamefont{Longacre}}
  \bibnamefont{et~al.}, \bibinfo{journal}{Phys. Lett. B}
  \textbf{\bibinfo{volume}{177}}, \bibinfo{pages}{223}
  (\bibinfo{year}{1986})\relax
\mciteBstWouldAddEndPuncttrue
\mciteSetBstMidEndSepPunct{\mcitedefaultmidpunct}
{\mcitedefaultendpunct}{\mcitedefaultseppunct}\relax
\EndOfBibitem
\bibitem[{\citenamefont{Martin and Ozmutlu}(1979)}]{Martin:1979gm}
\bibinfo{author}{\bibfnamefont{A.~D.} \bibnamefont{Martin}} \bibnamefont{and}
  \bibinfo{author}{\bibfnamefont{E.}~\bibnamefont{Ozmutlu}},
  \bibinfo{journal}{Nucl. Phys. B} \textbf{\bibinfo{volume}{158}},
  \bibinfo{pages}{520} (\bibinfo{year}{1979})\relax
\mciteBstWouldAddEndPuncttrue
\mciteSetBstMidEndSepPunct{\mcitedefaultmidpunct}
{\mcitedefaultendpunct}{\mcitedefaultseppunct}\relax
\EndOfBibitem
\bibitem[{\citenamefont{Oller and Oset}(1997)}]{Oller:1997ti}
\bibinfo{author}{\bibfnamefont{J.~A.} \bibnamefont{Oller}} \bibnamefont{and}
  \bibinfo{author}{\bibfnamefont{E.}~\bibnamefont{Oset}},
  \bibinfo{journal}{Nucl. Phys.} \textbf{\bibinfo{volume}{A620}},
  \bibinfo{pages}{438} (\bibinfo{year}{1997}), \bibinfo{note}{[Erratum: Nucl.
  Phys.A652,407(1999)]}\relax
\mciteBstWouldAddEndPuncttrue
\mciteSetBstMidEndSepPunct{\mcitedefaultmidpunct}
{\mcitedefaultendpunct}{\mcitedefaultseppunct}\relax
\EndOfBibitem
\bibitem[{\citenamefont{Danilkin and Vanderhaeghen}(2019)}]{Danilkin:2018qfn}
\bibinfo{author}{\bibfnamefont{I.}~\bibnamefont{Danilkin}} \bibnamefont{and}
  \bibinfo{author}{\bibfnamefont{M.}~\bibnamefont{Vanderhaeghen}},
  \bibinfo{journal}{Phys. Lett.} \textbf{\bibinfo{volume}{B789}},
  \bibinfo{pages}{366} (\bibinfo{year}{2019})\relax
\mciteBstWouldAddEndPuncttrue
\mciteSetBstMidEndSepPunct{\mcitedefaultmidpunct}
{\mcitedefaultendpunct}{\mcitedefaultseppunct}\relax
\EndOfBibitem
\bibitem[{\citenamefont{Danilkin
  et~al.}(2020{\natexlab{b}})\citenamefont{Danilkin, Deineka, and
  Vanderhaeghen}}]{Danilkin:2019opj}
\bibinfo{author}{\bibfnamefont{I.}~\bibnamefont{Danilkin}},
  \bibinfo{author}{\bibfnamefont{O.}~\bibnamefont{Deineka}}, \bibnamefont{and}
  \bibinfo{author}{\bibfnamefont{M.}~\bibnamefont{Vanderhaeghen}},
  \bibinfo{journal}{Phys. Rev.} \textbf{\bibinfo{volume}{D101}},
  \bibinfo{pages}{054008} (\bibinfo{year}{2020}{\natexlab{b}})\relax
\mciteBstWouldAddEndPuncttrue
\mciteSetBstMidEndSepPunct{\mcitedefaultmidpunct}
{\mcitedefaultendpunct}{\mcitedefaultseppunct}\relax
\EndOfBibitem
\bibitem[{\citenamefont{Deineka et~al.}(2019)\citenamefont{Deineka, Danilkin,
  and Vanderhaeghen}}]{Deineka:2019bey}
\bibinfo{author}{\bibfnamefont{O.}~\bibnamefont{Deineka}},
  \bibinfo{author}{\bibfnamefont{I.}~\bibnamefont{Danilkin}}, \bibnamefont{and}
  \bibinfo{author}{\bibfnamefont{M.}~\bibnamefont{Vanderhaeghen}},
  \bibinfo{journal}{Acta Phys. Polon.} \textbf{\bibinfo{volume}{B50}},
  \bibinfo{pages}{1901} (\bibinfo{year}{2019})\relax
\mciteBstWouldAddEndPuncttrue
\mciteSetBstMidEndSepPunct{\mcitedefaultmidpunct}
{\mcitedefaultendpunct}{\mcitedefaultseppunct}\relax
\EndOfBibitem
\bibitem[{\citenamefont{Bernabeu and Prades}(2008)}]{Bernabeu:2008wt}
\bibinfo{author}{\bibfnamefont{J.}~\bibnamefont{Bernabeu}} \bibnamefont{and}
  \bibinfo{author}{\bibfnamefont{J.}~\bibnamefont{Prades}},
  \bibinfo{journal}{Phys. Rev. Lett.} \textbf{\bibinfo{volume}{100}},
  \bibinfo{pages}{241804} (\bibinfo{year}{2008})\relax
\mciteBstWouldAddEndPuncttrue
\mciteSetBstMidEndSepPunct{\mcitedefaultmidpunct}
{\mcitedefaultendpunct}{\mcitedefaultseppunct}\relax
\EndOfBibitem
\bibitem[{\citenamefont{Oller and Roca}(2008)}]{Oller:2008kf}
\bibinfo{author}{\bibfnamefont{J.~A.} \bibnamefont{Oller}} \bibnamefont{and}
  \bibinfo{author}{\bibfnamefont{L.}~\bibnamefont{Roca}},
  \bibinfo{journal}{Eur. Phys. J.} \textbf{\bibinfo{volume}{A37}},
  \bibinfo{pages}{15} (\bibinfo{year}{2008})\relax
\mciteBstWouldAddEndPuncttrue
\mciteSetBstMidEndSepPunct{\mcitedefaultmidpunct}
{\mcitedefaultendpunct}{\mcitedefaultseppunct}\relax
\EndOfBibitem
\bibitem[{\citenamefont{Dai and Pennington}(2014{\natexlab{b}})}]{Dai:2014lza}
\bibinfo{author}{\bibfnamefont{L.-Y.} \bibnamefont{Dai}} \bibnamefont{and}
  \bibinfo{author}{\bibfnamefont{M.~R.} \bibnamefont{Pennington}},
  \bibinfo{journal}{Phys. Lett.} \textbf{\bibinfo{volume}{B736}},
  \bibinfo{pages}{11} (\bibinfo{year}{2014}{\natexlab{b}})\relax
\mciteBstWouldAddEndPuncttrue
\mciteSetBstMidEndSepPunct{\mcitedefaultmidpunct}
{\mcitedefaultendpunct}{\mcitedefaultseppunct}\relax
\EndOfBibitem
\bibitem[{\citenamefont{Uehara et~al.}(2009)}]{Uehara:2009cka}
\bibinfo{author}{\bibfnamefont{S.}~\bibnamefont{Uehara}} \bibnamefont{et~al.}
  (\bibinfo{collaboration}{Belle}), \bibinfo{journal}{Phys. Rev.}
  \textbf{\bibinfo{volume}{D79}}, \bibinfo{pages}{052009}
  (\bibinfo{year}{2009})\relax
\mciteBstWouldAddEndPuncttrue
\mciteSetBstMidEndSepPunct{\mcitedefaultmidpunct}
{\mcitedefaultendpunct}{\mcitedefaultseppunct}\relax
\EndOfBibitem
\bibitem[{\citenamefont{Marsiske et~al.}(1990)}]{Marsiske:1990hx}
\bibinfo{author}{\bibfnamefont{H.}~\bibnamefont{Marsiske}} \bibnamefont{et~al.}
  (\bibinfo{collaboration}{Crystal Ball}), \bibinfo{journal}{Phys. Rev.}
  \textbf{\bibinfo{volume}{D41}}, \bibinfo{pages}{3324}
  (\bibinfo{year}{1990})\relax
\mciteBstWouldAddEndPuncttrue
\mciteSetBstMidEndSepPunct{\mcitedefaultmidpunct}
{\mcitedefaultendpunct}{\mcitedefaultseppunct}\relax
\EndOfBibitem
\bibitem[{\citenamefont{Briceno
  et~al.}(2018{\natexlab{b}})\citenamefont{Briceno, Dudek, Edwards, and
  Wilson}}]{Briceno:2017qmb}
\bibinfo{author}{\bibfnamefont{R.~A.} \bibnamefont{Briceno}},
  \bibinfo{author}{\bibfnamefont{J.~J.} \bibnamefont{Dudek}},
  \bibinfo{author}{\bibfnamefont{R.~G.} \bibnamefont{Edwards}},
  \bibnamefont{and} \bibinfo{author}{\bibfnamefont{D.~J.}
  \bibnamefont{Wilson}}, \bibinfo{journal}{Phys. Rev. D}
  \textbf{\bibinfo{volume}{97}}, \bibinfo{pages}{054513}
  (\bibinfo{year}{2018}{\natexlab{b}}), \eprint{1708.06667}\relax
\mciteBstWouldAddEndPuncttrue
\mciteSetBstMidEndSepPunct{\mcitedefaultmidpunct}
{\mcitedefaultendpunct}{\mcitedefaultseppunct}\relax
\EndOfBibitem
\bibitem[{\citenamefont{Dudek et~al.}(2016)\citenamefont{Dudek, Edwards, and
  Wilson}}]{Dudek:2016cru}
\bibinfo{author}{\bibfnamefont{J.~J.} \bibnamefont{Dudek}},
  \bibinfo{author}{\bibfnamefont{R.~G.} \bibnamefont{Edwards}},
  \bibnamefont{and} \bibinfo{author}{\bibfnamefont{D.~J.}
  \bibnamefont{Wilson}}, \bibinfo{journal}{Phys. Rev.}
  \textbf{\bibinfo{volume}{D93}}, \bibinfo{pages}{094506}
  (\bibinfo{year}{2016})\relax
\mciteBstWouldAddEndPuncttrue
\mciteSetBstMidEndSepPunct{\mcitedefaultmidpunct}
{\mcitedefaultendpunct}{\mcitedefaultseppunct}\relax
\EndOfBibitem
\bibitem[{\citenamefont{Pascalutsa and
  Vanderhaeghen}(2010)}]{Pascalutsa:2010sj}
\bibinfo{author}{\bibfnamefont{V.}~\bibnamefont{Pascalutsa}} \bibnamefont{and}
  \bibinfo{author}{\bibfnamefont{M.}~\bibnamefont{Vanderhaeghen}},
  \bibinfo{journal}{Phys. Rev. Lett.} \textbf{\bibinfo{volume}{105}},
  \bibinfo{pages}{201603} (\bibinfo{year}{2010})\relax
\mciteBstWouldAddEndPuncttrue
\mciteSetBstMidEndSepPunct{\mcitedefaultmidpunct}
{\mcitedefaultendpunct}{\mcitedefaultseppunct}\relax
\EndOfBibitem
\bibitem[{\citenamefont{Pascalutsa et~al.}(2012)\citenamefont{Pascalutsa, Pauk,
  and Vanderhaeghen}}]{Pascalutsa:2012pr}
\bibinfo{author}{\bibfnamefont{V.}~\bibnamefont{Pascalutsa}},
  \bibinfo{author}{\bibfnamefont{V.}~\bibnamefont{Pauk}}, \bibnamefont{and}
  \bibinfo{author}{\bibfnamefont{M.}~\bibnamefont{Vanderhaeghen}},
  \bibinfo{journal}{Phys. Rev. D} \textbf{\bibinfo{volume}{85}},
  \bibinfo{pages}{116001} (\bibinfo{year}{2012})\relax
\mciteBstWouldAddEndPuncttrue
\mciteSetBstMidEndSepPunct{\mcitedefaultmidpunct}
{\mcitedefaultendpunct}{\mcitedefaultseppunct}\relax
\EndOfBibitem
\bibitem[{\citenamefont{Danilkin and Vanderhaeghen}(2017)}]{Danilkin:2016hnh}
\bibinfo{author}{\bibfnamefont{I.}~\bibnamefont{Danilkin}} \bibnamefont{and}
  \bibinfo{author}{\bibfnamefont{M.}~\bibnamefont{Vanderhaeghen}},
  \bibinfo{journal}{Phys. Rev. D} \textbf{\bibinfo{volume}{95}},
  \bibinfo{pages}{014019} (\bibinfo{year}{2017})\relax
\mciteBstWouldAddEndPuncttrue
\mciteSetBstMidEndSepPunct{\mcitedefaultmidpunct}
{\mcitedefaultendpunct}{\mcitedefaultseppunct}\relax
\EndOfBibitem
\bibitem[{\citenamefont{Dai and Pennington}(2017)}]{Dai:2017cvz}
\bibinfo{author}{\bibfnamefont{L.-Y.} \bibnamefont{Dai}} \bibnamefont{and}
  \bibinfo{author}{\bibfnamefont{M.}~\bibnamefont{Pennington}},
  \bibinfo{journal}{Phys. Rev. D} \textbf{\bibinfo{volume}{95}},
  \bibinfo{pages}{056007} (\bibinfo{year}{2017})\relax
\mciteBstWouldAddEndPuncttrue
\mciteSetBstMidEndSepPunct{\mcitedefaultmidpunct}
{\mcitedefaultendpunct}{\mcitedefaultseppunct}\relax
\EndOfBibitem
\bibitem[{\citenamefont{Aoyama et~al.}(2020)}]{Aoyama:2020ynm}
\bibinfo{author}{\bibfnamefont{T.}~\bibnamefont{Aoyama}} \bibnamefont{et~al.},
  \bibinfo{journal}{Phys. Rept.} \textbf{\bibinfo{volume}{887}},
  \bibinfo{pages}{1} (\bibinfo{year}{2020})\relax
\mciteBstWouldAddEndPuncttrue
\mciteSetBstMidEndSepPunct{\mcitedefaultmidpunct}
{\mcitedefaultendpunct}{\mcitedefaultseppunct}\relax
\EndOfBibitem
\bibitem[{\citenamefont{Danilkin et~al.}(2019)\citenamefont{Danilkin, Redmer,
  and Vanderhaeghen}}]{Danilkin:2019mhd}
\bibinfo{author}{\bibfnamefont{I.}~\bibnamefont{Danilkin}},
  \bibinfo{author}{\bibfnamefont{C.~F.} \bibnamefont{Redmer}},
  \bibnamefont{and}
  \bibinfo{author}{\bibfnamefont{M.}~\bibnamefont{Vanderhaeghen}},
  \bibinfo{journal}{Prog. Part. Nucl. Phys.} \textbf{\bibinfo{volume}{107}},
  \bibinfo{pages}{20} (\bibinfo{year}{2019})\relax
\mciteBstWouldAddEndPuncttrue
\mciteSetBstMidEndSepPunct{\mcitedefaultmidpunct}
{\mcitedefaultendpunct}{\mcitedefaultseppunct}\relax
\EndOfBibitem
\bibitem[{\citenamefont{Danilkin et~al.}(2013)\citenamefont{Danilkin, Lutz,
  Leupold, and Terschlusen}}]{Danilkin:2012ua}
\bibinfo{author}{\bibfnamefont{I.~V.} \bibnamefont{Danilkin}},
  \bibinfo{author}{\bibfnamefont{M.~F.~M.} \bibnamefont{Lutz}},
  \bibinfo{author}{\bibfnamefont{S.}~\bibnamefont{Leupold}}, \bibnamefont{and}
  \bibinfo{author}{\bibfnamefont{C.}~\bibnamefont{Terschlusen}},
  \bibinfo{journal}{Eur.Phys.J.} \textbf{\bibinfo{volume}{C73}},
  \bibinfo{pages}{2358} (\bibinfo{year}{2013})\relax
\mciteBstWouldAddEndPuncttrue
\mciteSetBstMidEndSepPunct{\mcitedefaultmidpunct}
{\mcitedefaultendpunct}{\mcitedefaultseppunct}\relax
\EndOfBibitem
\bibitem[{\citenamefont{Oller et~al.}(2008)\citenamefont{Oller, Roca, and
  Schat}}]{Oller:2007sh}
\bibinfo{author}{\bibfnamefont{J.~A.} \bibnamefont{Oller}},
  \bibinfo{author}{\bibfnamefont{L.}~\bibnamefont{Roca}}, \bibnamefont{and}
  \bibinfo{author}{\bibfnamefont{C.}~\bibnamefont{Schat}},
  \bibinfo{journal}{Phys. Lett.} \textbf{\bibinfo{volume}{B659}},
  \bibinfo{pages}{201} (\bibinfo{year}{2008})\relax
\mciteBstWouldAddEndPuncttrue
\mciteSetBstMidEndSepPunct{\mcitedefaultmidpunct}
{\mcitedefaultendpunct}{\mcitedefaultseppunct}\relax
\EndOfBibitem
\bibitem[{\citenamefont{Pennington}(2006)}]{Pennington:2006dg}
\bibinfo{author}{\bibfnamefont{M.~R.} \bibnamefont{Pennington}},
  \bibinfo{journal}{Phys. Rev. Lett.} \textbf{\bibinfo{volume}{97}},
  \bibinfo{pages}{011601} (\bibinfo{year}{2006})\relax
\mciteBstWouldAddEndPuncttrue
\mciteSetBstMidEndSepPunct{\mcitedefaultmidpunct}
{\mcitedefaultendpunct}{\mcitedefaultseppunct}\relax
\EndOfBibitem
\bibitem[{\citenamefont{Colangelo
  et~al.}(2017{\natexlab{b}})\citenamefont{Colangelo, Hoferichter, Procura, and
  Stoffer}}]{Colangelo:2017qdm}
\bibinfo{author}{\bibfnamefont{G.}~\bibnamefont{Colangelo}},
  \bibinfo{author}{\bibfnamefont{M.}~\bibnamefont{Hoferichter}},
  \bibinfo{author}{\bibfnamefont{M.}~\bibnamefont{Procura}}, \bibnamefont{and}
  \bibinfo{author}{\bibfnamefont{P.}~\bibnamefont{Stoffer}},
  \bibinfo{journal}{Phys. Rev. Lett.} \textbf{\bibinfo{volume}{118}},
  \bibinfo{pages}{232001} (\bibinfo{year}{2017}{\natexlab{b}})\relax
\mciteBstWouldAddEndPuncttrue
\mciteSetBstMidEndSepPunct{\mcitedefaultmidpunct}
{\mcitedefaultendpunct}{\mcitedefaultseppunct}\relax
\EndOfBibitem
\bibitem[{\citenamefont{Colangelo
  et~al.}(2017{\natexlab{c}})\citenamefont{Colangelo, Hoferichter, Procura, and
  Stoffer}}]{Colangelo:2017fiz}
\bibinfo{author}{\bibfnamefont{G.}~\bibnamefont{Colangelo}},
  \bibinfo{author}{\bibfnamefont{M.}~\bibnamefont{Hoferichter}},
  \bibinfo{author}{\bibfnamefont{M.}~\bibnamefont{Procura}}, \bibnamefont{and}
  \bibinfo{author}{\bibfnamefont{P.}~\bibnamefont{Stoffer}},
  \bibinfo{journal}{JHEP} \textbf{\bibinfo{volume}{04}}, \bibinfo{pages}{161}
  (\bibinfo{year}{2017}{\natexlab{c}})\relax
\mciteBstWouldAddEndPuncttrue
\mciteSetBstMidEndSepPunct{\mcitedefaultmidpunct}
{\mcitedefaultendpunct}{\mcitedefaultseppunct}\relax
\EndOfBibitem
\bibitem[{\citenamefont{Amaryan et~al.}(2020)}]{Amaryan:2020xhw}
\bibinfo{author}{\bibfnamefont{M.}~\bibnamefont{Amaryan}} \bibnamefont{et~al.},
  \bibinfo{journal}{arXiv: 2008.08215}  (\bibinfo{year}{2020})\relax
\mciteBstWouldAddEndPuncttrue
\mciteSetBstMidEndSepPunct{\mcitedefaultmidpunct}
{\mcitedefaultendpunct}{\mcitedefaultseppunct}\relax
\EndOfBibitem
\bibitem[{\citenamefont{Colangelo et~al.}(2014)\citenamefont{Colangelo,
  Hoferichter, Procura, and Stoffer}}]{Colangelo:2014dfa}
\bibinfo{author}{\bibfnamefont{G.}~\bibnamefont{Colangelo}},
  \bibinfo{author}{\bibfnamefont{M.}~\bibnamefont{Hoferichter}},
  \bibinfo{author}{\bibfnamefont{M.}~\bibnamefont{Procura}}, \bibnamefont{and}
  \bibinfo{author}{\bibfnamefont{P.}~\bibnamefont{Stoffer}},
  \bibinfo{journal}{JHEP} \textbf{\bibinfo{volume}{09}}, \bibinfo{pages}{091}
  (\bibinfo{year}{2014})\relax
\mciteBstWouldAddEndPuncttrue
\mciteSetBstMidEndSepPunct{\mcitedefaultmidpunct}
{\mcitedefaultendpunct}{\mcitedefaultseppunct}\relax
\EndOfBibitem
\bibitem[{\citenamefont{Pauk and Vanderhaeghen}(2014)}]{Pauk:2014rfa}
\bibinfo{author}{\bibfnamefont{V.}~\bibnamefont{Pauk}} \bibnamefont{and}
  \bibinfo{author}{\bibfnamefont{M.}~\bibnamefont{Vanderhaeghen}},
  \bibinfo{journal}{Phys. Rev.} \textbf{\bibinfo{volume}{D90}},
  \bibinfo{pages}{113012} (\bibinfo{year}{2014})\relax
\mciteBstWouldAddEndPuncttrue
\mciteSetBstMidEndSepPunct{\mcitedefaultmidpunct}
{\mcitedefaultendpunct}{\mcitedefaultseppunct}\relax
\EndOfBibitem
\end{mcitethebibliography}

\end{document}